\documentclass[a4paper,12pt]{article}
\usepackage{listings}
\usepackage[dvipsnames]{xcolor}
\usepackage{enumitem}

\usepackage{titlesec}
\titleformat{\section}
{\normalfont\fontsize{24}{15}\bfseries}{\thesection}{1em}{}

\titleformat{\subsection}
{\normalfont\fontsize{20}{15}\bfseries}{\thesubsection}{1em}{}

\titleformat{\subsubsection}
{\normalfont\fontsize{16}{15}\bfseries}{\thesubsubsection}{1em}{}

\usepackage{amsmath,amsfonts,amsthm,bm}

\newcommand{\myparagraph}[1]{\paragraph{#1}\mbox{}\\}

\usepackage[T1]{fontenc}
\usepackage[utf8]{inputenc}

\usepackage{helvet}

\usepackage[citestyle=numeric-comp,bibstyle=authoryear,giveninits=true,maxnames=15,sorting=none,isbn=false,url=false]{biblatex}
\DeclareNameAlias{author}{family-given} 
\AtEveryBibitem{\clearlist{language}}

\usepackage{xpatch}
\usepackage{xpatch}
\xpatchbibdriver{inproceedings}{\printfield{volume}}{}{}{}
\xpatchbibdriver{inproceedings}{\usebibmacro{chapter+pages}}{\printfield{volume}\newunit\newblock\usebibmacro{chapter+pages}}{}{}

\xpatchbibdriver{article}
  {\usebibmacro{title}%
   \newunit}
  {\usebibmacro{title}%
   \printunit{\addcomma\space}}
  {}
  {}
  
  \usepackage{xpatch}
\xpatchbibdriver{inproceedings}
  {\usebibmacro{title}%
   \newunit}
  {\usebibmacro{title}%
   \printunit{\addcomma\space}}
  {}
  {}

\usepackage{xpatch}
\xpatchbibmacro{volume+number+eid}{%
  \setunit*{\adddot}%
}{%
}{}{}

\DeclareFieldFormat[article]{number}{\mkbibparens{#1}}

\renewbibmacro*{journal+issuetitle}{%
  \usebibmacro{journal}%
  \setunit*{\addcomma\space}%
  \iffieldundef{series}
    {}
    {\newunit
     \printfield{series}%
     \setunit{\addspace}}%
  \usebibmacro{volume+number+eid}%
  \setunit{\addspace}%
  \usebibmacro{issue+date}%
  \setunit{\addcolon\space}%
  \usebibmacro{issue}%
  \newunit}

\renewbibmacro{in:}{}

\DeclareFieldFormat{labelnumberwidth}{\mkbibbrackets{#1}}
\defbibenvironment{bibliography}
  {\list
     {\printtext[labelnumberwidth]{%
    \printfield{prefixnumber}%
    \printfield{labelnumber}}}
     {\setlength{\labelwidth}{\labelnumberwidth}%
      \setlength{\leftmargin}{\labelwidth}%
      \setlength{\labelsep}{\biblabelsep}%
      \addtolength{\leftmargin}{\labelsep}%
      \setlength{\itemsep}{\bibitemsep}%
      \setlength{\parsep}{\bibparsep}}%
      }
  {\endlist}
  {\item}
\usepackage{xpatch}
\xpatchbibmacro{date+extrayear}{%
  \printtext[parens]%
}{%
  \setunit{\addperiod\space}%
  \printtext%
}{}{}

\newcommand*{\mkbiburlangle}[1]{Available at: #1}
\DeclareFieldFormat{url}{\mkbiburlangle{\url{#1}}}

\DeclareFieldFormat{urldate}{(Accessed: #1)}

\usepackage{graphicx}
\usepackage{pdfpages}
\usepackage{csquotes}
\usepackage[hidelinks, colorlinks=true,linkcolor=black,citecolor=blue]{hyperref}
\usepackage[english]{babel}
\usepackage{geometry}
\usepackage{amsmath}
\usepackage{enumerate}
\usepackage{fancyhdr}
\usepackage{xcolor}
\usepackage{booktabs}
\usepackage[justification=centering]{caption}
\usepackage{subcaption}
\usepackage{color}
\usepackage{soul}
\usepackage{wrapfig}
\usepackage{amssymb}
\usepackage{array}
\usepackage{float}
\usepackage{setspace}
\usepackage[export]{adjustbox}
\usepackage{multirow}
\usepackage{pdfpages}
\usepackage[toc,page,header]{appendix}
\usepackage{schemata}
\usepackage{adjustbox}
\usepackage{caption}
\usepackage{longtable}
\usepackage{cancel}
\usepackage{algorithm}
\usepackage[]{algpseudocode}
\usepackage{pdfpages}
\usepackage{rotating}

\usepackage{longtable}
\usepackage[intoc]{nomencl}
\makenomenclature
\usepackage{etoolbox}
\renewcommand\nomgroup[1]{%
  \item[\bfseries
  \ifstrequal{#1}{S}{Symbols}{%
  \ifstrequal{#1}{A}{Abbreviations}{%
  \ifstrequal{#1}{G}{Greek symbols}{}}}%
]}

\begin{document}

\renewcommand{\labelitemi}{$\circ$}
\renewcommand{\labelitemii}{$\circ$}

\pagestyle{fancy}
\fancyhead{} %limpio la cabecera
\fancyhead[L]{\setlength{\unitlength}{1in}
    \begin{picture}(0,0)
    \put(0,0){\includegraphics[width=0.05\textwidth]{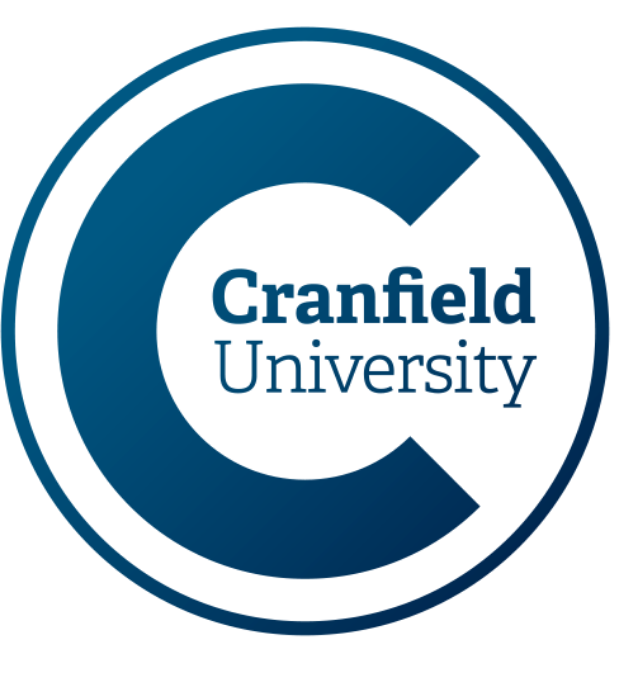}}
    \end{picture}}
\fancyhead[R]{\small{\textit {MSc. Computational Fluid Dynamics}}}
\renewcommand{\headrulewidth}{0.6mm}
\cfoot[]{\thepage}
\renewcommand{\footrulewidth}{0.2pt}
\renewcommand{\baselinestretch}{1.5}
%\maketitle

\begin{titlepage}
   \begin{center}
       \vspace*{0.5cm}
       
        \includegraphics[width=0.4\textwidth]{logo.PNG}
        
        \vspace*{1.0cm}
        \rule{\textwidth}{1pt}
        \vspace{0.1cm}
        
       \textbf{\Large{Assessment of CFD capability for prediction of the Coand\u{a} effect}}
       
       \vspace{0.3cm}
       \rule{\textwidth}{1pt}
 
       \vspace{1.5cm}
       
       \text{\large{SCHOOL OF AEROSPACE, TRANSPORT AND MANUFACTURING}}
       
       \vspace{0.5cm}
       
       \text{\large{MSc COMPUTATIONAL FLUID DYNAMICS}}
       
       \vspace{1.0cm}
       \text{\large{Academic Year: 2019-2020}}
      
       \vspace{2cm}
 
       \textbf{\large{AUTHOR: FLORENT MAURET}}
 
       \vspace{0.5cm}
       
       \textbf{\large{SUPERVISOR: Dr. TOM TESCHNER}}
       
       \vspace{0.5cm}
       
       \text{\large{August 27, 2020}}
       
       \vspace{0.5cm}
       
       \rule{\textwidth}{1pt}
       
       \vspace{0.5cm}
       
      \text{This thesis is submitted in partial fulfilment of the requirements for the degree of MSc}
      \vspace{0.5cm}
      
      \text{$\copyright$ Cranfield University 2020.  All rights reserved.  No part of this publication maybe}
      \text{reproduced without the written permission of the copyright owner.}

   \end{center}
\end{titlepage}

\pagenumbering{roman}

\clearpage
\section*{Abstract}
\addcontentsline{toc}{section}{Abstract}

    \noindent The tendency of a jet to stay attached to a flat or convex surface is called the Coand\u{a} effect and has many potential technical applications. The aim of this thesis is to assess how well Computational Fluid Dynamics can capture it. A Reynolds-Averaged Navier-Stokes approach with a 2-dimensional domain was first used to simulate an offset jet on a flat plane. Whether it was for $k-\omega$ SST or $k-\epsilon$ turbulence model, a good prediction of the flow was found. Since it is known that streamline curvature can have an important impact on the numerical results, a jet blown tangentially to a cylinder was then considered. Using the same approach as for the flat plane, with $k-\omega$ SST turbulence model, some of the flow features such as the separation location or velocity profiles near the jet exit were accurately predicted. However, the jet development was overall poorly captured. A Curvature Correction was then introduced in the turbulence model and if it did slightly improve the jet development, the negative impact on other quantities makes its benefits questionable. Due to the known presence of longitudinal and spanwise vortices in the flow, a Reynolds-Averaged Navier-Stokes approach with a 3-dimensional domain was attempted but was only able to reproduce the 2-dimensional results. Although if the longitudinal vortices are artificially generated at the inlet, their development is supported by the simulation. Finally, since the shortcomings of the numerical results obtained might be due to limitations of the Reynolds-Averaged Navier-Stokes approach, a Large Eddy Simulation was attempted. Unfortunately, due to time and computational restrictions a fully developed flow could not be obtained, the methodology and preliminary results are however presented.
\\
\\
\emph{\textbf{Keywords -} CFD, Coand\u{a} effect, cylinder, Reynolds-Averaged Navier Stokes (RANS), Large Eddy Simulation (LES)}

\clearpage
\section*{Acknowledgements}
\addcontentsline{toc}{section}{Acknowledgements}
I would like to sincerely thank my supervisor Dr. Tom Teschner for is support, guidance and availability throughout this thesis. I also thank my course director Dr. Zeeshan Rana, as well as, all the teaching and support staff from Cranfield University that accompanied me during this master. I would also like to extend my gratitude to Mrs. Sandra Durand, head of international relationships at Polytech Montpellier, that made this year possible. Last but not least, I am grateful for my family, my friends and my girlfriend for their continued support.

\clearpage
\begin{center}
\tableofcontents    
\end{center}

\nomenclature[A]{CC}{Curvature Correction}
\nomenclature[A]{CFD}{Computational Fluid Dynamics}
\nomenclature[A]{RANS}{Reynolds-Averaged Navier–Stokes
}
\nomenclature[A]{DES}{Detached Eddy Simulation
}
\nomenclature[A]{LES}{Large Eddy Simulation
}
\nomenclature[A]{HPC}{High Performance Computing
}
\nomenclature[A]{GCI}{Grid Convergence Index
}
\nomenclature[A]{DNS}{Direct Numerical Simulation
}
\nomenclature[A]{SIMPLE}{Semi-Implicit Method for Pressure Linked Equations}
\nomenclature[A]{SIMPLEC}{Semi-Implicit Method for Pressure Linked Equations-Consistent}
\nomenclature[A]{PISO}{Pressure-Implicit with Splitting of Operators}
\nomenclature[A]{SST}{Shear Stress Transport}

\nomenclature[G]{$\theta$}{Angle around the cylinder [$^{\circ}$]}
\nomenclature[G]{$\rho$}{Density
[$kg.m^{-3}$]}
\nomenclature[S]{$Re$}{Reynolds number [-]}
\nomenclature[S]{$C_p$}{Pressure coefficient [-]}
\nomenclature[S]{$C_f$}{Skin friction coefficient [-]}
\nomenclature[S]{$GCI_{i,i+1}$}{Grid convergence index between mesh i and i+1 [-]}
\nomenclature[S]{$q$}{Order of grid convergence [-]}
\nomenclature[S]{$r$}{Refinement ratio [-]}
\nomenclature[S]{$\mathbf{x}$}{Cartesian coordinates vector $(x,y,z)$ [$m$]}
\nomenclature[S]{$p$}{Pressure [$Pa$]}
\nomenclature[S]{$y^+$}{Non-dimensional spacing normal to the wall [-]}
\nomenclature[S]{$t$}{Time [$s$]}
\nomenclature[S]{$k$}{Turbulent kinetic energy [$m^2.s^{-2}$]}
\nomenclature[S]{$M$}{Mach number [-]}
\nomenclature[S]{$\mathbf{u}$}{Velocity vector $(u,v,w)$ [$m.s^{-1}$]}
\nomenclature[S]{$\mathbf{\overline{u}}$}{Filtered velocity vector (LES) / Mean velocity vector (RANS)  [$m.s^{-1}$]}
\nomenclature[S]{$\mathbf{u'}$}{Subgrid-scale velocity vector (LES) / Fluctuating velocity vector (RANS) [$m.s^{-1}$]}
\nomenclature[G]{$\nu$}{Kinematic viscosity [$m^2.s^{-1}$]}
\nomenclature[S]{$S_{ij}$}{Strain of rate tensor [$s^{-1}$]}
\nomenclature[G]{$\Delta$}{Filter width [$m$]}
\nomenclature[G]{$\tau_{ij}^S$}{Subgrid-scale stress tensor [$m^2.s^{-2}$]}
\nomenclature[S]{$k_{sgs}$}{Subgrid-scale turbulent kinetic energy [$m^2.s^{-2}$]}
\nomenclature[G]{$\nu_t$}{Turbulent kinematic viscosity [$m^2.s^{-1}$]}
\nomenclature[G]{$\epsilon_{sgs}$}{Subgrid-scale turbulence dissipation rate [$m^2.s^{-3}$]}
\nomenclature[G]{$\epsilon$}{Turbulence dissipation rate [$m^2.s^{-3}$]}
\nomenclature[G]{$\mu$}{Dynamic viscosity [$Pa.s$]}
\nomenclature[G]{$\mu_t$}{Turbulent dynamic viscosity [$Pa.s$]}
\nomenclature[G]{$\Omega_{ij}$}{Vorticity tensor [$s^{-1}$]}
\nomenclature[S]{$n$}{Normal to a cell face [-]}
\nomenclature[S]{$n$}{Normal to a cell face [-]}
\nomenclature[S]{$\mathbf{\dot{m}}$}{Mass flow rate vector [$kg.s^{-1}$]}
\nomenclature[S]{$H_j$}{Inlet height for flat plate  [$m$]}
\nomenclature[S]{$H_s$}{Inlet offset from bottom wall [$m$]}
\nomenclature[S]{$X_r$}{Reattachment length of bottom wall [$m$]}
\nomenclature[S]{$U_{jet}$}{Jet exit velocity [$m.s^{-1}$]}
\nomenclature[S]{$I$}{Turbulence intensity [-]}
\nomenclature[S]{$L$}{Turbulence length scale [$m$]}
\nomenclature[G]{$\tau_{\omega}$}{Streamwise wall shear stress [$Pa$]}
\nomenclature[S]{$U_{max}$}{Maximum streamwise velocity on the line normal to the wall [$m.s{-1}$]}
\nomenclature[S]{$y_2$}{Jet half width [$m$]}
\nomenclature[S]{$b$}{Inlet height for cylinder [$m$]}
\nomenclature[S]{$R$}{Cylinder radius [$m$]}
\nomenclature[G]{$\Delta x$}{Spacing in streamwise direction [-]}
\nomenclature[G]{$\Delta x+$}{Non dimensional spacing in streamwise direction [-]}
\nomenclature[G]{$\Delta z+$}{Non dimensional spacing in spanwise direction [$m$]}
\nomenclature[S]{$U$}{Streamwise velocity [$m.s^{-1}$]}
\nomenclature[S]{$R_t$}{Turbulent Reynolds number [-]}

\clearpage
\addcontentsline{toc}{section}{List of Figures}
\listoffigures

\clearpage

\addcontentsline{toc}{section}{List of Tables}
\listoftables

\clearpage
\printnomenclature[25mm]

\newpage

\pagenumbering{arabic}
%%%%%%%%%%%%%%%%%%%%%%%%%%%%%%%%%%%%%%%%%%%%%%%%%%%%%%%%%%%%%%%%%%%%
\section{Introduction and objectives}
\label{sec:intro}
\begin{figure}[h!]
\centering
\begin{subfigure}{0.49\textwidth}
  \centering
  \includegraphics[width=1\textwidth]{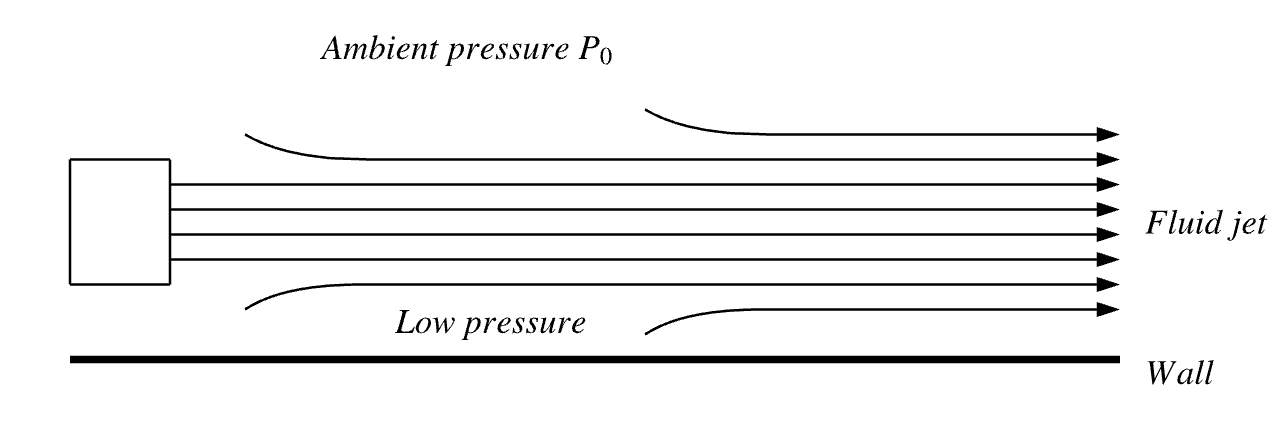}
  \caption{Low pressure region generation}
  \label{fig:principle_Coanda_a}
\end{subfigure}
\begin{subfigure}{0.49\textwidth}
  \centering
  \includegraphics[width=1\textwidth]{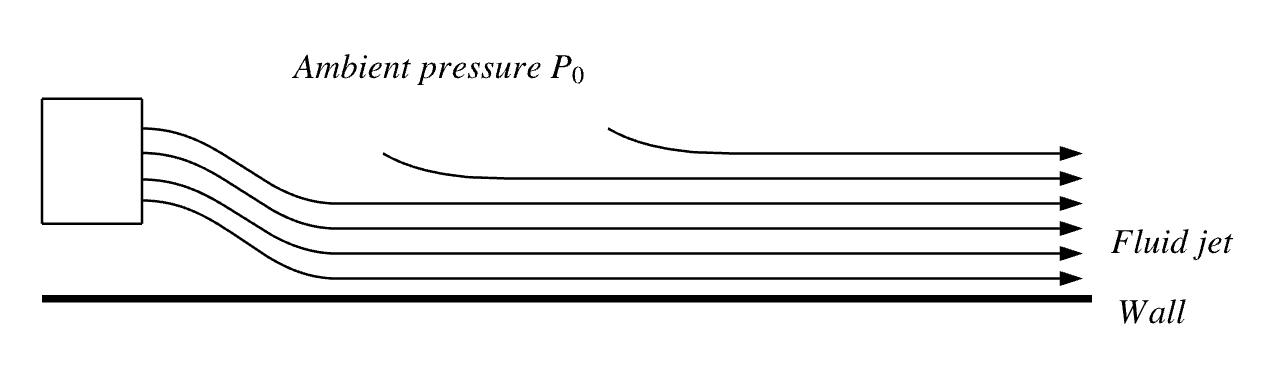}
  \caption{Attachment of fluid to the wall}
  \label{fig:principle_Coanda_b}
\end{subfigure}%

\caption{Illustration of the physical mechanisms behind the Coand\u{a} effect}
\label{fig:principle_Coanda}
\end{figure}

The Coand\u{a} effect can be defined as the tendency of a fluid jet to follow a flat or convex wall when blown close to them. This effect is named after the engineer Henri Coand\u{a} who was the first to recognize the potential of that effect in term of aircraft design \cite{coanda_procede_1934}. The physical mechanisms responsible for it are as follows. A fluid jet entrains the fluid in its immediate surroundings due to the effect of viscosity. When considering a jet blown close to a flat wall, this entrainment is restricted on the side of the wall due to the proximity between the two. This creates a difference of pressure, with a low-pressure region created between the wall and the jet as shown in Figure \ref{fig:principle_Coanda_a}. To accommodate that difference, the jet will bend and then adhere to the nearby wall surface as illustrated in Figure \ref{fig:principle_Coanda_b}. The effect is even more important when considering a curved surface as every small change in curvature will again create a difference of pressure through the same process.\\
\\

The jet deflected by the Coand\u{a} effect has the capability to suck in large quantities of fluids from the surroundings, up to 20 times the amount of fluid of the initial jet according to Reba \cite{reba_applications_1966}. In addition to that, the static wall pressure under the attached jet is actually below the ambient pressure and therefore generates aerodynamic forces as described by Gross and Fasel \cite{gross_coanda_2006}. These two properties have motivated the development of a wide range of applications making use of this effect. A famous example is the no tail rotor (NOTAR) helicopter that makes use of a Coand\u{a} flow  to generate the aerodynamic forces normally generated by the tail rotor and can, therefore, function without one, permitting an increased security and reducing noise. Another successful use of the Coand\u{a} effect is for Circulation Control Airfoils, where a Coand\u{a} flow is used to give an increased lift at low speed and therefore a better manoeuvrability. More information on these two aerospace applications, but also on the use of the Coand\u{a} effect in different fields such as medicine or air conditioning is given in the literature review Section \ref{subsec:applications}.\\
\\
To improve these applications, or develop new ones a better understanding of Coand\u{a} flows is still needed. Computational Fluid Dynamics (CFD) is a powerful tool that could permit that. In fact, the Navier-Stokes equations are capable of describing the motion of fluids, they can however only be solved analytically for really simple flows. By solving these governing equations for an ensemble of nodes on a grid and using the information from neighbour cells, CFD is theoretically capable of simulating all flows. This is already an advantage over experimental approach, were many flows might be difficult or impossible to set-up. In addition to that, a CFD simulation is generally a lot cheaper to carry out than an experiment and can easily be modified to test a wide range of parameters and flow conditions. However, when considering a complex flow and if high fidelity methods are used, getting results might take years, even with the current capabilities of high-performance computing (HPC). On the other hand, if too simple numerical models or too coarse grids are used the results might be disappointing, only capturing a few of the flows characteristics if any. The focus is therefore on finding a balance between computational cost and accuracy.\\
\\
The Coand\u{a} effect is a complex phenomenon and has been proven difficult to capture accurately using CFD tools, this will be further discussed in the literature review Section \ref{sec:literature}. The aim of this thesis is to investigate what models and set-ups are capable of capturing a Coand\u{a} flow accurately. First, a rather simple test case will be considered, with a jet blown on an offset flat plane, reproducing the experiment from Gao and Ewing \cite{gao_experimental_2007}. The flow will be solved using a 2 dimensional domain and solving the RANS equations. That configuration is a good point to start the investigation, but most applications using the Coand\u{a} effect such as the NOTAR helicopter or the Circulation Control Airfoil are using jets blown on curved surfaces. The experiment of Wygnanski et al. \cite{neuendorf_turbulent_2000,neuendorf_turbulent_1999, likhachev_streamwise_2001,cullen_role_2002, han_streamwise_2004, neuendorf_large_2004} considering a jet blown tangentially to a cylinder will therefore be replicated, again using a 2D RANS simulation. This is important since the streamline curvature in the flow might have a considerable impact on the accuracy of the numerical results. Since the presence of 3 dimensional structures in the flow around the cylinder has been found experimentally by Han et al. \cite{han_streamwise_2004}, the RANS equations will also be solved for a 3D domain in an attempt to improve the results obtained. Finally, still in the objective to improve the accuracy of the solution a LES simulation was attempted. Unfortunately, due to time and computational limitations a fully developed and statistically stationary flow could not be obtained, the methodology and preliminary results are however presented as they could be useful for eventual further work on the subject. All the simulations presented were done using the CFD solver OpenFOAM\textsuperscript{\textregistered}.\\
\\
The thesis will be decomposed in the following way. First, a literature review will be given on the different applications of the Coand\u{a} effect and the attempt both experimental and numerical to capture that effect. Then the governing equations for the different models used will be detailed. Following that the different numerical methods and schemes employed will be discussed, an introduction on OpenFoam will also be given, as well as, a discussion on how the available HPC capabilities were used. Finally, the results of the simulations will be presented and discussed.

\newpage
\section{Literature review}
\label{sec:literature}
This section gives a sample of applications making use of Coand\u{a} flows. Then an overview of the experimental and numerical results for a Coand\u{a} flow around curved surfaces are given. The choice was made to focus on curved surfaces, as they are the most commonly used for applications and are also the most challenging to simulate accurately. 

\subsection{Applications}
\label{subsec:applications}

\begin{figure}[h!]
\centering
\begin{subfigure}{0.6\textwidth}
  \centering
  \includegraphics[width=1\textwidth]{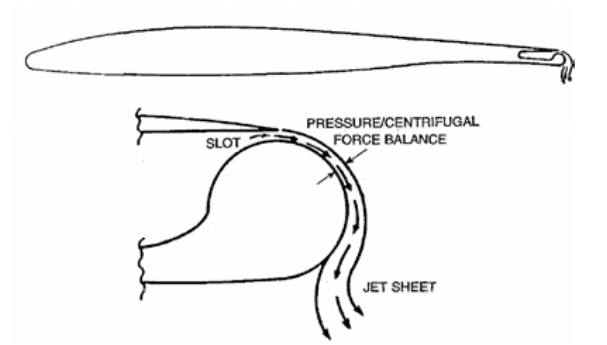}
  \label{fig:sub1}
\end{subfigure}%

\caption{2D slice of a basic Circulation Control Airfoil with zoom on the trailing edge (Reproduced from Englar \cite{englar_overview_2006})}
\label{fig:control_airfoil}
\end{figure}

\begin{figure}[h!]
\centering
\begin{subfigure}{0.69\textwidth}
  \centering
  \includegraphics[width=1\textwidth]{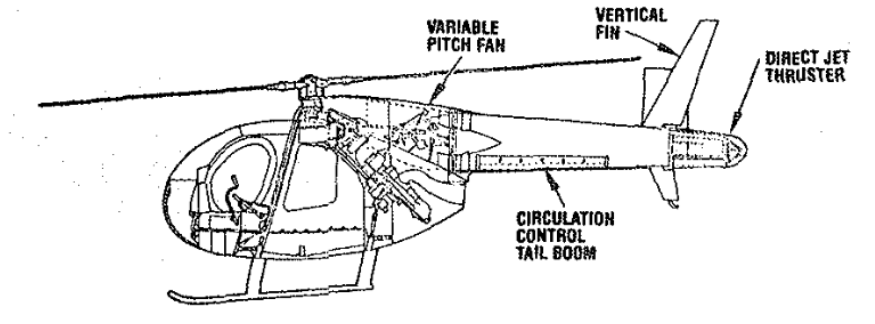}
  \label{fig:sub1}
\end{subfigure}%
\vspace{0.5mm}
\begin{subfigure}{0.29\textwidth}
  \centering
  \includegraphics[width=1\textwidth]{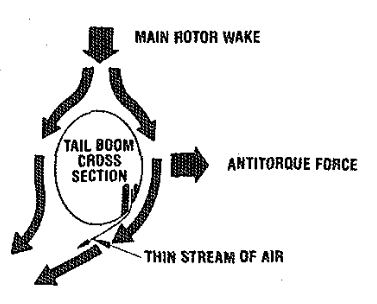}
  \label{fig:sub2}
\end{subfigure}

\caption{The no tail rotor helicopter (Reproduced from Sampatacos et al. \cite{sampatacos_notar_1983})}
\label{fig:NOTAR}
\end{figure}

As was discussed in the Introduction Section \ref{sec:intro}, the Coand\u{a} effect has been used in a wide variety of fields. This section aims at highlighting a few of them and describing what the effect permits to achieve. This section is far from exhaustive, for a more in-depth review, the paper from Lubert \cite{lubert_recent_2010} can be consulted.\\
\\
Let's first consider an aerospace application as it is one of the most significant domains where the Coand\u{a} effect has been used due to its capability to generate an increased lift. This is, for example, the case of a Circulation Control Airfoil, where a jet is blown over a rounded trailing edge and will, therefore, thanks to the Coanda effect "bent down" the air around the airfoil and result in aerodynamic lift. This increase in lift permitted by a blowing jet has been demonstrated experimentally \cite{abramson_two-dimensional_1977}. The main purpose of this device is to increase the maximum lift at low speed which gives increased manoeuvrability, as well as, offering the possibility to reduce take-off and landing speed. The Figure \ref{fig:control_airfoil} gives a 2D slice of a Circulation Control Airfoil, with a zoom on the trailing edge to show how the jet behaves. 
\\
The other example of application presented in the introduction is the NOTAR helicopter, that uses a Coand\u{a} flow to replace the tail rotor. In this kind of helicopter, an air jet is blown tangentially to the cylinder tail boom, thanks to the Coand\u{a} effect, this jet will adhere to the surface of the cylinder generating aerodynamic forces. These aerodynamic forces will produce the anti-torque effect normally given by the tail rotor. An illustration of that design is given in Figure \ref{fig:NOTAR}. The absence of a tail rotor permits a better safety and reduces the noise produced by the helicopter as discussed by Sampatacos et al. \cite{sampatacos_notar_1983}, unfortunately this comes at the cost of an increased complexity.\\

\newpage

\begin{figure}[h!]
\centering
\begin{subfigure}{0.9\textwidth}
  \centering
  \includegraphics[width=1\textwidth]{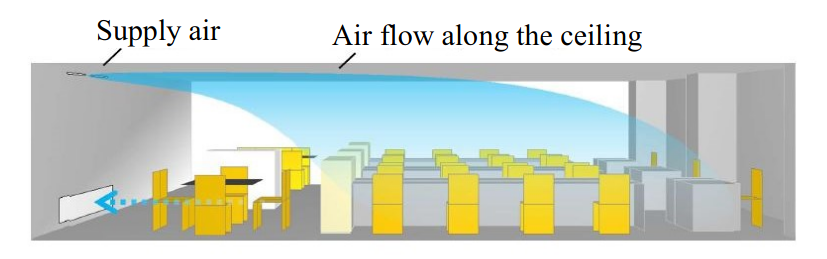}
  \label{fig:sub1}
\end{subfigure}%
\vspace{0.5mm}

\caption{Overview of a ductless air conditioning system (Reproduced from Igarashi et al. \cite{igarashi_effects_2019})}
\label{fig:aircond}
\end{figure}

In addition to aerospace, the Coand\u{a} effect finds applications in a lot of other fields. Air conditioning is an example of that, with some air conditioning systems making a beneficial use of it, as illustrated in Figure \ref{fig:aircond}. It can be seen that by blowing air close to the ceiling surface, the air will adhere to the ceiling thanks to the Coand\u{a} effect and will not need the use of duct to be conveyed around the room. This will save energy as the air will no longer have to overcome the frictional resistance inside the duct. 

All the applications presented so far are benefiting from the Coand\u{a} effect, however, this is not always the case. It is sometimes important to take into account the Coand\u{a} effect in order to avoid detrimental effects. This is, for example, the case for Color Doppler flow mapping, an imaging technique used in cardiology. Color Doppler is principally used to observe valvular regurgitations and other abnormal flows and assess the severity of the lesions that causes them. It has been argued that the Coand\u{a} effect "needs to be taken into account for an appropriate echocardiographic assessment" \cite{ginghina_coanda_2007}. In fact, the diagnostic is done by doing a jet area measurement, but depending on various factors (the jet location, the adjacent surface angle, the distance between that surface and the jet exit...), the Coand\u{a} effect might influence the jet size and therefore should be considered.

\subsection{Experimental results on the Coand\u{a} effect}
\begin{figure}[h!]
\centering
\includegraphics[width=0.6\textwidth]{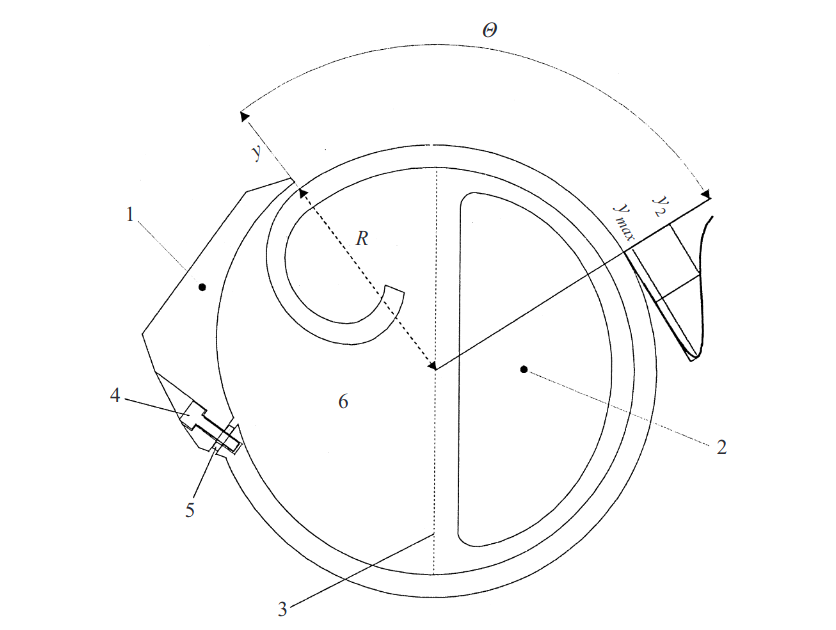}
\caption{Cross-section of the cylinder (Reproduced from Neuendorf et al. \cite{neuendorf_turbulent_1999})}
\label{fig:wygnanski}
\end{figure}

Many experimental studies have been carried out in relation with the Coand\u{a} effect over convex surfaces, both for a better understanding of a Coand\u{a} flow \cite{wilson_turbulent_1976,fujisawa_turbulence_1987} or to help the development of new applications making use of that effect \cite{oshima_experimental_1975}.\\
The most extensive experimental results focusing on the Coand\u{a} effect on convex surfaces are from Wygnanski et al. \cite{neuendorf_turbulent_2000,neuendorf_turbulent_1999,neuendorf_turbulent_1999, likhachev_streamwise_2001,cullen_role_2002, han_streamwise_2004, neuendorf_large_2004}. They carried out experiments on a wall jet flowing over a circular cylinder for different slot height and Reynolds number. At the difference from previous studies, they used a jet coming from inside the cylinder which is a lot more common in real-life applications. The device used for carrying out the experiments is presented in Figure \ref{fig:wygnanski}. They found a separation at around $\theta=230^{\circ}$. This separation was mainly caused by the entrainment of surrounding fluid by the jet, which appears to be responsible for both the Coand\u{a} effect and the separation. The flow presented two main regions of interest, the first one going up to $\theta=120^{\circ}$ was characterized by a constant surface pressure. The second region, going, up to the separation, had an adverse pressure gradient. In \cite{likhachev_streamwise_2001,han_streamwise_2004,cullen_role_2002,neuendorf_large_2004} the fact that the flow is not 2-dimensional was also demonstrated, with large streamwise vortices found, that grows when getting further away from the nozzle, as well as, spanwise vortices present around the outer shear layer of the jet.

The configuration considered is a good test case of a Coand\u{a} flow and the data from Wygnanski et al. \cite{neuendorf_turbulent_2000,neuendorf_turbulent_1999, likhachev_streamwise_2001,cullen_role_2002, han_streamwise_2004,neuendorf_large_2004} makes it well suited to use as a validation to test the capability of different CFD models to capture the flow accurately.

\subsection{Numerical results on the Coand\u{a} effect}
\begin{figure}[h!]
\centering
\includegraphics[width=8cm]{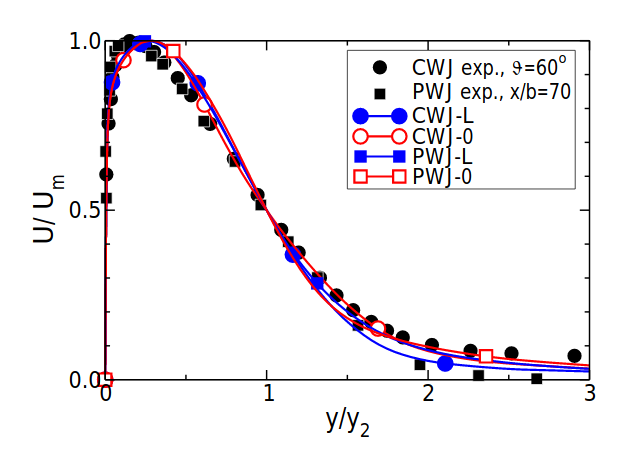}
\caption{Normalized mean streamwise velocity profile of DNS results, compared to experimental data at $\theta=60^{\circ}$ (Reproduced from Wernz et al. \cite{wernz_numerical_2005})}
\label{fig:DNS_velocity}
\end{figure}

In order to assess the capability of CFD models to capture the Coand\u{a} effect, various attempts have been made to reproduce the experimental results from Wygnanski et al. \cite{neuendorf_turbulent_2000,neuendorf_turbulent_1999, likhachev_streamwise_2001,cullen_role_2002, han_streamwise_2004,neuendorf_large_2004}, they will be reviewed here.\\
\\
Wernz et al. \cite{wernz_numerical_2003,wernz_numerical_2005}  attempted to reproduce the experimental results by using Direct Numerical Simulation (DNS). Promising results were obtained, with the presence of both spanwise and streamwise coherent structures found in the flow, as was observed in the experiment. In addition to that, the mean flow data were in good agreement with the experimental data as can, for example, be seen for the streamwise velocity at $\theta=60^{\circ}$ presented in Figure \ref{fig:DNS_velocity}. In this Figure, the line named CWJ exp is the experimental results found by Neuendorf and Wygnanski \cite{neuendorf_turbulent_1999}, while CWJ-L and CWJ-0 are DNS results for the flow around the cylinder with different forcing methods.

\begin{figure}[h!]
\centering

\begin{subfigure}{0.49\textwidth}
  \centering
  \includegraphics[width=1\textwidth]{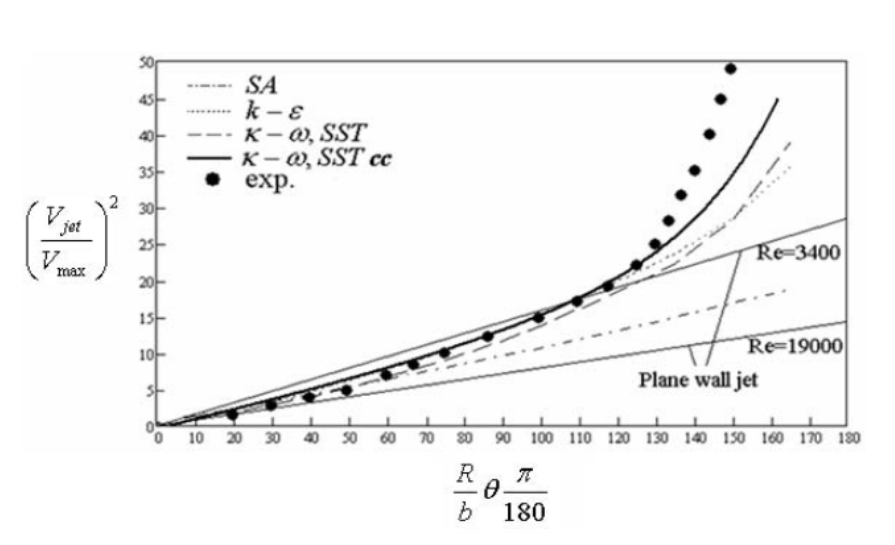}
  \caption{Jet velocity decay}
  \label{fig:sub1}
\end{subfigure}%
\vspace{0.5mm}
\begin{subfigure}{0.49\textwidth}
  \centering
  \includegraphics[width=1\textwidth]{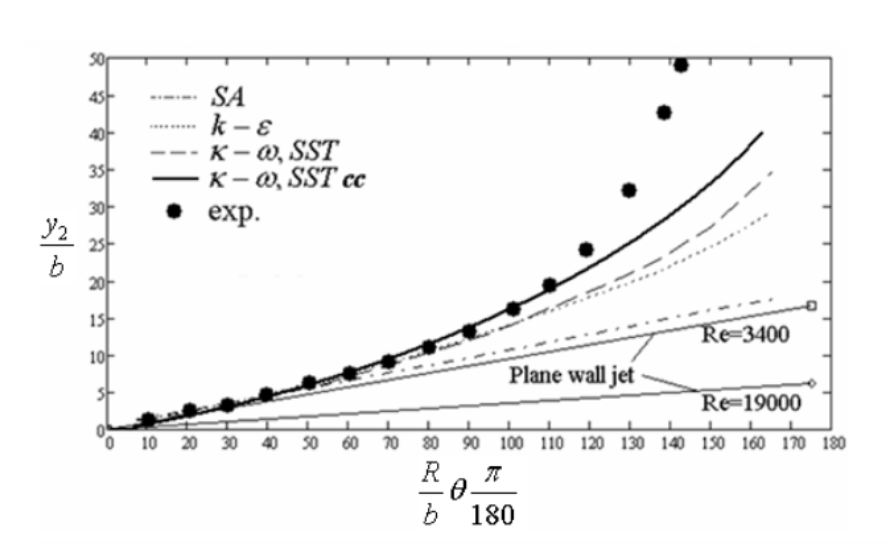}
  \caption{Jet spreading rate}
  \label{fig:sub2}
\end{subfigure}

\caption{Jet velocity decay and jet spreading rate, for various RANS simulations compared to experimental result (Reproduced from Frunzulica et al. \cite{frunzulica_method_2017})}
\label{fig:RANS_decay_review}
\end{figure}

If these results are promising, this methodology presents several flows. First, due to the prohibitive cost of DNS, only a part of the cylinder could be simulated and therefore the separation and the region around it could not be obtained, when it is one of the most challenging regions to capture accurately. In addition to that, some discrepancies can be observed in the velocity profile given Figure \ref{fig:DNS_velocity}, this could be due to many reasons, but the most significant might be that the mesh requirements for DNS could not be respected even for a small portion of the cylinder, and as was admitted by the authors: "the present Navier-Stokes simulations still do not qualify as fully resolved Direct Numerical Simulations and fall in the category of ”coarse-grid direct numerical simulations”" \cite{wernz_numerical_2005}. To conclude, if DNS is giving promising results, the prohibitive cost of such method makes it impossible to use for nearly all real-life applications, the need for a cheaper method is therefore needed.\\
\\
In order to reduce computational cost, Wernz et al.  \cite{wernz_numerical_2005}, attempted to copy the experiment by using a Large Eddy Simulation (LES) using the FSM methodology \cite{fasel_methodology_2002}. If some of the flow characteristics were captured by the simulation, the separation was predicted too early at $\theta=125^{\circ}$ against $\theta=220^{\circ}$ for the experiment. For the authors, this discrepancy might, at least partially, be due to a too thin spanwise extent.\\
Even if carried out on a different geometry, more recent LES have found good agreement for Coand\u{a} flow against experimental data. The work from Nishino et al. \cite{nishino_large-eddy_2010} can be cited, where good agreement against experiment was found for a Circulation Control Airfoil, showing the potential of that method. It, therefore, appears that LES might be a good alternative to solve Coand\u{a} flows, however even LES is generally too costly for most industrial applications due to the high Reynolds numbers and complex geometries involved. \\
\\
To considerably reduce computational cost Reynolds-Averaged Navier-Stokes (RANS) equations are an alternative and are actually the standard for most industrial applications. Gross and Fasel \cite{gross_coanda_2006} and Frunzulica et al. \cite{frunzulica_method_2017} used 2D RANS to simulate the jet experiment for a variety of turbulence models ($k - \omega$, $k - \omega$ SST, $k - \epsilon$, Spalart-Allmaras...). Important differences were found depending on the model employed, but $k-\omega$, $k-\omega$ SST and $k-\omega$ SST with Curvature Correction were capable of finding relatively accurate separation location and normalized velocity profiles. However none of the 2D RANS simulations seemed capable of accurately capturing the jet development as illustrated by the plot around the cylinder of the jet velocity decay and the jet spreading rate given Figure \ref{fig:RANS_decay_review}. A possible explanation of that shortcoming of RANS might the use of a 2D domain when the existence of 3D structures is demonstrated experimentally \cite{neuendorf_large_2004}. Gross and Fasel \cite{gross_coanda_2006} did use 3D RANS to study the evolution of streamwise vortical structures. However these structures were artificially generated at the nozzle exit, therefore, if the simulation is helpful to understand how these structures behave in a Coand\u{a} flow, it is not possible to compare the results to the experimental data from Wygnanski et al. \cite{neuendorf_turbulent_2000,neuendorf_turbulent_1999, likhachev_streamwise_2001,cullen_role_2002, han_streamwise_2004,neuendorf_large_2004}.

Other studies have tried different geometries than a cylinder with for example Mirkov and Rasuo \cite{mirkov_numerical_2015} using RANS equations with $k - \omega$ SST turbulence model to simulate the wall jet over a convex surface with non-constant curvature. This is of great interest for better understanding the Coand\u{a} effect, but also for applications that may benefit from using more complex geometries than a cylinder. However, the absence of experimental data for such geometries and the demonstrated difficulty of RANS equations to capture Coand\u{a} flows over convex surfaces accurately makes interpreting these results difficult.

\newpage
\section{Governing equations}

\subsection{Navier-Stokes equations}
The governing equations of fluid flows consist of the conservation laws for mass, momentum and energy. If the flow is Newtonian, which is widely regarded as true for the air considered in this thesis, then these equations are called Navier-Stokes equations. Since relatively low Mach number ($M<0.3$) are considered, the incompressible hypothesis can be done. This assumption permits to simplify the equations reflecting the conservation of mass and momentum and the one for the conservation of energy is actually only useful if the temperature field is needed which will not be the case here. Having only one phase and an incompressible flow also allows to neglect gravitational effects and there is no other source term considered. The following system of equation is, therefore, governing the flows simulated in this thesis \cite{wilcox_turbulence_2006}

\begin{equation}
    \frac{\partial u_i}{\partial x_i}=0,
    \label{equ:continuity}
\end{equation}

\begin{equation}
    \frac{\partial u_i}{\partial t}+u_j \frac{\partial u_i}{\partial x_i}=-\frac{1}{\rho}\frac{\partial p}{\partial x_i}+2\nu \frac{\partial S_{ij}}{\partial x_j}.
    \label{equ:momentum}
\end{equation}

With $S_{ij}$ being the strain of rate tensor,

\begin{equation}
    S_{ij}=\frac{1}{2}\left(\frac{\partial u_i}{\partial x_j}+\frac{\partial u_j}{\partial x_i}\right).
\end{equation}

\subsection{Turbulence modelling}
Turbulence is a complex phenomenon that occurs in a flow when a critical Reynolds number is exceeded. An exact definition of the phenomenon is difficult to give, the one from Hinze is generally considered and is as follows: "Turbulent fluid motion is an irregular condition of flow in which the various quantities show a random variation with time and space coordinates, so that statistically distinct average values can be discerned." \cite{hinze_turbulence_1975}.\\

Due to the omnipresence and impact of turbulence in nearly all practical flows, simulating them is important, but due to the complexity of the phenomenon it is a challenging part of CFD. In this section, a review of the three most common approaches (DNS, LES and RANS) will be briefly explained, with a particular focus on the models that have been used in this report.

\subsubsection{Direct Numerical Simulation (DNS)}
Direct Numerical Simulation is the method with the highest order of fidelity, but also the simplest approach conceptually. For this method, no turbulence model is used and the Navier-Stokes equations are simply solved in the three spatial dimensions and in time. If done properly, without numerical error and with perfect boundary conditions, this method should be capable to give the proper solution to the turbulence problem. \\
\\
However, this method generally comes at an extensive computational cost. In fact, the grid should be large enough to capture the largest eddies of the flow (that are comparable in size  to the characteristic flow scale), with cells small enough to capture the eddies at the Kolmogorov scale. The Kolmogorov scale is the smallest scale of eddies that can exist before the viscosity effectively dissipates them. The important ratio between these two scales is such that the point requirement for DNS is generally considered to be proportional to $Re_{Lf}^{\frac{9}{4}}$ \cite{pope_turbulent_2000}, with $Lf$ being the integral length scale, an approximation of the length of the largest eddies. In addition to that, to be accurate in time, the movement of a particle during a time step should be relatively small when compared to the grid spacing. Therefore, due to the small grid spacing used, the time step should also remain small.\\
\\
The need for really fine grid and small time steps makes DNS unsuitable to solve most engineering problems, that are often characterized by complex geometries and high Reynolds numbers. Instead, the method is generally used for simple geometries and low Reynolds numbers to get a better understanding of turbulent structures or as a source of data to validate other models.

\subsubsection{Large Eddy Simulation (LES)}
In a turbulent flow, the largest eddies are the ones containing the most energy and anisotropy. The Large Eddy Simulation methodology was therefore developed, were the largest eddies are directly computed while the smallest eddies, less critical, are modelled. Since the smallest eddies are modelled, the cells and by extension the time steps can be a lot bigger than what is used for DNS, considerably reducing the computational cost. To give an order of magnitude of how much time can be saved, the example of a flow over a backward-facing step, with a Reynolds number based on the step height of $5000$ is given by Wilcox \cite{wilcox_turbulence_2006}. With a similar agreement to the experiment, only $3\%$ of the grid points and $2\%$ of the computer time was needed for the LES when compared to the DNS. \\
\\
In order to separate the large and small eddies, a filtering is applied to all the flow variables that become the sum of their large and small scale contributions. By taking, for example, the velocity vector, with $\mathbf{\overline{u}}$ the filtered part and $\mathbf{u'}$ the subgrid-scale one, the following equation is obtained

\begin{equation}
    \mathbf{u}=\mathbf{\overline{u}}+\mathbf{u'}.
\end{equation}

The filtering process can be written as

\begin{equation}
    \mathbf{\overline{u}}(x,t)=\int G(\mathbf{x}-\bm{\xi},\Delta)\mathbf{u}(\bm{\xi},t)d\bm{\xi}.
\end{equation}

With G being the filter function, depending on the filter width $\Delta$ and that needs to respect the condition

\begin{equation}
    \int G(\mathbf{x}-\bm{\xi},\Delta)d\bm{\xi}=1.
\end{equation}

Various functions G exist, with the most common and used in OpenFOAM\textsuperscript{\textregistered} being the top hat function.\\
\\
This filtering is then applied to the Navier Stokes Equations (\ref{equ:continuity},\ref{equ:momentum}) and gives

\begin{equation}
    \frac{\partial \overline{u}_i}{\partial x_i}=0,
\end{equation}

\begin{equation}
    \frac{\partial \overline{u}_i}{\partial t}+\frac{\partial \overline{u}_i\overline{u}_j}{\partial x_j}=-\frac{1}{\rho}\frac{\partial \overline{p}}{\partial x_i}+\nu\frac{\partial^2 \overline{u}_i}{\partial x_jx_j}-\frac{\partial \tau^S_{ij}}{\partial x_j}.
\end{equation}

With $\tau^S_{ij}=\overline{u_iu_j}-\overline{u}_i\overline{u}_j$ being the subgrid-scale stress tensor, which describes the effects of the unresolved scales and therefore needs to be modelled. A wide variety of subgrid-scale models have been developed to do so, in that thesis the dynamic $k_{sgs}$ equation model was used and will be detailed here.

\myparagraph{Dynamic $k_{sgs}$ equation model}
This model was introduced by Kim and Menon and is based on the subgrid-scale kinetic energy $k_{sgs}=\frac{1}{2} \left(\overline{u_i^2}-\overline{u}_i^2\right)$ which is defined as

\begin{equation}
    \frac{\partial k_{sgs}}{\partial t}+\overline{u}_i\frac{\partial k_{sgs}}{\partial x_i}=-\tau^S_{ij}\frac{\partial \overline{u}_i}{\partial x_j}-\epsilon_{sgs}+\frac{\partial}{\partial x_i}\left(\nu_t\frac{\partial k_{sgs}}{\partial x_i} \right)
\end{equation}

The subgrid-scale stress tensor, eddy viscosity and subgrid-scale dissipation rate are modeled as

\begin{equation}
    \tau^S_{ij}=-2C_{\tau}k_{sgs}^{\frac{1}{2}}\overline{S}_{ij}\Delta+\frac{2}{3}\delta_{ij}k_{sgs},\ \ \ \nu_t=C_{\tau}k_{sgs}^{\frac{1}{2}}\Delta,\ \ \ \epsilon_{sgs}=C_{\epsilon}\frac{k_{sgs}^{\frac{2}{3}}}{\Delta}.
\end{equation}

The only unknown left, are the coefficients $C_{\tau}$ and $C_{\epsilon}$, that are obtained following a dynamic procedure. This is done by using a filter width $\widehat{\Delta}$ to create a test scale field that is used to compute the coefficients, generally $\widehat{\Delta}=2 \Delta$ is used. The dynamic procedure will not be fully detailed here, for more information please refer to the original paper from Kim and Menon \cite{kim_new_1995} or for a more concise version the paper from Kim and Menon \cite{kim_application_1997}.

\subsubsection{Reynolds-Averaged Navier-Stokes (RANS) equations}
RANS is the turbulence model the most commonly used in the industry due to its low computational cost, that makes it suitable to simulate the complex geometries and high Reynolds number that are often considered. Here all the turbulence scales are modelled. The method is based on the Reynolds averaging in time of the flow quantities. For example, the decomposition of the velocity field vector $\mathbf{u}$ in a mean $\mathbf{\overline{u}}$ and a fluctuating part $\mathbf{u'}$ is given as follows

\begin{equation}
    \mathbf{u}=\mathbf{\overline{u}}+\mathbf{u'}.
\end{equation}

This decomposition should not be mistaken with the one done for LES, as it has a completely different meaning and does not follow the same properties. By introducing it in the mass and momentum Equations (\ref{equ:continuity}, \ref{equ:momentum}), the following RANS equations are obtained,

\begin{equation}
    \frac{\partial \overline{u}_i}{\partial x_i}=0,
\end{equation}

\begin{equation}
    \rho \frac{\partial \overline{u}_i}{\partial t}+\rho \overline{u}_j\frac{\partial \overline{u}_i}{\partial x_j}=-\frac{\partial \overline{p}}{\partial x_i}+\frac{\partial}{\partial x_j}\left(2\mu\overline{S}_{i,j}-\rho\overline{u_i'u_j'} \right).
\end{equation}

A new term $-\rho\overline{u_i'u_j'}$, called the Reynolds stress tensor appears, which is composed of 6 unknowns that cannot be resolved directly as no new equations are gained in the process. Various models have been introduced to model this term, the most common approach being based on the Boussinesq approximation, that considers that the Reynolds stress tensor can be written in the following way 

\begin{equation}
    -\rho\overline{u_i'u_j'}=2\mu_t \overline{S}_{ij}-\frac{2}{3}\rho k \delta_{ij}.
\end{equation}

With $\overline{S}_{i,j}$ being the mean strain rate tensor and $\mu_t$ the turbulent viscosity that needs to be computed in order to get the Reynolds stress tensor. The models making use of that assumption are called eddy viscosity models, the ones used in this thesis are going to be described.

\myparagraph{$k-\epsilon$}
The $k-\epsilon$ model is a two-equation model solving the turbulent kinetic energy $k$ and the rate of dissipation of turbulent kinetic energy $\epsilon$. This model makes the assumption that the flow is fully turbulent and is known to perform well for free-shear layer flows. However, the model does not predict well flow with strong adverse pressure gradient. In this thesis the exact formulation used is the Launder-Sharma $k-\epsilon$ that was introduced by Launder and Sharma \cite{launder_application_1974}. This variation was chosen as it is the $k-\epsilon$ low Reynolds model available in OpenFOAM\textsuperscript{\textregistered}. Low Reynolds model means that it can be used to model the whole boundary layer and need a grid fine enough near the wall ($y+<1$ is generally advised). The compact equations of the model are given by \cite{atila_comparative_2016} and are going to be presented here. First, the turbulent viscosity needed to get the Reynolds stress tensor is defined as 

\begin{equation}
    \mu_t=\rho C_{\mu} f_{\mu}\frac{k^2}{\epsilon}.
\end{equation}

The equations for k and $\epsilon$ are given as

\begin{equation}
    \frac{\partial \rho \overline{u}_i k}{\partial x_i}=\frac{\partial}{\partial x_i}\left(\left[\mu+\frac{\mu_t}{\sigma_k}\right]\frac{\partial k}{\partial x_i}\right)+\rho G_k +\rho (\epsilon + D),
\end{equation}

\begin{equation}
    \frac{\partial \rho \overline{u}_i \epsilon}{\partial x_i}=\frac{\partial}{\partial x_i}\left(\left[\mu+\frac{\mu_t}{\sigma_{\epsilon}}\right]\frac{\partial \epsilon}{\partial x_i}\right)+\rho C_1 f_1\frac{\epsilon}{k}G_k -\rho C_2 f_2\frac{\epsilon^2}{k}+E.
\end{equation}

With $G_k$, $E$ and $D$ being equal to

\begin{equation}
    G_k=-\overline{u_i'u_j'}\frac{\partial \overline{u}_i}{\partial x_j},\ \ \ E=2\mu \nu_t \frac{\partial^2 \overline{u}}{\partial y^2},\ \ \ D=2\nu\frac{\partial k^{0.5}}{\partial y}. 
\end{equation}

The constants of the model are: $C_\mu=0.09$, $C_1=1.44$, $C_2=1.92$, $\sigma_k=1.0$ and $\sigma_\epsilon=1.3$. Finally, the damping functions are defined as

\begin{equation}
f_{\mu}=e^{-3.4/(1+R_t/50)^2},\ \ \ f_1=1.0,\ \ \ f_2=1-0.3e^{-R_t^2}.  
\end{equation}

With $R_t=k^2/\epsilon \nu$ being the turbulent Reynolds number.

\myparagraph{$k-\omega$ SST}
The $k-\omega$ SST model is also a two-equation model solving the turbulent kinetic energy $k$ and the specific rate of dissipation $\omega$. The objective of this model is to keep the accuracy of the $k-\omega$ model in the near-wall region, while benefiting from the free-stream independence of the $k-\epsilon$ model. As often, various implementations exist and the one used here is the 2003 variant presented by Menter et al. \cite{menter_ten_2003}. This formulation is now going to be described, starting by the turbulent viscosity defined as 

\begin{equation}
    \mu_t=\frac{\rho a_1k}{max(a_1\omega,|S|F_2)}.
\end{equation}

$|S|$ is the invariant measure of the strain rate. The two equations for the turbulent properties of the flow $k$ and $\omega$ are

\begin{equation}
    \frac{\partial(\rho k)}{\partial t}+\frac{\partial(\rho \overline{u}_i k)}{\partial x_i}=\tilde{P_k}-\beta^* \rho k \omega +\frac{\partial}{\partial x_i}\left[(\mu+\sigma_k \mu_t) \frac{\partial k}{\partial x_i} \right],
    \label{equ:continuity_SST}
\end{equation}

\begin{equation}
\frac{\partial(\rho \omega)}{\partial t}+\frac{\partial(\rho \overline{u}_i \omega)}{\partial x_i}=\alpha \rho |S|^2 - \beta \rho \omega^2 +\frac{\partial}{\partial x_i}\left[(\mu+\sigma_\omega \mu_t) \frac{\partial \omega}{\partial x_i} \right] + 2(1-F_1) \rho \sigma_{\omega2}\frac{1}{\omega}\frac{\partial k}{\partial x_i}\frac{\partial \omega}{\partial x_i}.
\label{equ:momentum_SST}
\end{equation}

$P_k$ is the production of the turbulence kinetic energy and is introduced with a limiter as follows

\begin{equation}
    P_k=\mu_t\frac{\partial \overline{u}_i}{\partial x_j}\left(\frac{\partial \overline{u}_i}{\partial x_j}+\frac{\partial \overline{u}_j}{\partial x_i}\right),\ \ \ 
    \tilde{P_k}=min\left(P_k,10 \beta^* \rho k \omega   \right).
\end{equation}

The blending functions $(F_1, F_2)$ are given as follows

\begin{equation}
    F_1=tanh\left(\left[min\left(max\left(\frac{\sqrt{k}}{\beta^* \omega Y},\frac{500 \nu}{Y^2 \omega} \right),\frac{4 \rho \sigma_{\omega 2}k}{CD_{k\omega}Y^2} \right)   \right]^4\right),
\end{equation}
    
\begin{equation}
    CD_{k\omega}=max\left(2\rho \sigma_{\omega2}\frac{1}{\omega}\frac{\partial k}{\partial x_i}\frac{\partial \omega}{\partial x_i},10^{-10} \right),
\end{equation}

\begin{equation}
F_2=tanh \left( \left[max\left(\frac{\sqrt{k}}{\beta^* \omega Y},\frac{500 \nu}{Y^2 \omega} \right)\right]^2 \right),
\end{equation}

Y is the distance to the nearest wall. The constant for this model are the constant from the $k-\omega$ model ($\alpha_1=5/9$, $\beta_1=3/40$, $\sigma_{k1}=0.85$, $\sigma_{\omega1}=0.5$), the constant from the $k-\epsilon$ model ($\alpha_2=0.44$,  $\beta_2=0.0828$,  $\sigma_{k2}=1$, $\sigma_{\omega2}=0.856$) and the constants proper to the SST model ($\beta^*=0.09$, $a_1=0.31$). The remaining constants needed are computed from the constant of $k-\omega$ and $k-\epsilon$ in the following way

\begin{equation}
    \phi=\phi_1 F_1 + \phi_2 (1-F_1).
\end{equation}

\myparagraph{$k-\omega$ SST with Curvature Correction}

The $k-\omega$ SST capability to give accurate results have been tested extensively. However as is generally the case for eddy viscosity models, it is not capable of capturing the effect of streamline curvature or system rotation. For the cylinder considered in this work, the lack of streamline curvature consideration might be an issue. Therefore a version of the model with Curvature Correction to take that into account has also been tested. The correction used is the Spalart-Shur correction presented for the SST turbulence model by Smirnov and Menter \cite{smirnov_sensitization_2009}. The model stay close to the original with two modifications. The term $\tilde{P_k}$ is replaced by $P_k f_{r1}$ in Equation \ref{equ:continuity_SST} and in Equation \ref{equ:momentum_SST} the term $\alpha \rho |S|^2$ is replaced by $\alpha \rho |S|^2 f_{r1}$. The new term $f_{r1}$ is a function permitting the sensitization of the model to streamline curvature and rotation of the system, it is defined as

\begin{equation}
    f_{r1}=max\left(min\left((1+c_{r1})\frac{2r^*}{1+r^*}[1-c_{r3}\ tan^{-1}(c_{r2}\tilde{r})],1.25\right),0\right)
\end{equation}

The arguments of the function are

\begin{equation}
    r^*=\frac{|S|}{\Omega},
\end{equation}

\begin{equation}
    \tilde{r}=2\Omega_{jk}S_{jk}\left(\frac{DS_{ij}}{DT}+(\varepsilon_{imn}S_{jn}+\varepsilon_{jmn}S_{in})\Omega^{rot}_m \right)\frac{1}{\Omega\  max(S^2,0.09\omega)^{\frac{3}{2}}}.
\end{equation}

With $\Omega_{ij}$ the vorticity tensor, $\Omega^{rot}_m$ the components of the system rotation vector and $\varepsilon_{imn}$ the tensor of Levi–Civita.
Finally, the constants $c_{r1}$, $c_{r2}$ and $c_{r3}$ are equal to $1.0$, $2.0$ and $1.0$.

\newpage
\section{Numerics and software}
\subsection{OpenFOAM\textsuperscript{\textregistered}}
The CFD solver chosen for this thesis is OpenFOAM\textsuperscript{\textregistered}
 v1912, which is a free and open-source software coded in C++. Multiple reasons have motivated that choice, first OpenFOAM\textsuperscript{\textregistered} has an extensive range of features, with an important amount of solvers, algorithms and models already officially implemented in the solver and new addition that can easily be made whether by coding it yourself or using a code already made by a really active community. In addition to that, the user has a lot of liberties over the parameters, and although this is not user friendly, it offers a better control that what is possible in other commercial software such as ANSYS\textsuperscript{\textregistered} Fluent. The fact that it is free is also a great advantage for OpenFOAM\textsuperscript{\textregistered} as commercial CFD software can be quite expensive, especially for multi-processors licenses. If the solver is still not widely used in the industry due to a lack of robustness and it's less user-friendly usage it is attracting more and more interest. A final and more personal reason for using that software, is the fact that in my opinion using OpenFOAM\textsuperscript{\textregistered} was a much better experience than ANSYS\textsuperscript{\textregistered} Fluent with which I am already quite familiar, and doesn't permit as much control.\\
 \\
 In order to help eventual future work on the subject, all the C++ files used for setting the simulations for the 2D $k-\omega$ SST RANS and the LES on the cylinder are given in Appendix \ref{appendixA} and \ref{appendixB}. Obviously some parameters were changed to run the other simulations carried out, but this gives a good overview of the set-up that were used.
 
 \subsection{Finite Volume Method}
 In order to numerically simulate a fluid flow, the partial differential equations characterizing it need to be transformed in a system of linear algebraic equations. Various methods have been developed to do so, with the most commonly used in CFD being the Finite Volume Method. This is mainly due to the exact respect of the conservation laws (eg: mass, momentum, energy...) permitted by the approach. Without surprise, OpenFOAM\textsuperscript{\textregistered} also uses this discretization methodology, a brief introduction to it is, therefore, going to be given. For more information on the subject, the books from Blazek \cite{blazek_computational_2007} or Moukalled et al. \cite{moukalled_finite_2016} can be consulted.\\
 \\
 The first step of the method is to subdivide the spatial domain into a number of cells, forming what is called a grid or mesh. Then the governing equations are integrated over each cell of the grid. This is going to be demonstrated with the momentum Equation \ref{equ:momentum}, for a problem considered as steady for simplicity, which gave
 
     \begin{equation}
    \frac{\partial u_iu_j}{\partial x_j} =-\frac{1}{\rho}\frac{\partial p}{\partial x_i}+\nu \frac{\partial^2 u_i}{\partial x_j\partial x_j}.
\end{equation}

The integral form of that equation over a cell noted $C$ can be written as

\begin{equation}
    \int_{\partial V_c} u_i u_j n_j dS = -\int_{\partial V_c} \frac{p_i}{\rho}n_i dS + \int_{\partial V_c} \nu \frac{\partial u_i}{\partial x_j}n_j dS,
\end{equation}

 with $n$ being the normal of the surface of the element $C$. Then the surface integrals are replaced by summation of the flux terms over the faces of element $C$. The number of integration points on each face needs to be chosen, it is generally taken as 1 and this is what is going to be considered here. The equation obtained is then
 
 \begin{equation}
     \sum_f (u_iu_jn_j)_fS=-\sum_f \left(\frac{p_i}{\rho}n_i\right)_fS+\sum_f \left(\nu\frac{\partial u_i}{\partial x_j}n_j\right)_fS,
     \label{equ:momentum_fvm}
 \end{equation}
 
 f referring to a given face of the element $C$ and S the surface over that face, considered here as uniform. Finally, a linear algebraic system of equations is obtained by expressing the fluxes in terms of the values at the cell $C$ and neighbouring cell centres. This choice to have the variables stocked at the cell centres is not the only option, but is the most popular and the one used by OpenFOAM\textsuperscript{\textregistered}. The exact way that the fluxes are computed depends on the term that is considered and the scheme used and will be not be discussed further in this introduction to the method. 
 
 \subsection{Algorithms for incompressible flows}
 The Finite Volume Method permits to discretize the Navier-Stokes equations. When considering an incompressible flow, as is done throughout this thesis, 4 unknowns are considered and with the momentum and continuity equations, 4 scalar equations are available which should be enough to solve the problem. However since the pressure does not appear directly in the continuity equations, the system cannot be solved by an iterative mean. Various methods have been developed to deal with this problem, the most widely used being the so-called pressure correction methods. This type of methods relies on using the continuity and momentum equations to get an equation for the pressure. Let's consider therefore a generalized form of the Navier-Stokes equations
 \begin{equation}
     \nabla \cdot \mathbf{u}=0,
     \label{eq:continuity_matrix}
 \end{equation}
 
 \begin{equation}
     \mathcal{M}\mathbf{u}=-\nabla p.
 \end{equation}
 
 With $\mathcal{M}$ being a matrix which coefficients are obtained by discretizing the momentum equation. $\mathcal{M}$ can be decomposed in a diagonal $\mathcal{A}$ and non-diagonal $\mathcal{H}$ components, and the momentum equation can then be written as
 
 \begin{equation}
     \mathcal{A} \mathbf{u}-\mathcal{H}=\nabla p.
     \label{eq:momentum_decomp}
 \end{equation}
 
 Equation \ref{eq:momentum_decomp} can be rearranged in the following way,
 
 \begin{equation}
     \mathbf{u}=\mathcal{A}^{-1}\mathcal{H}-\mathcal{A}^{-1}\nabla p.
     \label{eq:matrix_momentum}
 \end{equation}
 
 By substituting Equation \ref{eq:matrix_momentum} into Equation \ref{eq:continuity_matrix} the following pressure equation is obtained 
 
 \begin{equation}
     \nabla \cdot (\mathcal{A}^{-1} \Delta p)=\nabla \cdot (\mathcal{A}^{-1} \mathcal{H}).
     \label{eq:pressure_correction}
 \end{equation}
 
 If the four equations needed are now obtained, the pressure and velocity are still coupled and a special treatment is needed to solve the system. The most common way is by the use of a segregated approach, with for example the SIMPLE algorithm. A brief description of how SIMPLE handle the coupling between pressure and velocity is going to be given.  If this algorithm is the most common approach, variations of SIMPLE have been developed, principally in an attempt to improve the convergence rate. A full account of these algorithms will not be given here, but SIMPLEC and PISO will be briefly introduced as they have been used in this work. A more complete discussion of these algorithms can be found in Moukalled et al. \cite{moukalled_finite_2016}.  
 
 \subsubsection{SIMPLE algorithm}
 The main steps of the SIMPLE algorithm developed by Patankar and Spalding \cite{patankar_calculation_1972} to solve the Navier-Stokes equations on a collocated grid are given here.\\
 
\ \ \ 1. Start with an initial guessed value for the pressure $p$, velocity $\bold{u}$ and mass flow rate $\bold{\dot{m}}$ fields.\\

\ \ \ 2. Solve the discretized momentum equation to obtain a new velocity field $\bold{u^*}$.\\

\ \ \ 3. By using Rhie-Cho interpolation get the mass flow rate field $\bold{\dot{m}^*}$ at the cells faces. This step permits to use the SIMPLE algorithm initially thought for a staggered grid on a collocated grid and avoid checkerboard problems.\\

\ \ \ 4. Solve the pressure correction Equation \ref{eq:pressure_correction} to obtain the pressure correction field $p_c$. \\

\ \ \ 5. Update the fields in order to obtain continuity-satisfying fields as shown below

\begin{equation}
    p^{*}=p+p_c,\ \ \ \bold{u^{**}}=\bold{u^*}+\bold{u_c},\ \ \ \bold{\dot{m}^{**}}=\bold{\dot{m}^*}+\bold{\dot{m}_c}.
\end{equation}
With $\bold{u_c}$ and $\bold{m_c}$ that can be obtained using $p_c$.\\

\ \ \ 6. Finally, if a chosen convergence condition is not met go back to step 1 by taking $p^{*}$,  $\bold{u^{**}}$ and $\bold{\dot{m}^{**}}$ as initial guesses.\\
\\
This algorithm can easily be extended to handle transient problems, but this is generally not recommended as the convergence is not efficient. In OpenFOAM\textsuperscript{\textregistered}, SIMPLE is only coded to resolve steady-state flows. Under-relaxation is generally needed in order to get a converged solution.

  \subsubsection{SIMPLEC algorithm}
  The SIMPLEC algorithm presented by Doormal and Raithby \cite{doormaal_enhancements_1984} is a slightly modified version of the SIMPLE algorithm. In SIMPLE, in order to make the equations manageable, some terms of the velocity correction equation are neglected. These terms reach zero at convergence and therefore do not impact the final solution. In the SIMPLEC algorithm, the terms neglected are slightly less significant. Therefore the two algorithms have the same steps and give the same final solution, but SIMPLEC has a higher rate of convergence and can, therefore, permit to save computational time. 
  
   \subsubsection{PISO algorithm}
   
   The PISO algorithm developed by Issa \cite{issa_solution_1986}, is another variation of the SIMPLE algorithm, it is generally used for transient flows. The steps of the algorithm are as follows.\\
   
   \ \ \ 1-5. Similar to the steps of SIMPLE.\\
   
   \ \ \ 6. With $\bold{u}^{**}$ and $p^*$ assemble and solve the momentum equation to obtain a new velocity field $\bold{u}^{***}$.\\
   
   \ \ \ 7. By using Rhie-Cho interpolation get the mass flow rate field $\bold{\dot{m}^{***}}$ at the cells faces.\\
   
   \ \ \ 8. Solve the pressure correction Equation \ref{eq:pressure_correction} to obtain the pressure correction field $p_{cc}$.\\
   
   \ \ \ 9. Similarly as for step 5 of SIMPLE, update the fields in order to obtain the continuity-satisfying fields $\bold{u^{****}}$, $\bold{\dot{m}^{****}}$ and $p^{**}$.\\
   
   \ \ \ 10. If a given number of iterations is not reached go back to step 2 but use the fields obtained in step 9.\\
   
   \ \ \ 11. Finally, if a chosen convergence condition is not met go back to step 1 by taking $p^{**}$,$\bold{u^{****}}$ and $\bold{\dot{m}^{****}}$ as initial guesses.\\
   
   This version of the algorithm is presented for the case of a steady problem. PISO was designed and is generally used for transient problems. In this case, when the convergence condition is reached at step 11, the fields obtained are the solution at the current time of the simulation. To advance in time, a time step is added, and the algorithm goes back to step 1 by using the results from step 11 as initial guesses. At the difference from SIMPLE, under-relaxation is not needed and the convergence is obtained by using a time step small enough, generally, the condition to have a maximum courant number in the domain smaller than 1 is given. With some time-stepping schemes, especially implicit ones, a higher courant number might be used while still maintaining convergence, but this might deteriorate the accuracy of the solution depending on the problem considered. Again for a fully converged solution SIMPLE and PISO should give the same results, but in general, PISO is a lot faster for transient problems.
   
 \subsection{Parallel computing and HPC}
In order to reduce computation time, a common practice in CFD is to use multiple processors as they can theoretically divide the time taken by a sequential simulation by the number of processors used. In practice, however, the speed up from using multiple processors is lower due to the communication needed between the processors. In addition to that, making a solver portable to parallel computing is far from being an easy task. Fortunately, OpenFOAM\textsuperscript{\textregistered} does have well-developed codes for parallelisation that are rather easy to use. \\
\\
In this thesis all the simulations presented were run in parallel whether by using a personal computer for the cheaper simulations or the HPC capabilities of Cranfield University for the more computationally demanding ones. The domains were decomposed by using the scotch algorithm, that automatically decomposes the domain and attempt to minimise the number of processors boundaries and should, therefore, reduce the communication time to a minimum. This method was chosen, for this reduction in communication time, but also since it does not need any input from the user and was found to divide the number of cells nearly equally between the processors, which is another indicator that the domain is decomposed in an optimized way.\\
\\

\newpage
\section{Results and discussion}

\subsection{Jet attaching to offset flat plate}
\label{sec:jet_offset_results}
Before carrying out a numerical study on the impact of the Coand\u{a} effect for a jet blown on a cylinder, a more fundamental test case was considered, namely a jet blown close to an adjacent flat plane. Extensive investigation on this type of flow can be found in the literature, for various configurations and focusing on various aspects of the flow. These investigations have been analytical \cite{bourque_reattachment_1960,sawyer_two-dimensional_1963}, experimental \cite{nasr_comparison_1997,nasr_turbulent_1998,gao_experimental_2007}  and numerical \cite{nasr_turbulent_1998,fu_insights_2016}. 
\\

\begin{figure}[h!]
\centering
\includegraphics[width=15cm]{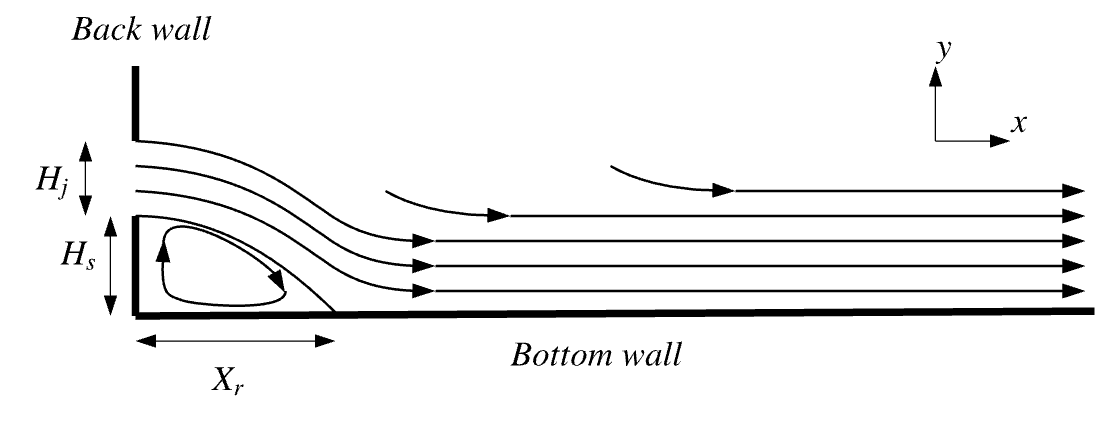}
\caption{Schematic of the approximate streamlines of a jet attaching to an offset flat plane}
\label{fig:schematic_offset}
\end{figure}

In this thesis, the experiment from Gao and Ewing \cite{gao_experimental_2007} was replicated numerically. The choice of this set of data was made because the geometry considered, with a small offset distance, has been less investigated, and also since the quantities given are mostly the same as for the cylinder experiment from Wygnanski et al. \cite{neuendorf_turbulent_2000,neuendorf_turbulent_1999, likhachev_streamwise_2001,cullen_role_2002, han_streamwise_2004, neuendorf_large_2004}, making it better suited to compare the results that will be obtained. An illustration of the flow attaching to the flat plane is given in Figure \ref{fig:schematic_offset}.

\subsubsection{Simulation strategy}
Following the set up used by Gao and Ewing \cite{gao_experimental_2007} a horizontal jet was blown from an inlet of height $H_j=3.8\times 10^{-2}\ m$ situated at a distance $H_s=2.28\times 10^{-2}\ m$ from the plane bottom wall. The average inlet velocity was $U_{jet}=18.4\ m.s^{-1}$, giving a Reynolds number based on the inlet height of 44000, by considering $\nu=1.59\times10^{-5}\ m^2.s^{-1}$.\\
\\
The flow being at relatively low velocity, the flow was considered as incompressible. The Reynolds-Averaged Navier-Stokes equations are solved using the SIMPLEC algorithm. The gradients are computed using a centred second order scheme (named Gauss linear in OpenFoam). The divergence of the velocity is computed using Upwind second order (Gauss linearUpwind), while the turbulent quantities are computed with first order Upwind (Gauss upwind). Two turbulence models have been tested, the $k-\omega$ SST and the Launder-Sharma $k-\epsilon$.

\newpage
\subsubsection{Mesh}
\label{sec:mesh_offset}
Three 2D structured hexahedral grids with different level of refinement have been generated. In order to minimize the impact of the boundary on the results a computational domain extending $235H_j$ in the $x$-direction and $470H_j$ in the $y$-direction was used. A wall-resolved strategy is implemented and a $y+<1$ with a relatively small expansion ratio was therefore used. In order to capture the region around the jet reattachment accurately, the same spacing and expansion ratio was also used for the $x$-direction. The parameters of the different grids are given in Table \ref{offset_parameters}.

\begin{table}[H]
\centering
\begin{tabular}{|c|c|c|c|}
\hline
               & Fine & Medium & Coarse \\ \hline
$y+$           &0.7      &0.85        &1
\\ \hline
Expansion rate &1.029      &1.042        &1.061      \\      \hline
Grid points    &136224 &      70104 &        35854        \\ \hline
\end{tabular}
\caption{Parameters for definition of the grids around the offset jet blown on a flat plane}
\label{offset_parameters}
\end{table}

\begin{figure}[h!]
\centering
\includegraphics[width=12cm]{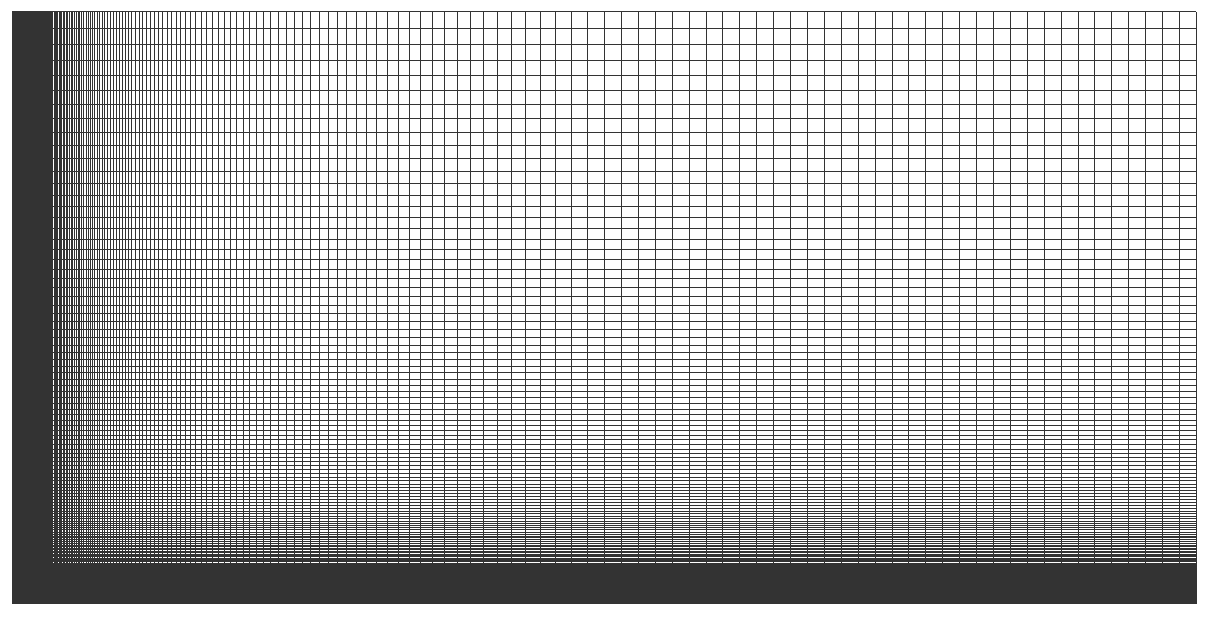}
\caption{Fine grid for the jet attaching to the offset plane}
\label{fig:grid_offset}
\end{figure}

The values of $y+$ given in Table \ref{offset_parameters} were obtained using an estimation based on flat-plate boundary layer theory for turbulent flow that can be found in White \cite{white_fluid_2016}. This is useful since, before running a simulation, it is difficult to get a precise value for $y+$, but this quantity is crucial to ensure the accuracy of the simulation. After reaching convergence, a more precise value of the non-dimensional spacing can be obtained and it is important to check that the condition $y+<1$ is well respected. The following maximum values for $y+$ were obtained when using the $k-\omega$ SST model, 0.28 for the fine grid, 0.35 for the medium and 0.41 for the coarse. The condition on $y+$ is therefore well respected. The Figure \ref{fig:grid_offset} presents the fine mesh used, the clustering of cells around the jet exit can be observed.

\subsubsection{Boundary and initial conditions}
The boundary conditions in the domain are as follows: the jet exit is an inlet, the back and bottom walls are no-slip walls and the top and right boundaries are outlets. Initially, the fluid is considered at rest with a pressure $p=0$ and a velocity vector equal to $(0,0,0)$ in the internal domain.\\

\begin{figure}[h!]
\centering
\includegraphics[width=8cm]{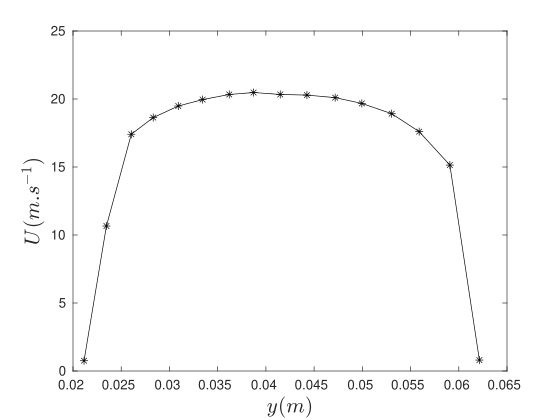}
\caption{Velocity profile at inlet for jet attaching to offset flat plane}
\label{fig:offset_inlet}
\end{figure}

The velocity profile at the inlet was interpolated from the experimental profile given by Gao and Ewing \cite{gao_experimental_2007} and is presented in Figure \ref{fig:offset_inlet}.

When it comes to the turbulent quantities, for the $k-\omega$ SST model, following the recommendation from Menter \cite{menter_zonal_1993} for wall resolved simulation, all turbulent quantities on the walls are set to 0, except $\omega$ that is estimated as follows

\begin{equation}
    \omega=10\frac{6\mu}{\beta_1 \Delta y^2}=3.3 \times 10^9.
    \label{equ:omega_wall}
\end{equation}

With $\beta_1=0.075$ and $\Delta y$ the distance from the wall to the first cell centre.\\
The initial value in the domain and the inlet boundary condition are based on the estimate of isotropic turbulence and given as follows

\begin{equation}
    k=1.5(|U_{jet}|I)^2,
    \label{equ:k_isotropic}
\end{equation}

\begin{equation}
    \omega=\frac{C_{\mu}^{-0.25}k^{0.5}}{L}.
    \label{equ:omega_isotropic}
\end{equation}

Where $C_{\mu}$ is a constant equal to 0.09, I is the turbulence intensity and L the turbulence length scale. A low turbulence intensity of $I=0.05\%$ is considered, for the turbulent length scale $L=0.07 H_j$ was used. This length scale is equal to $7\%$ of the hydraulic diameter of the inlet and is actually only correct for fully developed pipe flow, but was deemed accurate enough to initialize the turbulent quantities.\\
\\
For the $k-\epsilon$ model, $k$ is given in the same way as for the previous model while $\epsilon$ is estimated by

\begin{equation}
    \epsilon=\frac{C_{\mu}^{0.75}k^{1.5}}{L}.
\end{equation}

\subsubsection{Grid convergence study}
In order to assess the independence of the results from the grid, the grid convergence methodology introduced by Roache \cite{roache_verification_1998} will be carried out on the three grids presented in section \ref{sec:mesh_offset} for the average pressure coefficient $C_p$ on the bottom wall, using the $k-\omega$ SST turbulence model. The pressure coefficient $C_p$ is defined by Anderson \cite{anderson_fundamentals_2017} as

\begin{equation}
    C_p=\frac{p-p_{\infty}}{\frac{1}{2}\rho_{\infty}U_{jet}}.
    \label{eq:cp}
\end{equation}

With the freestream pressure $p_{\infty}$ taken as $0\ Pa$, the density $\rho_{\infty}$ as $1\ kg.m^{-3}$ since it is the default value of OpenFOAM\textsuperscript{\textregistered} for incompressible fluids. This yields the average $C_p$ on the wall given in Table \ref{tab:cp_offset}

\begin{table}[H]
\centering
\begin{tabular}{|c|c|c|c|}
\hline
              & Fine  & Medium & Coarse \\ \hline
Average $C_p$ & 7.96e-3 & 7.95e-3  & 8.03e-3  \\ \hline
\end{tabular}
\caption{Average value of $C_p$ on the bottom flat wall for the different grids}
\label{tab:cp_offset}
\end{table}

The results of the GCI analysis based on the values given in Table \ref{tab:cp_offset} are presented in Table \ref{tab:gci_offset}, with the grids numbered from 1 (fine) to 3 (coarse). The high decrease between $GCI_{12}$ and $GCI_{23}$ proves that further refining the mesh will only result is small improvement of the solution. In addition to that, the value of $GCI_{12}$ is relatively small, showing that we are close to the Richardson extrapolated value between fine and medium grids. Finally, the GCI ratio is close to 1, proving that the solutions are in the asymptotic range of convergence, which is important for the methodology used to be valid. Based on those results the choice was done to use the fine grid for the rest of the simulations.

\begin{table}[H]
\centering
\begin{tabular}{|c|c|c|c|c|c|}
\hline
              & r   & q    & $GCI_{12}$ (\%) & $GCI_{23}$ (\%) & $GCI ratio$ \\ \hline
Average $C_p$ & 1.4 & 7.18& 0.011            & 0.12            & 1.0009                          \\ \hline
\end{tabular}
\caption{GCI analysis for the average $C_p$ on the bottom flat wall}
\label{tab:gci_offset}
\end{table}

\begin{figure}[H]
    \centering    
    \includegraphics[width=0.49\linewidth]{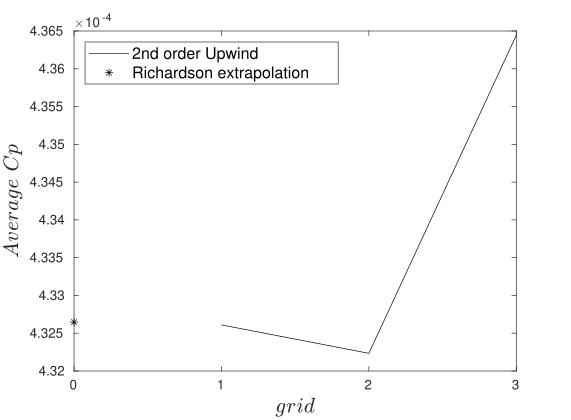}
    \caption{Average $C_p$ on the bottom wall for fine (1) to coarse (3) grid and Richardson extrapolation}
    \label{fig:gci_plot_offset}
\end{figure}

To get a more visual representation on the evolution of the solution for the different grids compared to the extrapolated value between the fine and medium grids the Figure \ref{fig:gci_plot_offset} is given.

\subsubsection{Flow around the flat plate}
\begin{figure}[h!]
\centering
\begin{subfigure}{0.5\textwidth}
  \centering
  \includegraphics[width=0.99\textwidth]{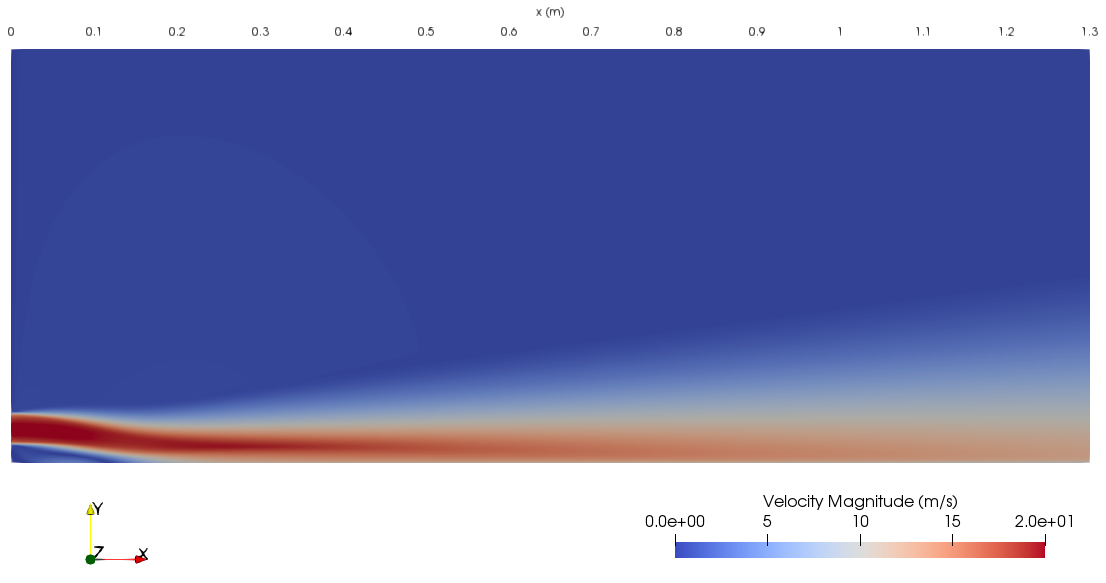}
  \caption{Velocity contours - $k-\omega$ SST}
  \label{fig:sub1}
\end{subfigure}%
\begin{subfigure}{0.5\textwidth}
  \centering
  \includegraphics[width=0.99\textwidth]{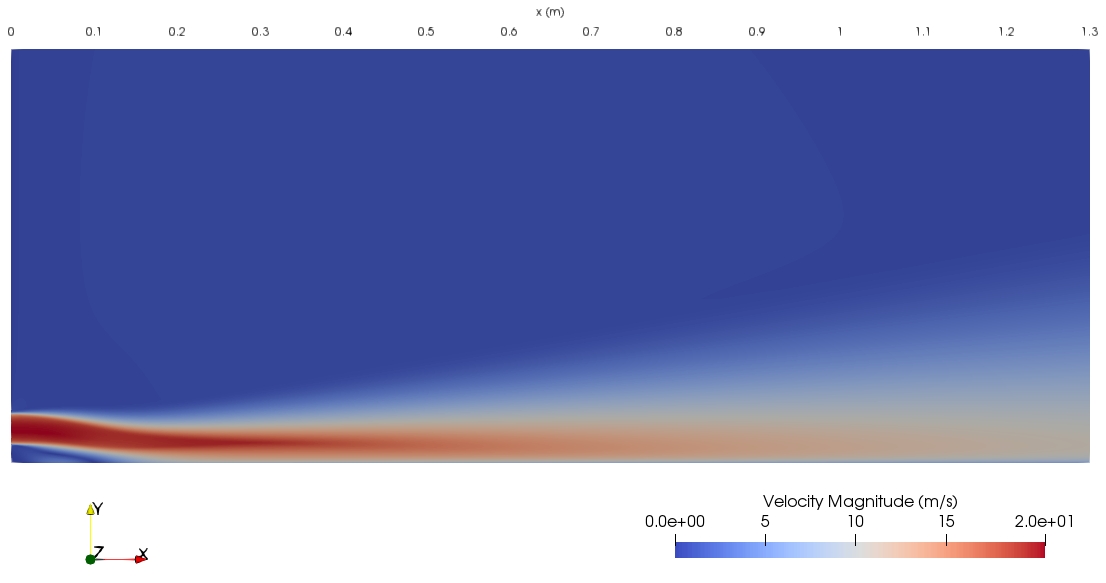}
  \caption{Velocity contours - $k-\epsilon$}
  \label{fig:sub2}
\end{subfigure}

\begin{subfigure}{0.5\textwidth}
  \centering
  \includegraphics[width=0.99\textwidth]{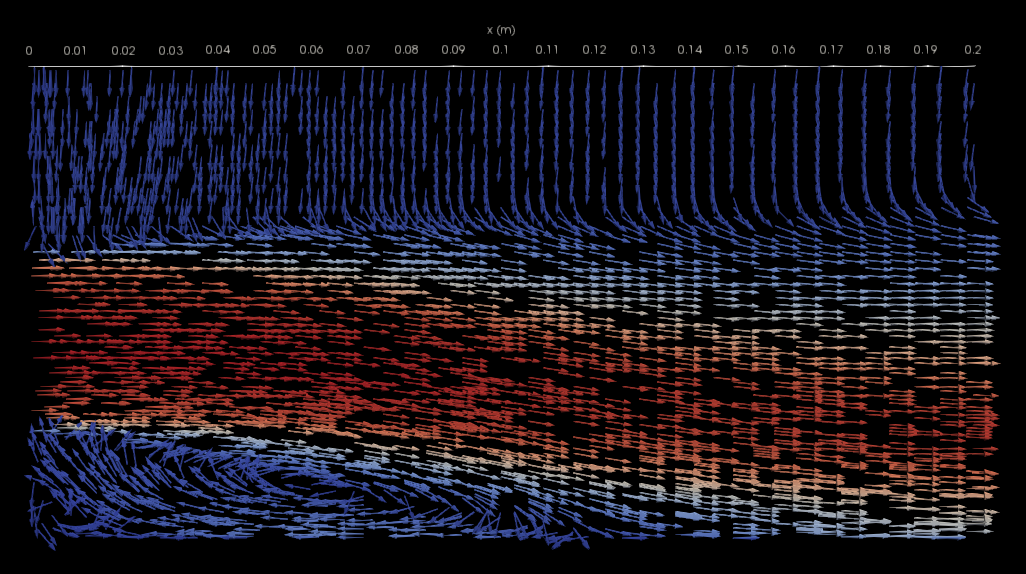}
  \caption{Velocity vectors - $k-\omega$ SST}
  \label{fig:sub1}
\end{subfigure}%
\begin{subfigure}{0.5\textwidth}
  \centering
  \includegraphics[width=0.99\textwidth]{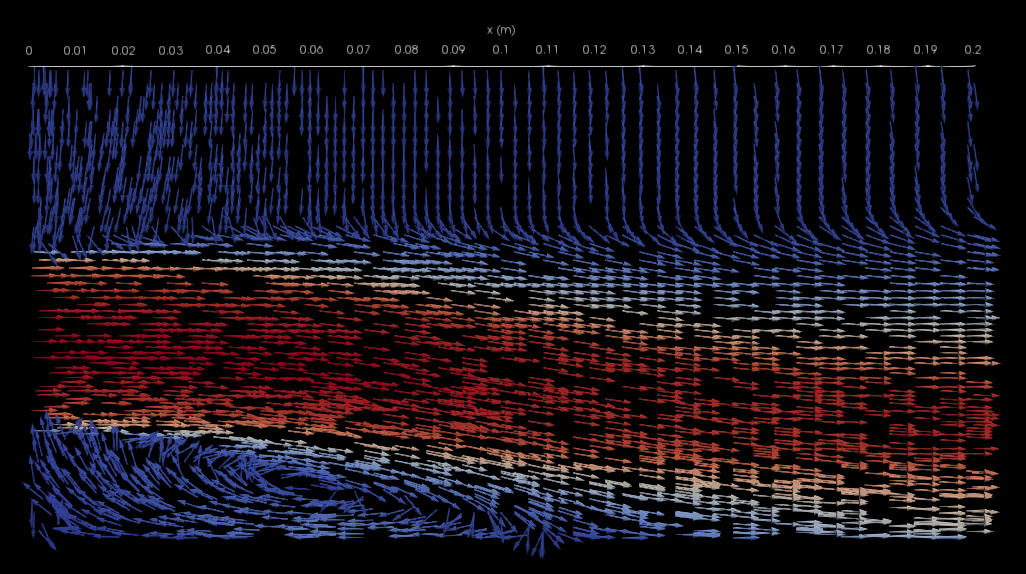}
  \caption{Velocity vectors - $k-\epsilon$}
  \label{fig:sub2}
\end{subfigure}

\caption{Velocity contours and vectors around the flat plane near the jet exit obtained from RANS calculations}
\label{fig:2D_RANS_contours_offset}
\end{figure}

The flow obtained numerically is going to be discussed and compared to the one obtained experimentally by Gao and Ewing \cite{gao_experimental_2007}. First, the velocity contours and velocity vectors around the jet exit are given in Figure \ref{fig:2D_RANS_contours_offset}. The impact of the Coand\u{a} effect can be clearly observed with the jet blown horizontally that bends and attaches itself to the flat plane. It is known that an offset wall jet can be considered in three regions. Near the jet exit is the converging region were the jet has not yet attached to the wall and characterized by a recirculation. Then comes the reattachment region were the boundary layer on the wall develop and finally a wall jet region, were the jet spread out and behave has a standard wall jet. These three regions appear to be well captured by both simulations, in particular, the velocity vectors permit to observe the recirculation zone and the attachment of the jet to the wall. In addition, the entrainment of surrounding fluid by the jet can be observed. When it comes to the impact of the turbulence model chosen, if some slight differences can be observed, the results visually appears as really similar.

\begin{figure}[h!]
\centering
\begin{subfigure}{0.5\textwidth}
  \centering
  \includegraphics[width=0.99\textwidth]{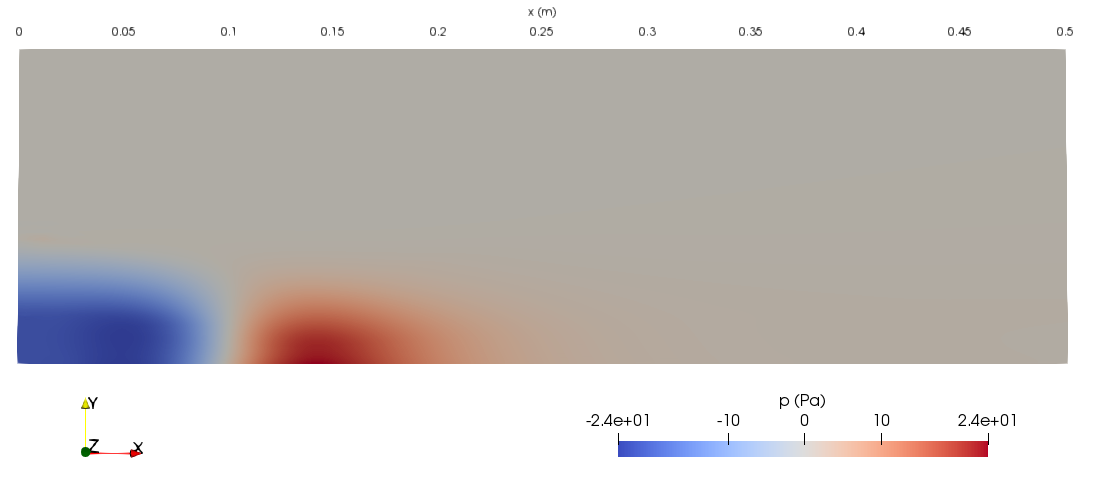}
  \caption{$k-\omega$ SST}
  \label{fig:sub1}
\end{subfigure}%
\begin{subfigure}{0.5\textwidth}
  \centering
  \includegraphics[width=0.99\textwidth]{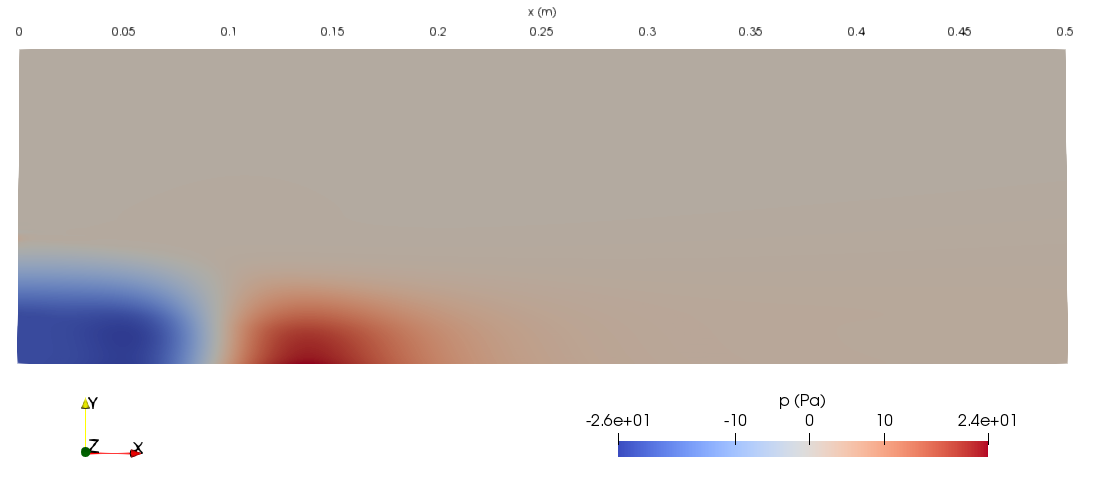}
  \caption{$k-\epsilon$}
  \label{fig:sub2}
\end{subfigure}

\caption{Pressure contours around the flat plane near the jet exit obtained from RANS calculations}
\label{fig:2D_RANS_contours_offset_pressure}
\end{figure}

The pressure contours are given in Figure \ref{fig:2D_RANS_contours_offset_pressure}, a negative pressure region can be observed below the jet exit. This region is due to the entrainment of fluid that cannot be replaced as detailed in the introduction Section \ref{sec:intro} and is responsible for the bending of the jet. After the reattachment of the jet, a positive pressure region is created due to the interaction between the jet and the wall.

\newpage

\begin{figure}[H]
    \centering    
    \includegraphics[width=0.49\linewidth]{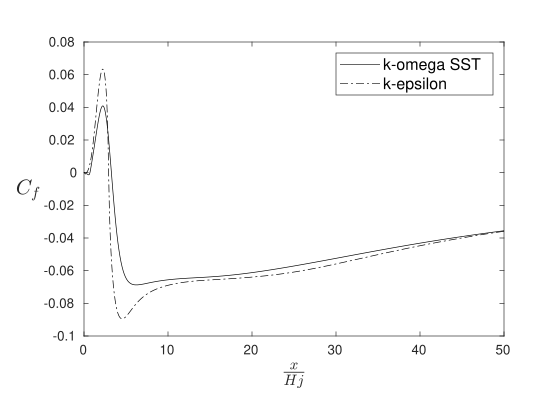}
    \caption{Streamwise $C_f$ on the flat plate obtained from RANS calculations}
    \label{fig:cf_offset}
\end{figure}

The reattachment length of the jet on the flat plane is now going to be discussed with a comparison between the experimental results from Gao and Ewing \cite{gao_experimental_2007} and the numerical ones. In order to compute the reattachment point, the skin friction coefficient was used. In fact, it is known that the reattachment or separation occurs when the streamwise wall Shear Stress $\tau_{\omega}$ is equal to 0 \cite{cebeci_calculation_1972}. The skin coefficient $C_f$ is a dimensionless wall shear stress and is generally used to get the reattachment or separation location. The expression of $C_f$ is defined as

\begin{equation}
    C_f=\frac{\tau_{\omega}}{\frac{1}{2}\rho_{\infty}U_{jet}},
\end{equation}

with the freestream density $\rho_\infty$ taken as $1\ kg.m^{-3}$. The Figure \ref{fig:cf_offset} gives the plot of $C_f$ on the bottom wall for both turbulence models. Using that, the reattachment length is found to be $X_r=0.124\ m$ for $k-\omega$ SST and $X_r=0.111\ m$ for $k-\epsilon$. Taking into account experimental incertitude, Gao and Ewing \cite{gao_experimental_2007} gave a reattachment included in the interval $X_r=[0.116,0.126]\ m$. The $X_r$ for the $k-\omega$ SST model is in that interval, while the $X_r$ for $k-\epsilon$ is found to be slightly lower than the experimental result. This under-prediction is not unexpected as other authors found the same trend with that particular turbulence model used for a jet blown on an offset plane, this is, for example, the case for Fu et al. \cite{fu_insights_2016}.

\begin{figure}[H]
\centering
\begin{subfigure}{9cm}
  \centering
  \includegraphics[width=8.9cm]{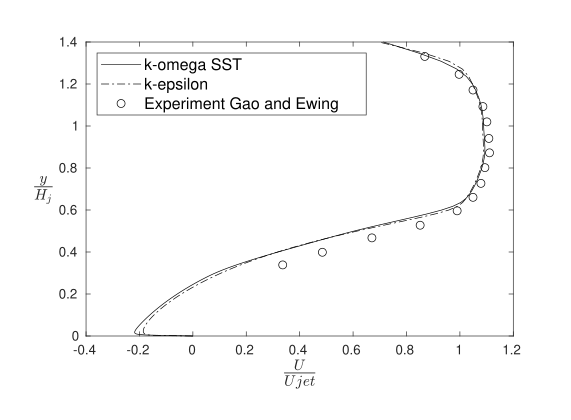}
  \caption{$\frac{x}{H_j}=2$}
  \label{fig:sub1}
\end{subfigure}%
\begin{subfigure}{9cm}
  \centering
  \includegraphics[width=8.9cm]{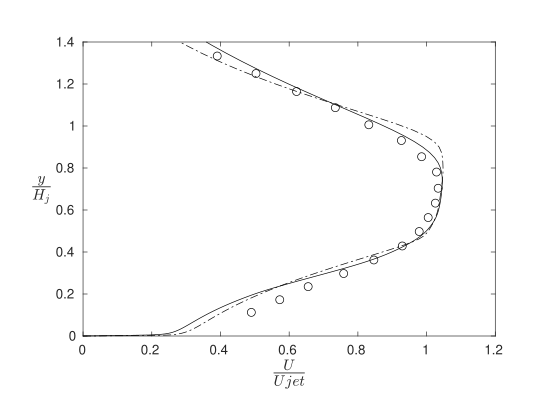}
  \caption{$\frac{x}{H_j}=4$}
  \label{fig:sub2}
\end{subfigure}

\begin{subfigure}{9cm}
  \centering
  \includegraphics[width=8.9cm]{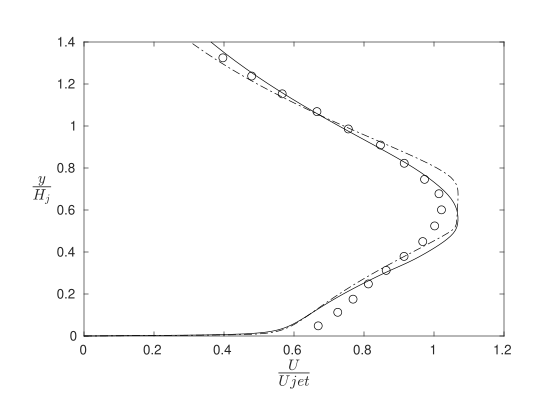}
  \caption{$\frac{x}{H_j}=6$}
  \label{fig:sub2}
\end{subfigure}

\caption{Comparison of normalized streamwise velocity profiles for the offset jet blown on a flat plate at various $x$ locations obtained from RANS calculations with experimental data \cite{gao_experimental_2007}}
\label{fig:offset_velocity_profiles}
\end{figure}

\newpage
The normalized velocity profiles at various downstream locations are given in Figure \ref{fig:offset_velocity_profiles}. At $x/H_j=2$ the presence of the recirculation zone in the converging region can be well observed with a negative velocity near the bottom wall. Between $x/H_j=4$ and 6 the flow is in the reattachment region, it can be seen that the y location of the maximum velocity is approaching the bottom wall due to the development of the boundary layer. When it comes to the comparison between numerical and experimental results, good agreement can be observed, with $k-\omega$ SST appearing to be slightly more accurate than $k-\epsilon$. The slight discrepancies observed could be due to various factors such as the error introduced by the turbulence models, the experimental uncertainty, the grid... It can, however, be concluded that at least when considering $x/H_j \le 6$ the flow seems to be captured accurately, especially by using the $k-\omega$ SST model.

\begin{figure}[H]
\centering
\begin{subfigure}{9cm}
  \centering
  \includegraphics[width=8.9cm]{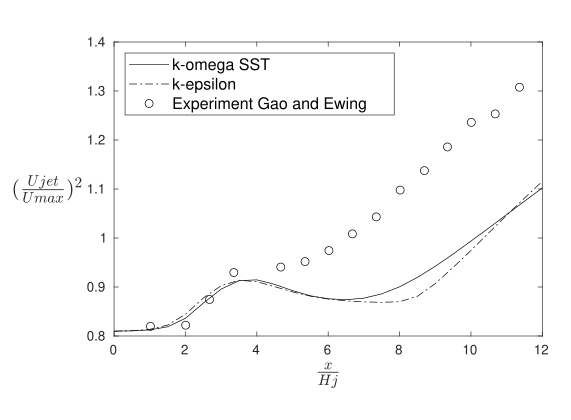}
  \caption{Velocity decay}
  \label{fig:sub1}
\end{subfigure}%
\begin{subfigure}{9cm}
  \centering
  \includegraphics[width=8.9cm]{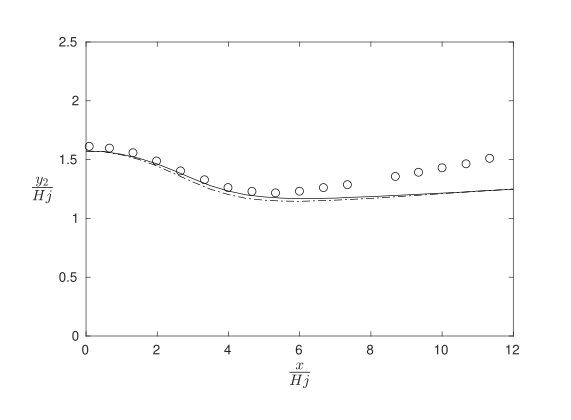}
  \caption{Jet half-width}
  \label{fig:sub2}
\end{subfigure}

\caption{Comparison of normalized jet decay and jet half-width for a jet blown on an offset flat plane obtained from RANS calculations with experimental data \cite{gao_experimental_2007}}
\label{fig:offset_decay_spread}
\end{figure}

\newpage
To get a more global assessment on how well the flow is captured numerically, the development of the jet needs to be compared for the whole length of the experimental domain. To do so, two quantities are going to be plotted at different locations, $U_{max}$ and $y_2$. $U_{max}$ is the maximum velocity at a given location $x$, while $y_2$ is the jet half-width, which is for a given $x$, the distance from the wall where the velocity is only half of $U_{max}$. Plot of the streamwise variations of these two quantities is given in Figure \ref{fig:offset_decay_spread}. When considering the maximum velocity it can be seen that the main trends found in the experiment are also found numerically. In fact, from the jet exit to approximately the reattachment location $U_{max}$ is decreasing as the jet decay. However, after that, the decay will stop, for a transition period due to the attachment of the jet to the wall. Then the decay will start again and appear to be linear as is expected for a plane wall jet. Non-negligible discrepancies can, however, be observed when compared to the experiment, with an under-prediction of the decay for both turbulence models. To give an idea of the error between experiment and simulation, the maximum differences to the experiment were computed and found to be approximately $19.5\%$ for $k-\omega$ SST and $21.7\%$ for $k-\epsilon$. When it comes to the jet half-width, it can be seen that it gradually decreases in a non-linear fashion, reach a minimum and then starts to increase linearly. Again this trend is well captured by both turbulence models, but in the linear increase of $y_2$ a non-negligible under-prediction of the rate of spread can again be observed. This under-prediction generated a maximum error of approximately $16.3\%$ for $k-\omega$ SST and $16.4\%$ for $k-\epsilon$.

\begin{figure}[H]
  \centering
  \includegraphics[width=8.9cm]{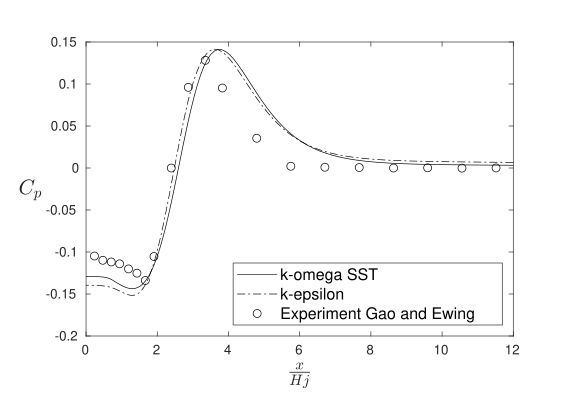}
\caption{Comparison of wall pressure coefficient on the flat plane obtained from RANS calculations with experimental data \cite{gao_experimental_2007}}
\label{fig:offset_Cp}
\end{figure}

The pressure coefficient on the bottom wall in the streamwise direction is plotted in Figure \ref{fig:offset_Cp}. Without surprise, the pressure behaves in a similar fashion on the wall as in the pressure contours given in Figure \ref{fig:2D_RANS_contours_offset_pressure}, with a low-pressure region near the jet exit followed by a high-pressure region. If the general trend of the experimental pressure coefficient is also found numerically, some discrepancies are however found. Especially, the length of the high-pressure region found just after the reattachment of the jet is overpredicted compared to the experiment. 

\begin{figure}[h!]
\centering
\begin{subfigure}{0.51\textwidth}
  \centering
  \includegraphics[width=0.99\textwidth]{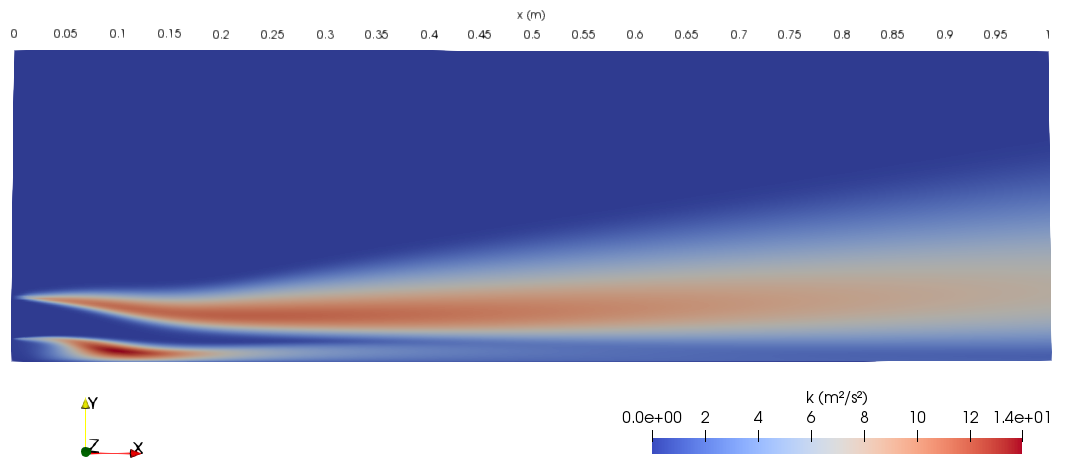}
  \caption{$k-\omega$ SST}
  \label{fig:sub1}
\end{subfigure}%
\begin{subfigure}{0.51\textwidth}
  \centering
  \includegraphics[width=0.99\textwidth]{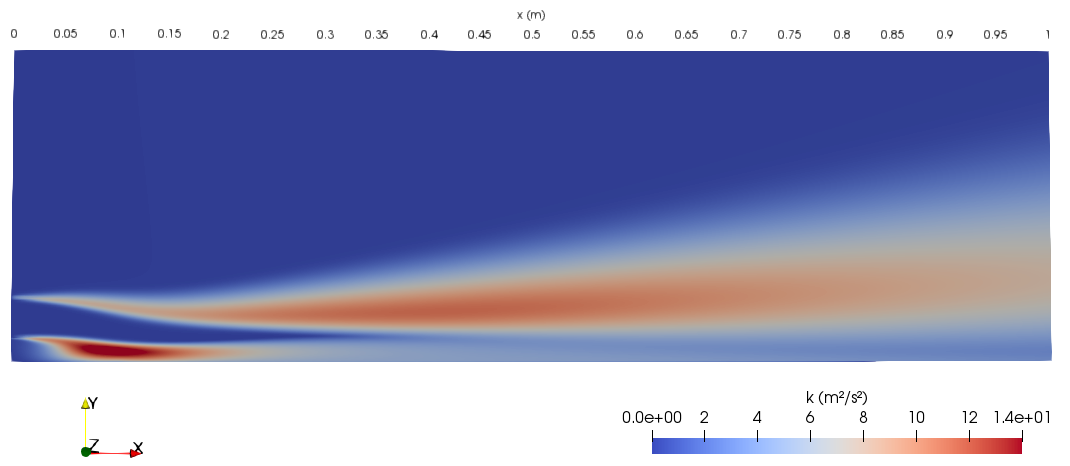}
  \caption{$k-\epsilon$}
  \label{fig:sub2}
\end{subfigure}

\caption{Turbulent kinetic energy contours around the flat plane near the jet exit obtained from RANS calculations}
\label{fig:offset_k}
\end{figure}

In the experimental results, the RMS streamwise fluctuating velocities are given to assess the presence of turbulence in the flow. Due to the steady nature of the solution, RANS cannot permit to obtain that quantity. Instead, Figure \ref{fig:offset_k} gives the contours of the turbulence kinetic energy $k=1/2(\overline{u'^2}+\overline{v'^2}+\overline{w'^2})$. It can be seen that upstream of the attachment point the highest turbulence are found in the inner and outer shear layer of the jet. After the reattachment, the turbulence propagates in all the jet, while its intensity decreases. This behaviour is similar to what was captured by Fu et al. \cite{fu_insights_2016}.

\subsubsection{Conclusion on the jet attaching to an offset flat plate}
It appears that both turbulence models used are capable of capturing relatively accurately the flow considered. In fact, the flow behaves as expected with the jet bending and attaching to the flat plane due to the Coand\u{a} effect. In addition, the reattachment length and the velocity profiles at least up to $x/H_j \le 6$ are in good agreement with the experiment. When going further downstream some discrepancies in the jet development can be observed, they remain however relatively low and the main trends of the flow are still captured. It, therefore, appears that using a 2D RANS simulation is a viable option to simulate the Coand\u{a} effect when considering a jet blown close to an offset flat plane. When it comes to what turbulence model is performing better, $k-\omega$ SST gives more accurate results near the jet exit but also further downstream. This last fact was surprising as $k-\epsilon$ is known to perform well in planar free-shear layer, which is the flow considered in the wall jet region. $k-\omega$ SST was however designed to replicate the free-stream independence of $k-\epsilon$ and seems to be successful here.

\newpage
\subsection{Jet blown on a cylinder}
\begin{figure}[h!]
\centering
\includegraphics[width=10cm]{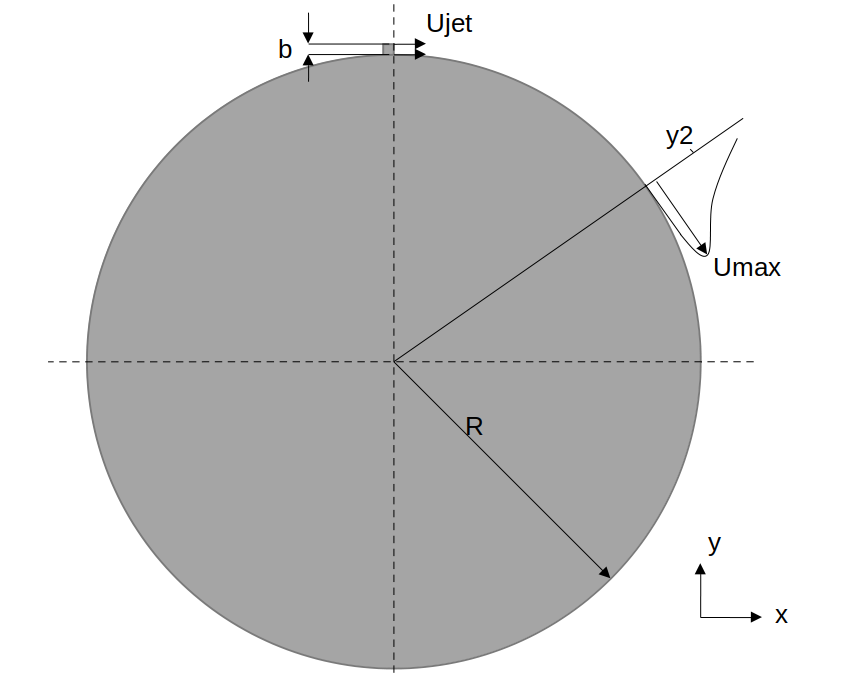}
\caption{Schematic of the cylinder}
\label{fig:schematic_cylinder}
\end{figure}

It has been found in Section \ref{sec:jet_offset_results} that a 2D RANS simulation was mostly satisfactory to capture the impact of the Coand\u{a} effect for an offset jet blown on a flat plane. Introducing important curvature in the geometry could however change that and it has actually been detailed in the literature review Section \ref{sec:literature} that numerically capturing a Coand\u{a} flow around a cylinder is a challenging task. In this part of the report, this is what is going to be attempted, by reproducing the results from the experiment of Wygnanski et al. \cite{neuendorf_turbulent_2000,neuendorf_turbulent_1999, likhachev_streamwise_2001,cullen_role_2002, han_streamwise_2004, neuendorf_large_2004}. The schematic of the cylinder is given in Figure \ref{fig:schematic_cylinder}. The radius of the cylinder is $R=0.1016\ m$, the jet exit velocity is $U_{jet}=48\ m.s^{-1}$ and the inlet height is $b=2.34 \times 10^{-3}\ m$. The fluid considered is air with a viscosity $\nu=1.5 \times 10^{-5}\ m^2.s^{-1}$, which gave a Reynolds number based on the inlet height of 7488. It can be noted that here, at the difference from the offset jet, $U_{jet}$ refers to the maximum velocity at the inlet and not the average.

\subsubsection{Results for 2-dimensional RANS simulations}

\myparagraph{Simulation strategy}
\label{sec:2d_rans_strategy}
Similarly to the offset jet, the relatively low velocity encountered permit to make the incompressible hypothesis. The equations are solved using the SIMPLEC algorithm. The gradients are computed using a centred second order scheme. For the divergence, U is computed using second order Upwind, but $\omega$ and $k$ are computed with first order Upwind. This choice was made since using second order scheme give more accurate results, but using second order Upwind for the turbulent quantities, was found to yield unphysical results with the flow staying attached to the cylinder for more than $360^{\circ}$ (results not included in this report).\\
\\
When it comes to the turbulence model chosen, it was seen in Section \ref{sec:jet_offset_results} that $k-\omega$ SST was performing better than $k-\epsilon$. Since now, a flow with a separation and an important streamline curvature is considered, both criteria known to make $k-\epsilon$ inaccurate, only $k-\omega$ SST will be kept. To try to account the streamline curvature, a modified version of $k-\omega$ SST with Curvature Correction will also be considered. It can be noted that the Curvature Correction implementation for $k-\omega$ SST is not part of the official OpenFOAM\textsuperscript{\textregistered} distribution and was obtained in a public repository developed by Ancolli \cite{colli_ancollikomegasstcc_2019}. When considering the results given by that model, it is therefore important to keep in mind that the validation of the code cannot be guaranteed. These two turbulence models appear as the most promising according to Frunzulica et al. \cite{frunzulica_method_2017} and will, therefore, be the only ones considered.    

\myparagraph{Mesh}
\label{sec:mesh}
In order to get the best mesh quality possible and since the geometry is simple, a structured grid strategy was adopted. Three hexahedral meshes with different level of refinement were generated. All the meshes extend up to 50 R away from the cylinder in order to reduce the impact of the boundary of the domain and as was advised by Gross et al. \cite{gross_coanda_2006}. Since a wall-resolved strategy is employed, a $y+<1$ was used. Another cell clustering was created at the top of the nozzle as it is another critical region for the flow considered. Different parameters of the grids are given in Table \ref{tab:2d_grids}.

\begin{table}[H]
\centering
\begin{tabular}{|c|c|c|c|}
\hline
               & Fine & Medium & Coarse \\ \hline
$y+$           &0.74      &0.85        &1
\\ \hline
Expansion rate &1.1      &1.15        &1.24      \\ \hline
$\Delta x$     &6.2e-4      &8.7e-4        & 1.2e-3       \\ \hline
Grid points    &229152 &      117120 &        59850        \\ \hline
\end{tabular}
\caption{Parameters for definition of the grids around the cylinder for the 2D RANS simulations}
\label{tab:2d_grids}
\end{table}

Again the values of $y+$ were obtained based on the flat-plate boundary layer theory for turbulent flow and are therefore only an assumption. After running the simulation for the $k-\omega$ SST model, the actual values of the non-dimensional spacing was found to be 0.31 for the fine grid, 0.36 for the medium and 0.42 for the coarse. The condition on $y+$ is therefore well respected. To permit a better grasp of the meshing strategy, the coarse grid is presented in Figure \ref{fig:grid_2dRANS} for different levels of zoom.

\begin{figure}[h!]
\centering
\begin{subfigure}{0.45\textwidth}
  \centering
  \includegraphics[width=1\textwidth]{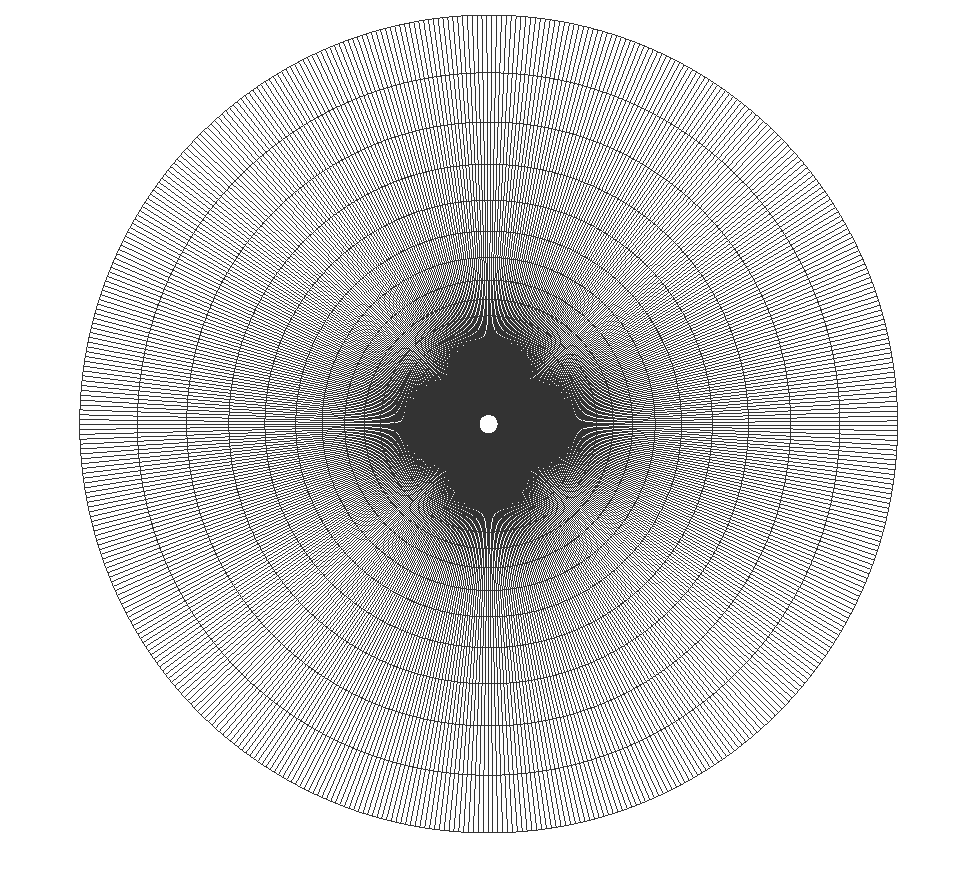}
  \caption{Total grid}
  \label{fig:sub1}
\end{subfigure}%
\caption{Coarse grid used for the 2D RANS simulations of the coand\u{a} flow around the cylinder}
\label{fig:grid_2dRANS}
\end{figure}

\begin{figure}[h!]\ContinuedFloat
\begin{subfigure}{0.49\textwidth}
  \centering
  \includegraphics[width=1\textwidth]{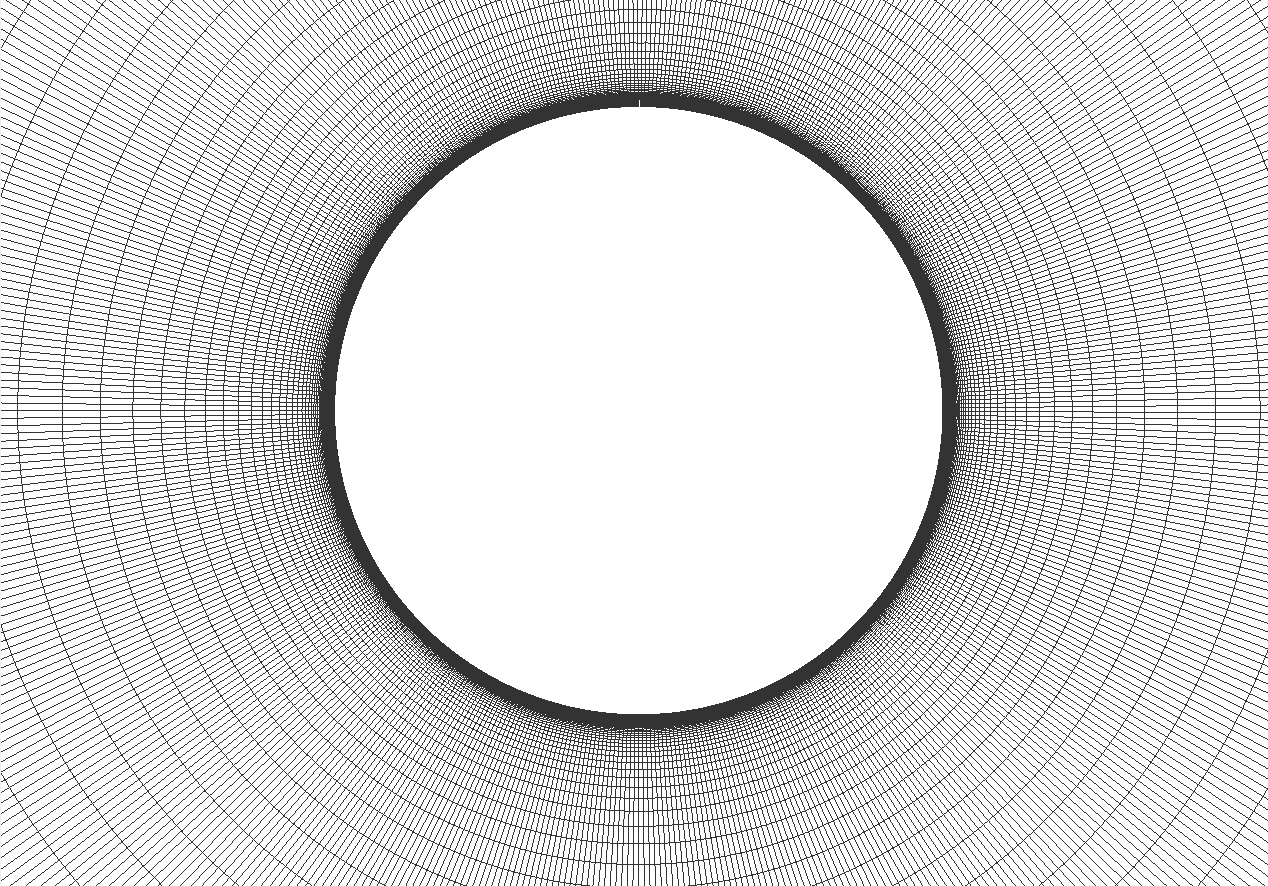}
  \caption{Grid around cylinder}
  \label{fig:sub1}
\end{subfigure}%
\vspace{0.5mm}
\begin{subfigure}{0.49\textwidth}
  \centering
  \includegraphics[width=1\textwidth]{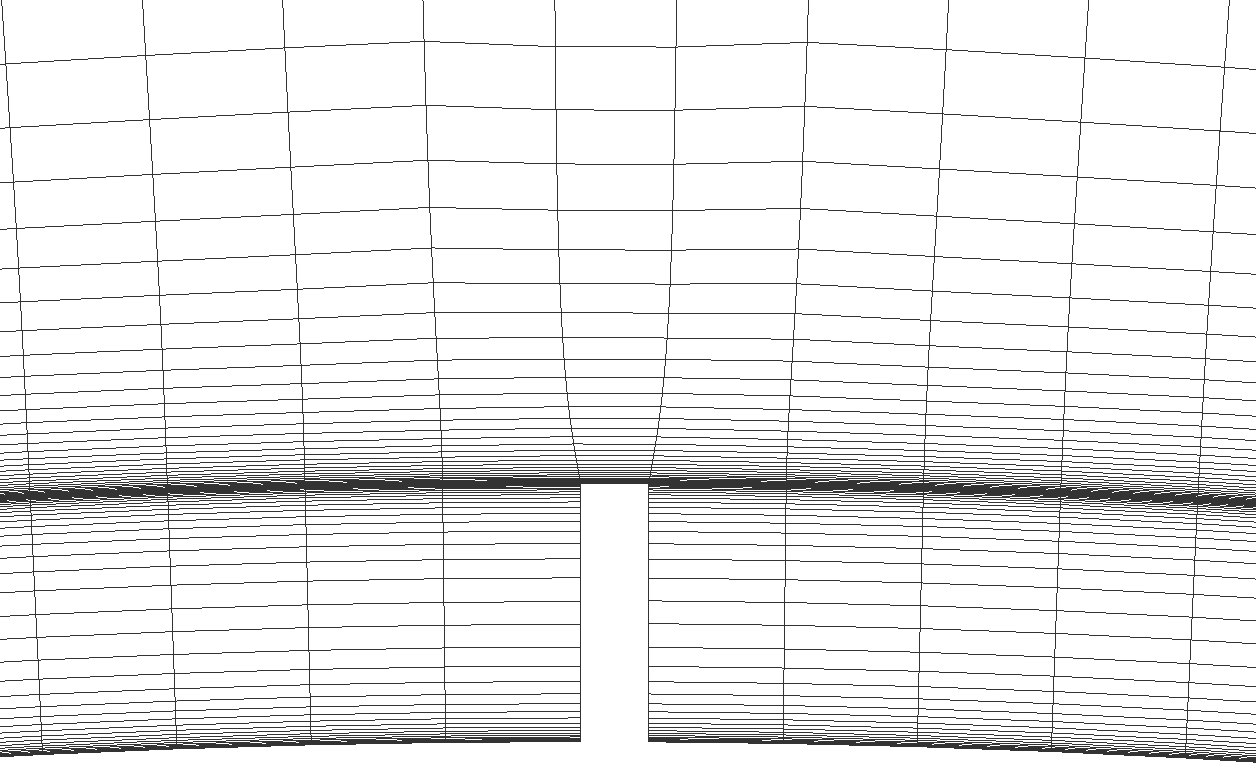}
  \caption{Grid around nozzle exit}
  \label{fig:sub2}
\end{subfigure}

\caption[]{Coarse grid used for the 2D RANS simulations of the coand\u{a} flow around the cylinder (Continued)}
\end{figure}

\begin{figure}[h!]
\centering
\includegraphics[width=8cm]{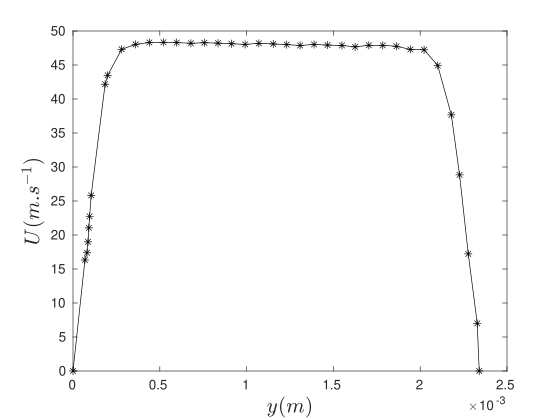}
\caption{Velocity profile at inlet for the jet blown tangentially to a cylinder}
\label{fig:velocity_inlet}
\end{figure}

\myparagraph{Boundary and initial conditions}
\label{sec:boundary}
The boundary conditions were considered as follows. The nozzle exit is an inlet, the extent of the computational domain an outlet and the cylinder a no-slip wall. The fluid is initially considered at rest with $p=0$ and a velocity vector of $(0,0,0)$ in the internal domain.\\
\\
For the prescribed velocity profile at the jet exit, it was decided not to rely on a top-hat velocity profile assumption as done by other authors \cite{gross_coanda_2006,frunzulica_method_2017}, due to the importance of that profile for the development of the flow. Instead, the mean velocity profile given by Neuendorf and Wygnanski \cite{neuendorf_turbulent_1999} for the calibration, was scaled to be at the desired jet velocity of $48m.s^{-1}$ and used as the velocity profile at the inlet. This profile is given in figure \ref{fig:velocity_inlet}.

In order to resolve the flow as accurately as possible, a wall-resolved strategy is employed as the $y+<1$ indicates. Following the recommendation from Menter \cite{menter_zonal_1993}, all turbulent quantities on the cylinder are set to 0, except $\omega$ that is estimated by Equation \ref{equ:omega_wall}.
The initial and inlet boundary conditions are based on the estimate of isotropic turbulence and given Equations \ref{equ:k_isotropic} and \ref{equ:omega_isotropic}. The turbulence intensity I and the turbulence length scale L, are not available in the experiment and will be considered as $I=2.3\times10^{-5}$ and $L=1.16 \times 10^{-8}\ m$
as used by Gross et al. \cite{gross_coanda_2006}.

\myparagraph{Grid convergence study}
Again to assess the independence of the results from the grid, the grid convergence methodology introduced by Roache \cite{roache_verification_1998} will be carried out on the three grids presented in section \ref{sec:mesh} for the average pressure coefficient $C_p$ on the cylinder, using the $k-\omega$ SST turbulence model.\\
The pressure coefficient $C_p$ is defined in Equation \ref{eq:cp}. Similarly to the previous GCI $p_{\infty}$ will be taken as $0\ Pa$ and $\rho_{\infty}$ as $1\ kg.m^{-3}$. The average $C_p$ on the cylinder for the different grids are given in Table \ref{tab:cp}.

\begin{table}[H]
\centering
\begin{tabular}{|c|c|c|c|}
\hline
              & Fine  & Medium & Coarse \\ \hline
Average $C_p$ & -1.65e-2 & -1.71e-2  & -1.94e-2  \\ \hline
\end{tabular}
\caption{Average value of $C_p$ on the cylinder for the different grids}
\label{tab:cp}
\end{table}

The results of the GCI analysis based on these coefficients are given in Table \ref{tab:gci}. Similarly, as for the previous study, the grids are numbered from 1 (fine) to 3 (coarse). The high decrease between $GCI_{12}$ and $GCI_{23}$ proves that the result is near to be grid-independent. In addition, the value of $GCI_{12}$ is relatively small, showing that we are close to the Richardson extrapolated value between fine and medium grids. Finally, the GCI ratio is close to 1, proving that the solutions are in the asymptotic range of convergence. All these indicate that using the fine grid for the rest of the calculations will give results that are nearly independent from the grid. 

\begin{table}[H]
\centering
\begin{tabular}{|c|c|c|c|c|c|}
\hline
              & r   & q    & $GCI_{12}$ (\%) & $GCI_{23}$ (\%) & $GCI ratio$ \\ \hline
Average $C_p$ & 1.4 & 3.98 & 1.60            & 5.91            & 0.965                          \\ \hline
\end{tabular}
\caption{GCI analysis for the average $C_p$ on the cylinder}
\label{tab:gci}
\end{table}

\begin{figure}[H]
    \centering    
    \includegraphics[width=0.49\linewidth]{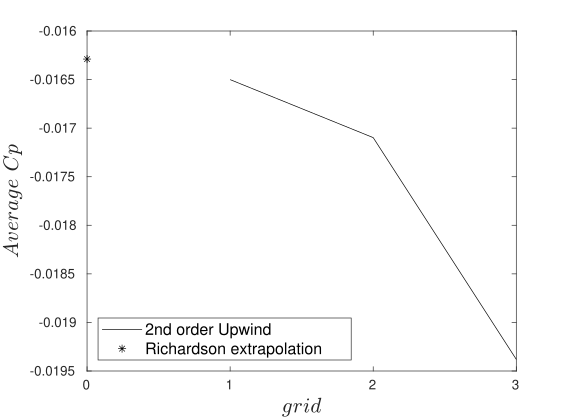}
    \caption{Average $C_p$ on the cylinder for fine (1) to coarse (3) grids and Richardson extrapolation}
    \label{fig:gci_plot}
\end{figure}

To get a more visual representation of the evolution of $C_p$ for the different grids, compared to the extrapolated value based on the two finer ones, the Figure \ref{fig:gci_plot} is given.

\begin{figure}[h!]
\centering
\begin{subfigure}{0.5\textwidth}
  \centering
  \includegraphics[width=0.99\textwidth]{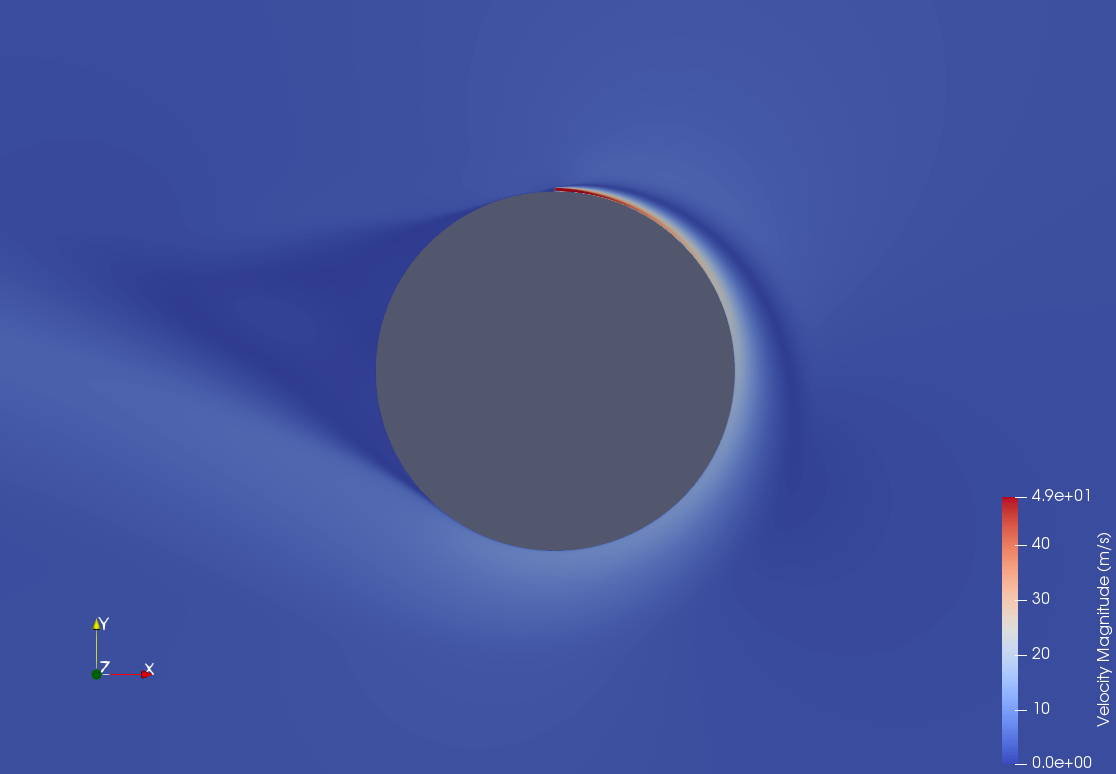}
  \caption{$k-\omega$ SST}
  \label{fig:sub1}
\end{subfigure}%
\begin{subfigure}{0.5\textwidth}
  \centering
  \includegraphics[width=0.99\textwidth]{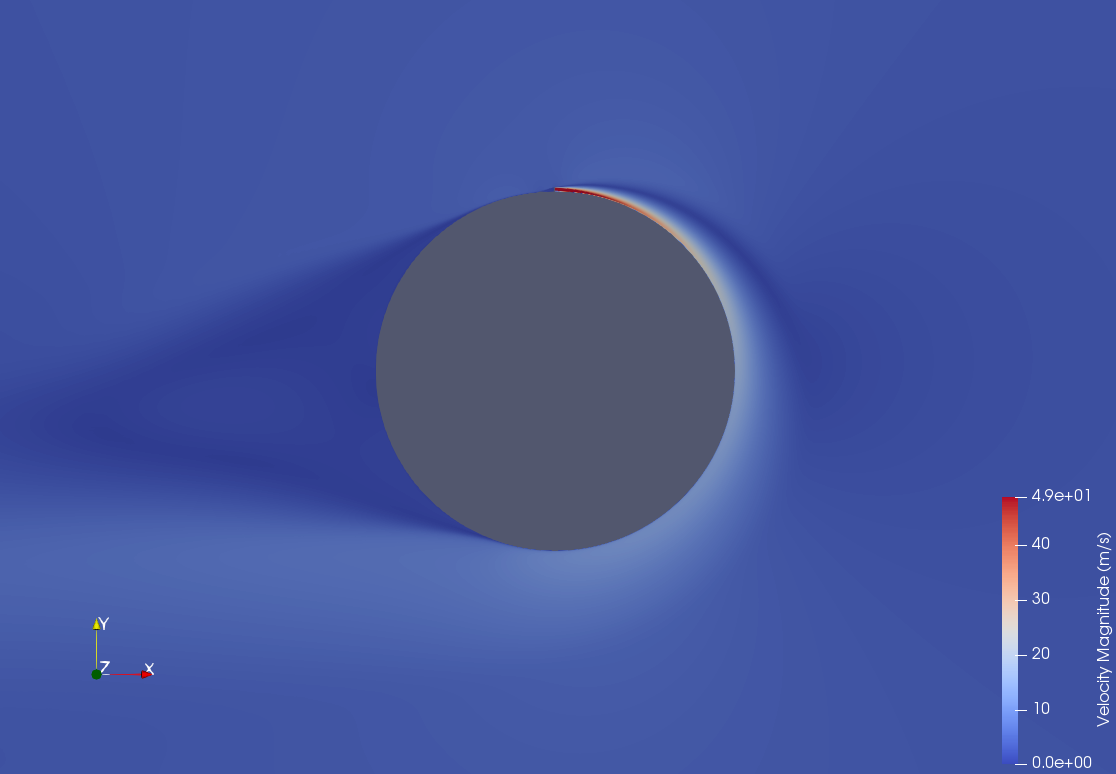}
  \caption{$k-\omega$ SST with Curvature Correction}
  \label{fig:sub2}
\end{subfigure}
\caption{Velocity contours around the cylinder obtained from 2D RANS calculations}
\label{fig:2D_RANS_contours}
\end{figure}

\begin{figure}[H]
\begin{subfigure}{0.5\textwidth}
  \centering
  \includegraphics[width=0.99\textwidth]{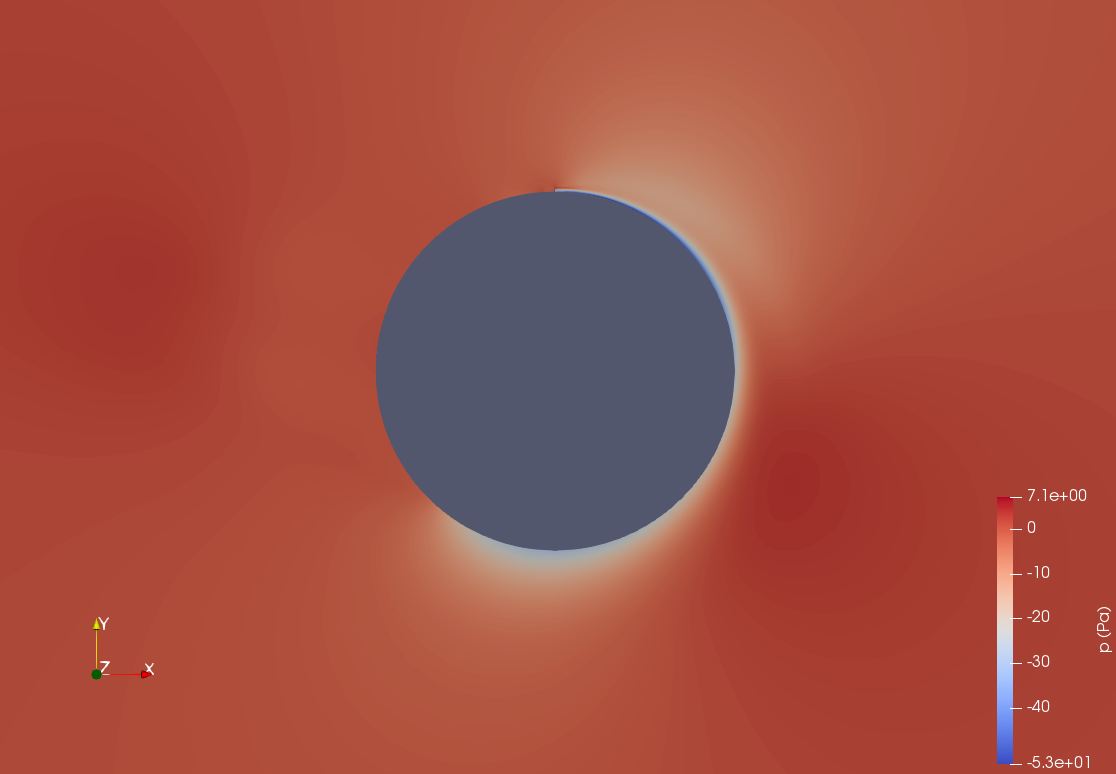}
  \caption{$k-\omega$ SST}
  \label{fig:sub1}
\end{subfigure}%
\begin{subfigure}{0.5\textwidth}
  \centering
  \includegraphics[width=0.99\textwidth]{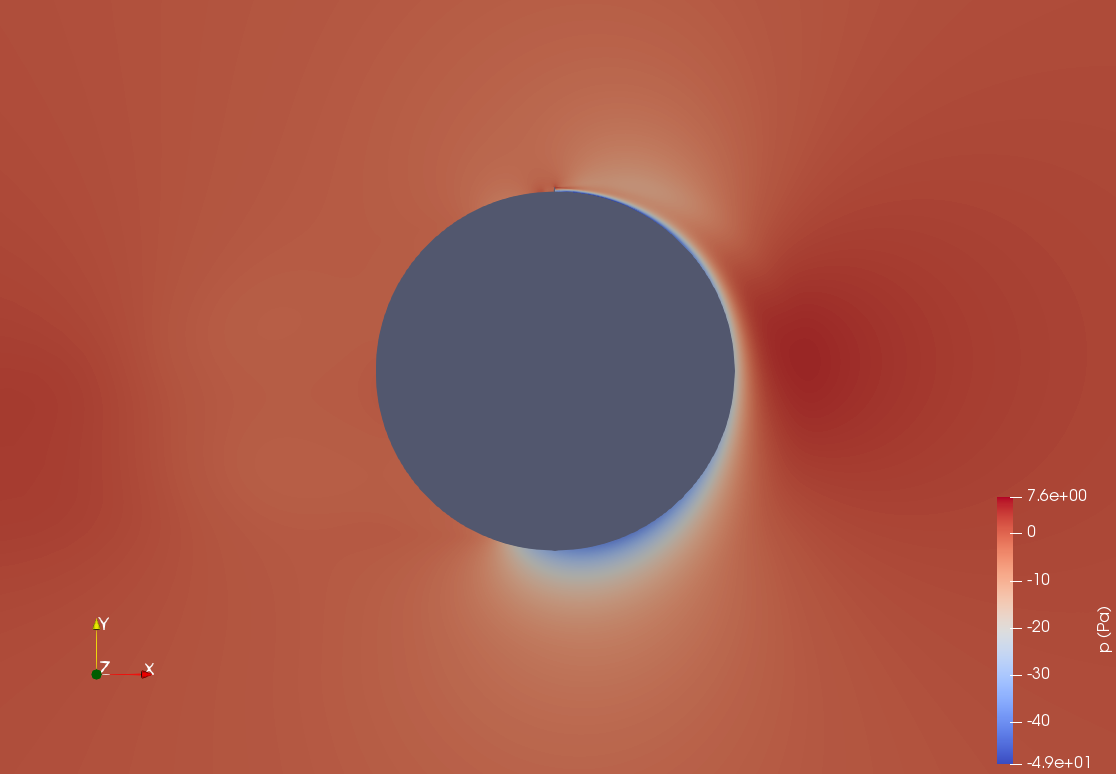}
  \caption{$k-\omega$ SST with Curvature Correction}
  \label{fig:sub2}
\end{subfigure}

\caption{Pressure contours around the cylinder obtained from 2D RANS calculations}
\label{fig:2D_RANS_contours_pressure}
\end{figure}

\myparagraph{Flow around the cylinder}
\label{sec:flow_2d_rans}
In this section, the flow around the cylinder simulated with the $k-\omega$ SST turbulence model, with and without Curvature Correction will be given and compared to the experimental results obtained by Wygnanski et al. \cite{neuendorf_turbulent_2000,neuendorf_turbulent_1999, likhachev_streamwise_2001,cullen_role_2002, han_streamwise_2004, neuendorf_large_2004}. First, the velocity contours around the cylinder for both turbulence models are given in Figure \ref{fig:2D_RANS_contours}. The impact of the Coand\u{a} effect can be well observed in this test case, with the jet blown in the positive $x$-direction, staying attached for more than $180^{\circ}$ on the cylinder and therefore having its direction reversed. The sensibility of the simulation to the numerical parameters chosen can also be seen, with important differences between the two contours, especially around the separation region.\\
\\
 In Figure \ref{fig:2D_RANS_contours_pressure} the pressure contours are presented. The lower pressure region created under the jet responsible for the generation of lift by the Coand\u{a} effect can be well observed. It can be seen that this region is occupying a much more important section of the wall than for the offset jet on a flat plane. This is due to the curvature of the geometry that recreates the mechanisms responsible for the Coand\u{a} effect for every small change in curvature as detailed in the introduction Section \ref{sec:intro} and explains why curved geometries are generally used for applications of the Coand\u{a} effect making use of the aerodynamic forces generated by it (e.g: NOTAR helicopter, Circulation Control Airfoil).

\begin{figure}[H]
    \centering    
    \includegraphics[width=0.49\linewidth]{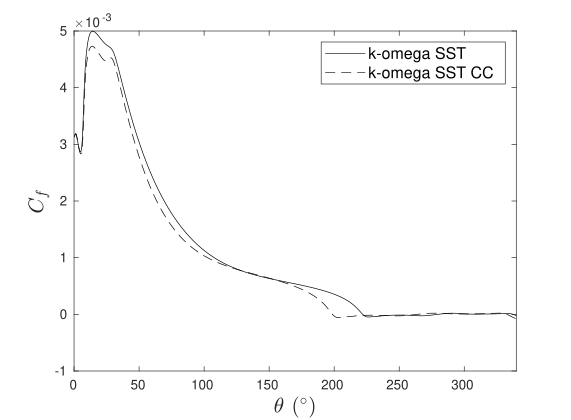}
    \caption{Streamwise $C_f$ around the cylinder obtained from 2D RANS calculations}
    \label{fig:skin_friction}
\end{figure}

The separation location of the jet with the cylinder surface is an important aspect of the flow that will now be discussed. Wygnanski et al. experimentally found a separation location at around $\theta \approx 220^\circ$ \cite{neuendorf_turbulent_1999}. Similarly, as for the offset jet on a flat plane, the separation angles found numerically were computed by using the location where the streamwise skin friction coefficient around the cylinder reaches 0. The Figure \ref{fig:skin_friction} gives the plot of $C_f$ around the cylinder for both turbulence models. Using this plot the separation is found at $\theta=221.9^\circ$ for $k-\omega$ SST and at $\theta=198.7^\circ$ for $k-\omega$ SST with Curvature Correction. It can be concluded that the simulation using $k-\omega$ SST model has a separation in really good agreement with the experiment. Surprisingly using a Curvature Correction gave a less accurate result, although still relatively close to the experiment.

\begin{figure}[H]
\centering
\begin{subfigure}{9cm}
  \centering
  \includegraphics[width=8.9cm]{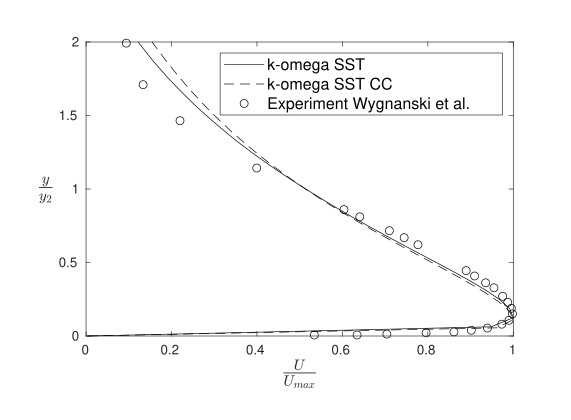}
  \caption{$50^{\circ}$}
  \label{fig:sub1}
\end{subfigure}%
\begin{subfigure}{9cm}
  \centering
  \includegraphics[width=8.9cm]{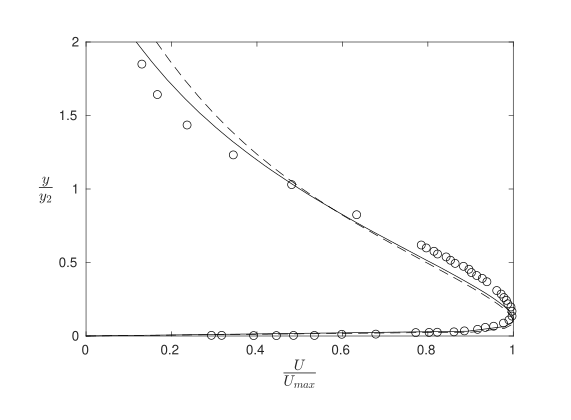}
  \caption{$90^{\circ}$}
  \label{fig:sub2}
\end{subfigure}

\begin{subfigure}{9cm}
  \centering
  \includegraphics[width=8.9cm]{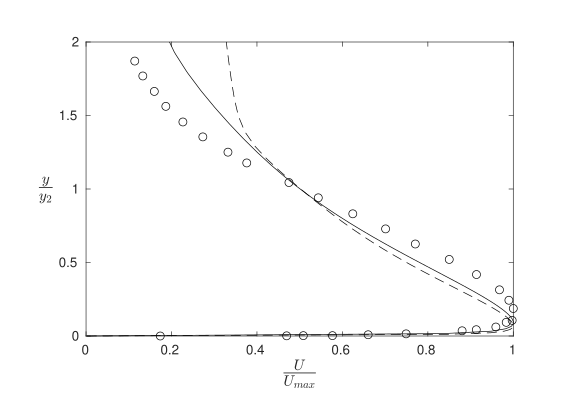}
  \caption{$140^{\circ}$}
  \label{fig:sub1}
\end{subfigure}%
\begin{subfigure}{9cm}
  \centering
  \includegraphics[width=8.9cm]{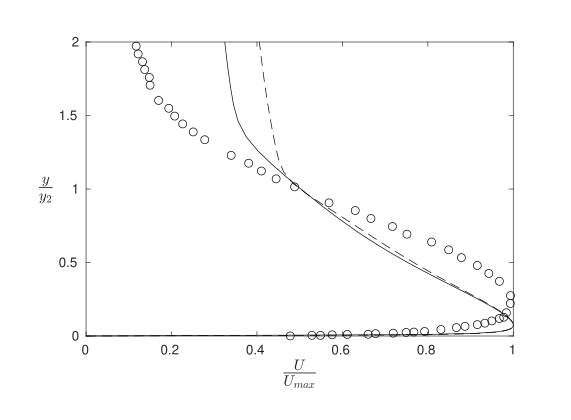}
  \caption{$180^{\circ}$}
  \label{fig:sub_180_vel}
\end{subfigure}

\caption{Comparison of normalized streamwise velocity profiles around the cylinder at various angles obtained from the 2D RANS calculations with experimental data \cite{neuendorf_turbulent_1999}}
\label{fig:2D_velocity_profiles}
\end{figure}

Another important aspect of the flow that will now be considered is the velocity profiles around the cylinder. The Figure \ref{fig:2D_velocity_profiles} gives the normalized velocity profiles obtained at different angles compared to the experimental results from Neuendorf and Wygnanski \cite{neuendorf_turbulent_1999}. The profiles are normalized by the maximum velocity at the given angle $U_{max}$ and the jet half-thickness $y_2$, which is the thickness where the velocity $U$ is half of $U_{max}$.\\
\\
It can be seen that the profiles at $\theta=50^{\circ}$ and $90^{\circ}$ are in relatively good agreement with the experiment. However, at $\theta=140^{\circ}$ some notable discrepancies start to be observed and at $\theta=180^{\circ}$ this is even more true. It is documented by Neuendorf and Wygnanski \cite{neuendorf_turbulent_1999}, that the flow around the cylinder can be decomposed into two distinct regions. The first one going up to $\theta=120^{\circ}$ characterized by a nearly constant surface pressure and self-similar velocity profiles. While the second region is characterized by an adverse pressure gradient that will lead to the separation of the flow from the wall. Based on the results presented, it appears that if the simulations capture relatively well the first region, it is not the case for the second one, and the presence of an adverse pressure gradient might be in cause even if $k-\omega$ SST is known to handle that pretty well. Another explanation could be the impact of the streamline curvature on the flow that is poorly captured. However, it can surprisingly be noted that using a Curvature Correction in our simulation is actually making the prediction of the velocity profiles less accurate.

\begin{figure}[H]
\centering
\begin{subfigure}{9cm}
  \centering
  \includegraphics[width=8.9cm]{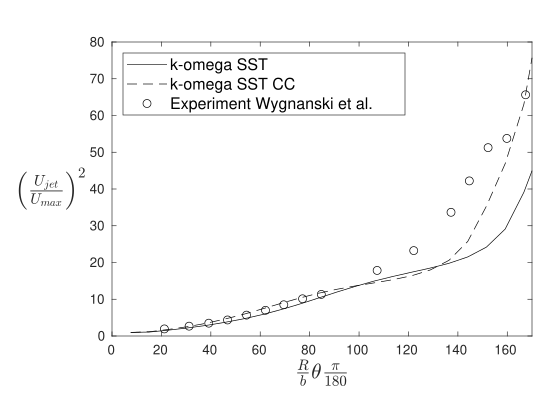}
  \caption{Jet velocity decay}
  \label{fig:sub1}
\end{subfigure}%
\begin{subfigure}{9cm}
  \centering
  \includegraphics[width=8.9cm]{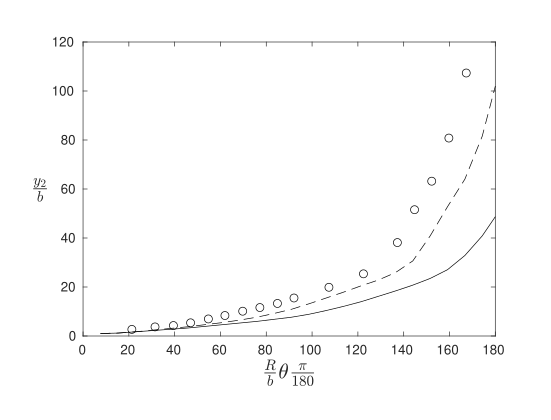}
  \caption{Jet half-width}
  \label{fig:sub2}
\end{subfigure}

\caption{Comparison of normalized velocity decay and jet half-width around the cylinder obtained from 2D RANS calculations with experimental data \cite{neuendorf_turbulent_1999}}
\label{fig:2D_RANS_decay}
\end{figure}

To get a better understanding of the flow around the whole cylinder, the normalized jet half-thickness and the normalized decay of the maximum velocity are given in Figure \ref{fig:2D_RANS_decay}. It can be observed that both quantities tend to be under-predicted by the simulation. This trend has generally been found in previous work trying to simulate this set-up using 2D RANS \cite{gross_coanda_2006,frunzulica_method_2017}. For these particular quantities, the Curvature Correction does appear to permit some improvement. As was done for the offset jet on a flat wall, the maximum error for both quantities was computed. For the jet velocity decay, the maximum error was approximately $37.2\%$ with Curvature Correction and $51.9\%$ without. When it comes to the jet half-width an error of $39.1\%$ was found for the model with Curvature Correction and $68.8\%$ without. The errors with the offset jet presented in Section \ref{sec:jet_offset_results} were below $20\%$ for $k-\omega$ SST, showing how much a change of geometry has affected the jet development prediction.\\
\\
With such poor prediction of the jet development, it is surprising that the separation locations found were so close to the experiment. In fact, with an under-prediction of the decay and rate of spread of the jet, it would be expected to have a separation further downstream than in the experiment. However as can be observed in Figure \ref{fig:sub_180_vel}
it appears that the entrainment of surrounding fluid by the jet is overpredicted numerically. According to Neuendorf and Wygnsnaki \cite{neuendorf_turbulent_1999}, the jet entrainment of surrounding fluid is the main motor of the separation. It, therefore, appears that the under-prediction of the jet development and the over-prediction of the entrainment compensate each other and ultimately gave a satisfying separation location. This situation is obviously not ideal as a change in flow condition or geometry might break that balance and give poor results.

\begin{figure}[H]
\centering
\begin{subfigure}{9cm}
  \centering
  \includegraphics[width=8.9cm]{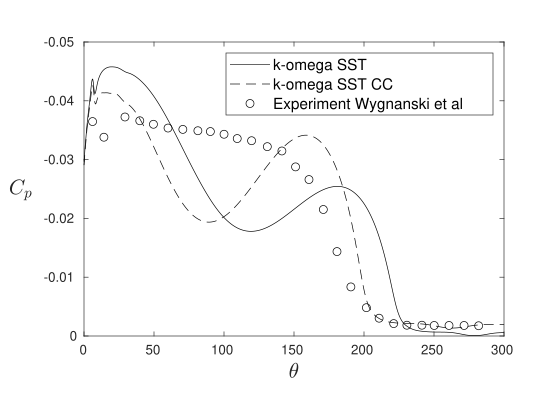}
\end{subfigure}%
\caption{Comparison of wall pressure coefficient on the cylinder obtained from 2D RANS calculations with experimental data \cite{neuendorf_turbulent_1999}}
\label{fig:cp_cylinder}
\end{figure}

The streamwise variation of the pressure coefficient on the cylinder wall is presented in Figure \ref{fig:cp_cylinder}. As detailed previously in the first region of the flow, below $\theta=120^{\circ}$, the surface pressure is nearly constant. This was however not found numerically, with an increase in pressure actually found in that region for both models used. Surprisingly this did not affect the velocity profiles too much, as good agreement with the experimental ones were found in that region. The adverse pressure gradient downstream of $\theta=120
^{\circ}$ cannot be said to have been well predicted either, around the separation, the drop of pressure to 0 is however found.

\begin{figure}[H]
\begin{subfigure}{0.5\textwidth}
  \centering
  \includegraphics[width=0.99\textwidth]{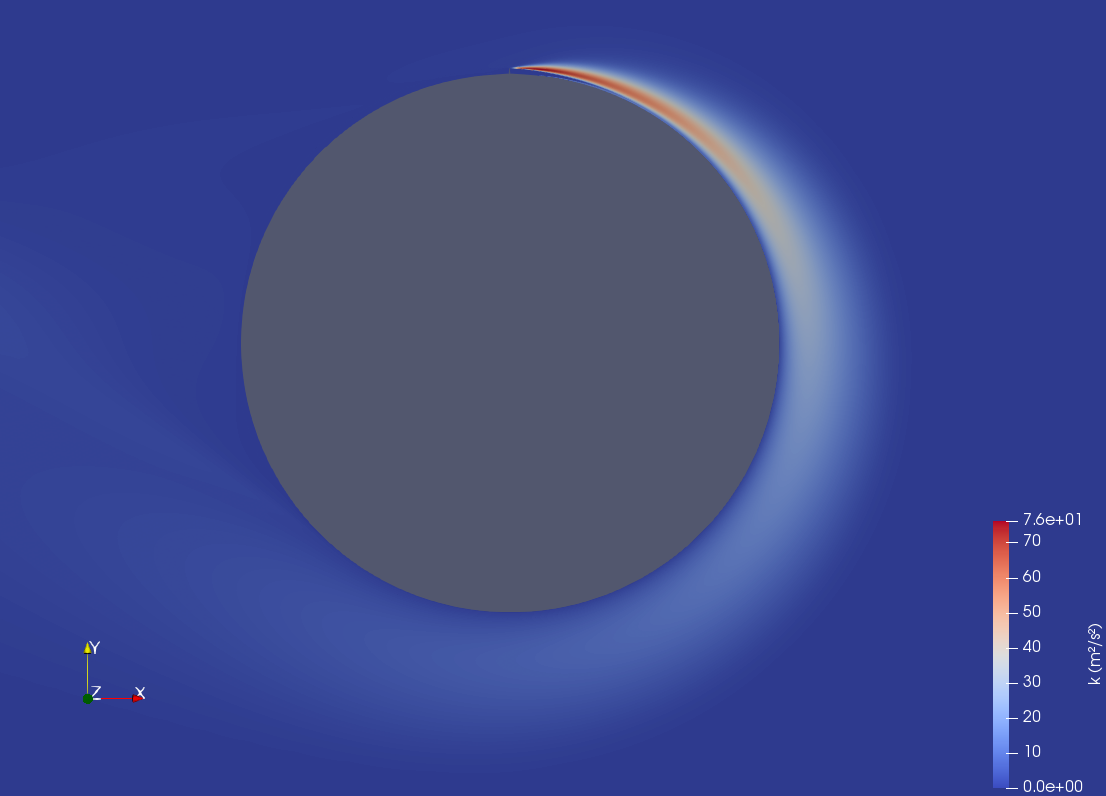}
  \caption{$k-\omega$ SST}
  \label{fig:sub1}
\end{subfigure}%
\begin{subfigure}{0.5\textwidth}
  \centering
  \includegraphics[width=0.99\textwidth]{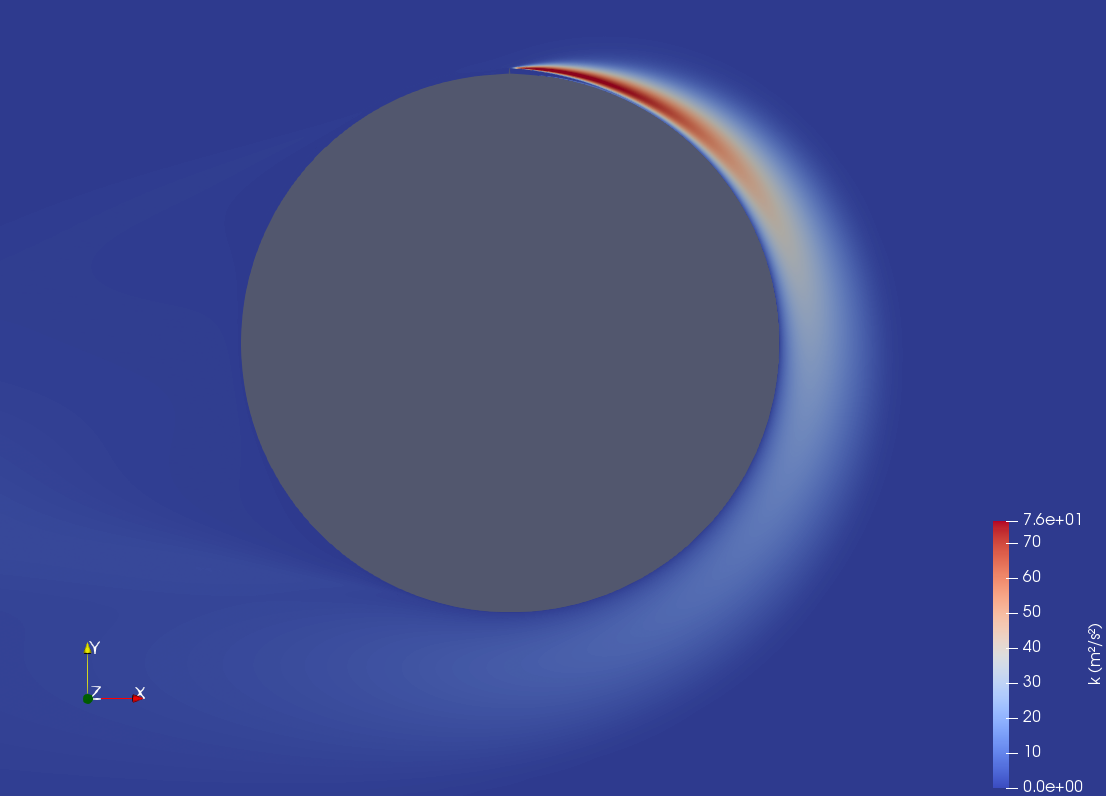}
  \caption{$k-\omega$ SST with Curvature Correction}
  \label{fig:sub2}
\end{subfigure}

\caption{Turbulent kinetic energy contours around the cylinder obtained from 2D RANS calculations}
\label{fig:tke_cylinder}
\end{figure}

The contours of the turbulent kinetic energy $k$ are presented in Figure \ref{fig:tke_cylinder}. It is observed that the region of maximum turbulence is in the outer shear layer of the jet near the inlet. The turbulence then diffuses in all the jet while being dissipated. This is similar to what was found for the offset jet, with the difference that no inner shear layer is present here and there is therefore not a second region of high turbulence. The fact that the Curvature Correction introduces a higher level of turbulence can be noted.

\myparagraph{Conclusion on 2D RANS}
It was shown that 2D RANS using a $k-\omega$ SST turbulence model with and without Curvature Correction is capable to capture to some extent the impact of the Coand\u{a} effect for a flow around a cylinder as can easily be seen in Figure \ref{fig:2D_RANS_contours}. Some of the flow characteristics were in good agreement with the experimental results obtained by Wygnanski et al. \cite{neuendorf_turbulent_2000,neuendorf_turbulent_1999, likhachev_streamwise_2001,cullen_role_2002, han_streamwise_2004, neuendorf_large_2004}, namely the separation location and the velocity profiles at relatively low angles.  However, when going further downstream, the velocity profiles are presenting some important discrepancies, the jet decay and the jet half-thickness are both under-predicted proving that the jet development is not well captured. The fact that the jet development is so poorly captured makes the accurate separation location found rather surprising, and it is uncertain if for different flow conditions this will be the case. It, therefore, appears that introducing a streamline curvature in the flow makes the predictions of the 2D RANS simulation unreliable. Various factors could explain that, first if 2D RANS is an attractive model due to its low computational cost, it might be insufficient to capture the flow considered. In addition to that, the existence of 3D structures in the flow have been demonstrated by experimental work \cite{cullen_role_2002,han_streamwise_2004} and can obviously not be captured by a 2-dimensional domain. In the next section, an attempt will be made to use the same RANS set-up, with $k-\omega$ SST turbulence model in a 3D domain and see the impact it has on the results. \\
\\
When it comes to the impact of the Curvature Correction. If the correction gave some improved prediction of the jet decay and jet half-thickness, the velocity profiles and the separation were actually more accurate without it. Since the benefit of the correction are questionable and due to the lack of validation for its implementation in OpenFOAM\textsuperscript{\textregistered}, it was chosen to not use it for the following 3D simulations.

\subsubsection{Results for 3-dimensional RANS simulations}

As the 3D structures present in the flow around the cylinder might have an important impact on it, the RANS simulation carried out with a 2-dimensional domain was done using a 3-dimensional domain in an attempt to improve the results obtained. 

\myparagraph{Simulation strategy}
 The same flow conditions, algorithms and numerical schemes than for the 2D RANS simulation detailed in Section \ref{sec:2d_rans_strategy} were used. The turbulence model used is $k-\omega$ SST, as it could not be concluded that the Curvature Correction permitted definitive improvement and its implementation in OpenFOAM\textsuperscript{\textregistered} was not validated.

\myparagraph{Mesh}
The fine grid presented in Section \ref{sec:mesh} was extruded in the spanwise direction with 48 cells evenly distributed on a spanwise extent of $60\ mm$ to obtain the 3D mesh. The choice of spanwise extent was done by taking into account the results from Wernz et al. \cite{wernz_numerical_2003,wernz_numerical_2005}, in these papers, they carried out a DNS for a domain with a spanwise extent of $40\ mm$ and $80\ mm$ and a LES with an extent of $20\ mm$. The LES results were disappointing as it appears that the domain chosen was too thin, the DNS with an extent of $40\ mm$ gave promising results even if the longitudinal vortices far downstream were too large for the domain and finally the DNS with an extent of $80\ mm$ was mostly satisfying. In an attempt to be as accurate as possible while keeping computational cost low, an extent of $60\ mm$ was chosen in this thesis. \\
\\
To be rigorous, a grid convergence study should be carried out for the 3D simulation too. However, due to time and computational restraints and since a GCI was carried out for the 2D version of the grid, this was not done here. The value of $y+$ found for the 2D grid should be approximately the same for the 3D grid and was indeed found to be 0.30, the $y+<1$ condition is therefore respected. The final grid is composed of 10.9 million hexahedral cells.

\myparagraph{Boundary and initial conditions}
The boundary conditions for the inlet, outlet and cylinder are the same as what was used for the 2D case and described in Section \ref{sec:boundary}. For the top and bottom surfaces in the spanwise direction, cyclic boundary conditions were used.

In order to save computational time, the results obtained for the 2D simulation were extrapolated on the 3D domain and used as an initial condition.

\myparagraph{Flow around the cylinder}

\begin{figure}[h!]
\centering
\begin{subfigure}{0.5\textwidth}
  \centering
  \includegraphics[width=0.99\textwidth]{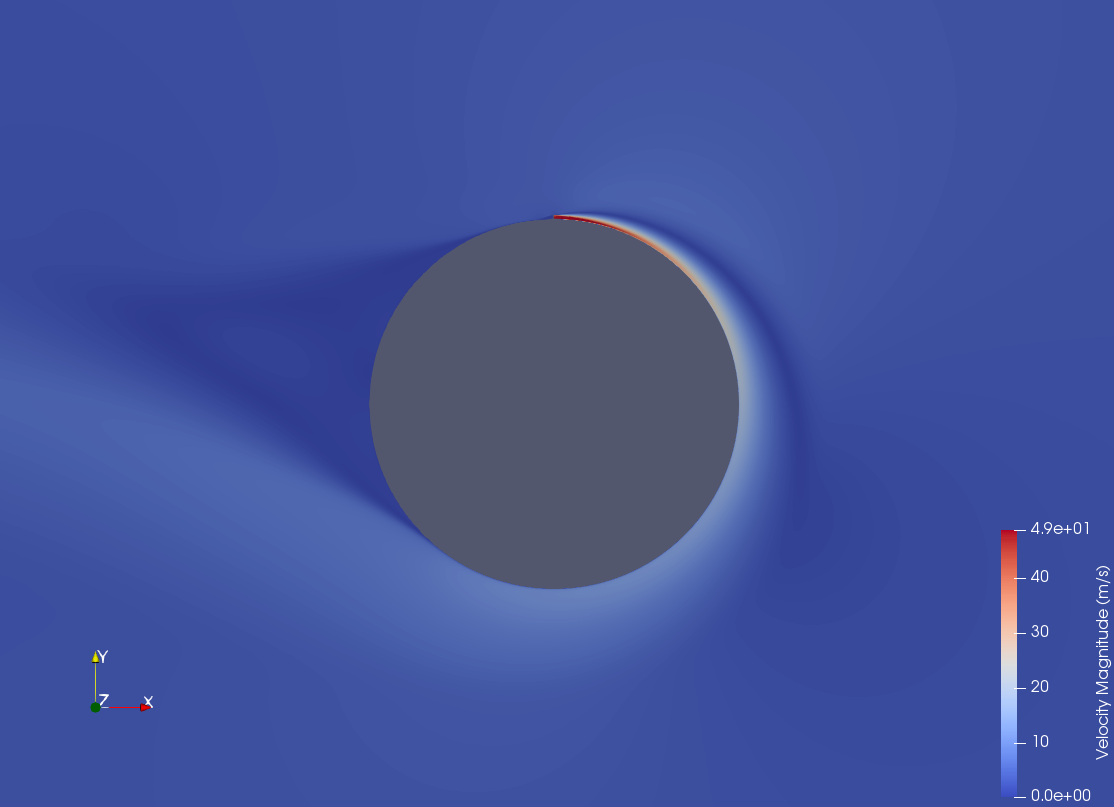}
  \caption{Velocity}
  \label{fig:sub1}
\end{subfigure}%
\begin{subfigure}{0.5\textwidth}
  \centering
  \includegraphics[width=0.99\textwidth]{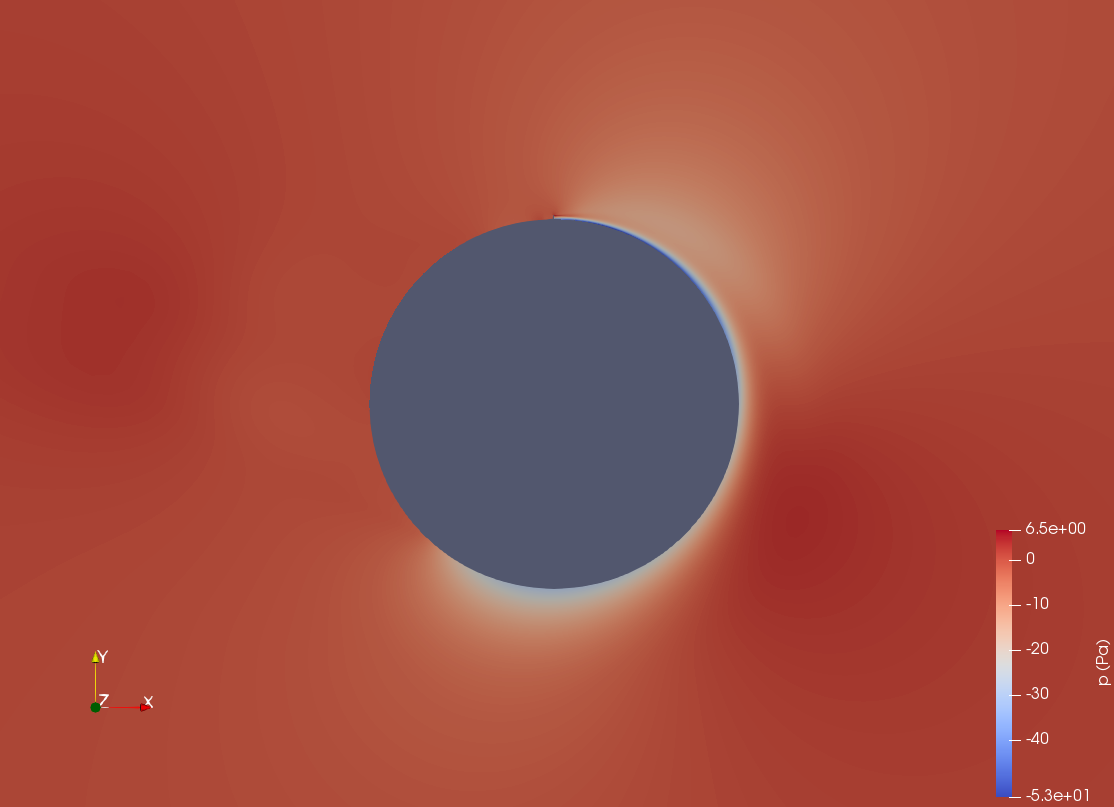}
  \caption{Pressure}
  \label{fig:sub2}
\end{subfigure}

\caption{Pressure and Velocity contours around the cylinder on the plane at $z=0.03m$ obtained from 3D RANS calculations}
\label{fig:3D_RANS_contours}
\end{figure}

To visualize the result of the flow around the cylinder, first the pressure and velocity contours on the plane of normal z and passing by the point $(0,0,0.03)$ is given in Figure \ref{fig:3D_RANS_contours}. It can be seen that visually the results appear to be similar to what was obtained for the 2D simulation presented in Figure \ref{fig:2D_RANS_contours} and \ref{fig:2D_RANS_contours_pressure}.

\newpage
A more quantitative comparison is now going to be done between 2D and 3D results. A few precisions need to be given on how the 3D data was transformed in 2D plot in order to be compared. First, at the difference from the previous simulations, some of the residuals were presenting periodic oscillations, that might be due to some unsteady behaviour captured by the RANS simulation. In order to still get meaningful results, the fields were averaged over 1000 iterations. In addition to that, the results were also averaged in space by taking 10 uniformly spaced slices in the spanwise direction. \\

\begin{figure}[H]
\centering
\begin{subfigure}{9cm}
  \centering
  \includegraphics[width=8.9cm]{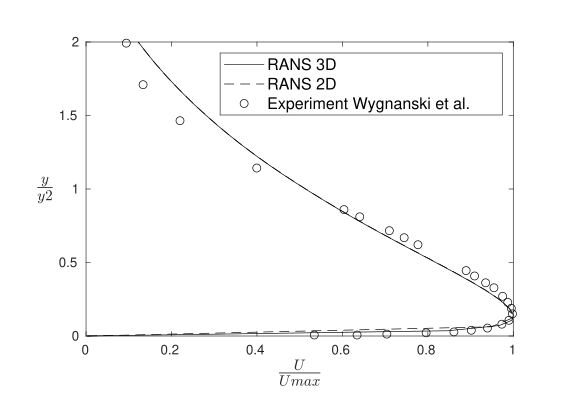}
  \caption{$\theta=50^{\circ}$}
  \label{fig:sub1}
\end{subfigure}%
\begin{subfigure}{9cm}
  \centering
  \includegraphics[width=8.9cm]{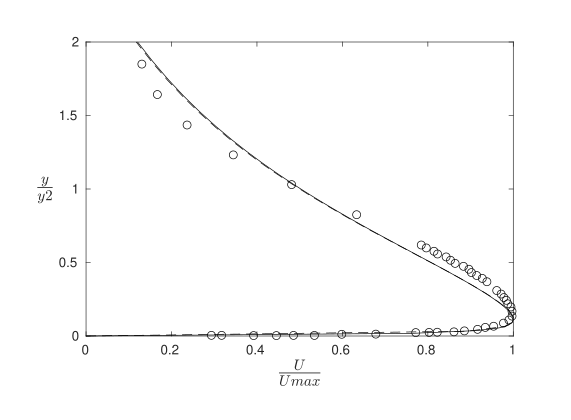}
  \caption{$\theta=90^{\circ}$}
  \label{fig:sub2}
\end{subfigure}

\begin{subfigure}{9cm}
  \centering
  \includegraphics[width=8.9cm]{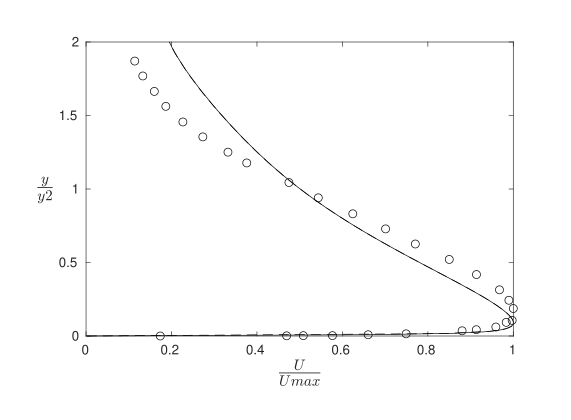}
  \caption{$\theta=140^{\circ}$}
  \label{fig:sub1}
\end{subfigure}%
\begin{subfigure}{9cm}
  \centering
  \includegraphics[width=8.9cm]{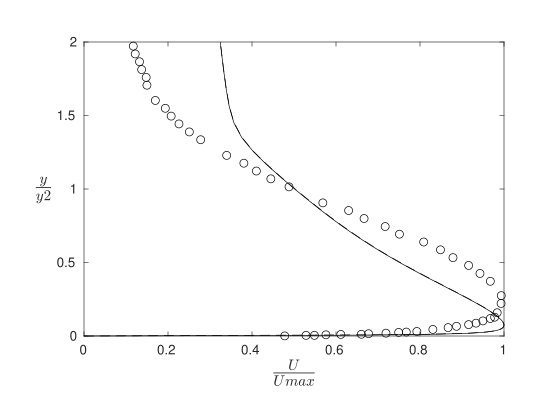}
  \caption{$\theta=180^{\circ}$}
  \label{fig:sub2}
\end{subfigure}

\caption{Comparison of normalized streamwise velocity profiles around the cylinder obtained from 2D and 3D RANS calculations with experimental data \cite{neuendorf_turbulent_1999}}
\label{fig:2d_against_3d_rans_1}
\end{figure}

\begin{figure}[H]
\begin{subfigure}{9cm}
  \centering
  \includegraphics[width=8.9cm]{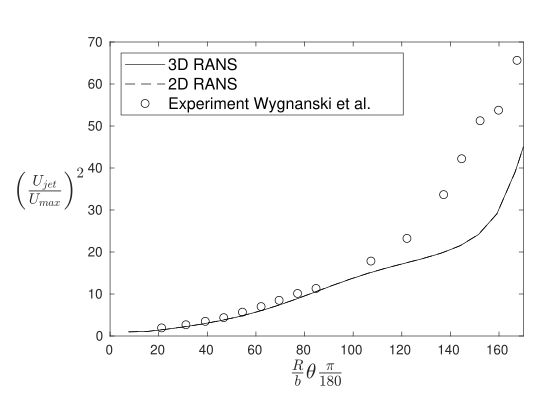}
  \caption{Jet velocity decay}
  \label{fig:sub1}
\end{subfigure}%
\begin{subfigure}{9cm}
  \centering
  \includegraphics[width=8.9cm]{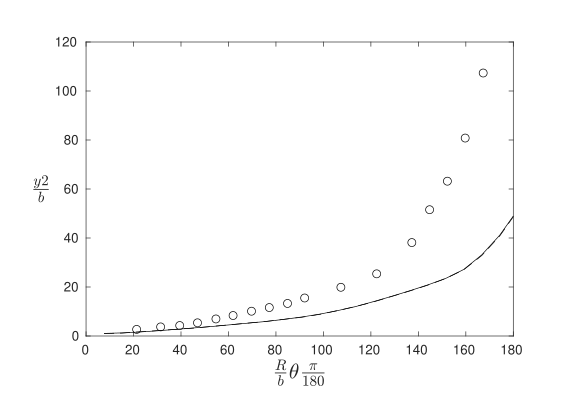}
  \caption{Jet half-width}
  \label{fig:sub2}
\end{subfigure}

\caption{Comparison of normalized jet decay and jet half-width around the cylinder obtained from 2D and 3D RANS calculations with experiment data \cite{neuendorf_turbulent_1999}}
\label{fig:2d_against_3d_rans_2}
\end{figure}

\begin{figure}[H]
\centering
  \includegraphics[width=8.9cm]{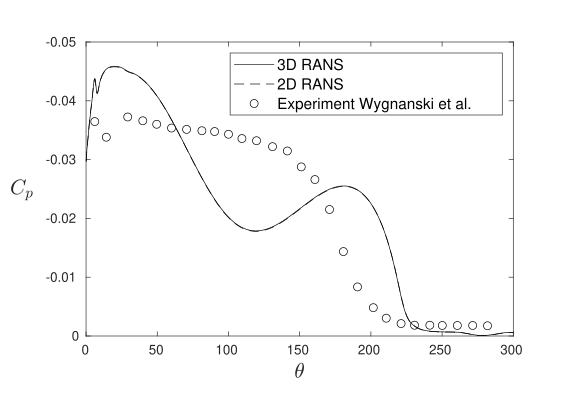}
  \label{fig:sub1}

\caption{Comparison of wall pressure coefficient around the cylinder obtained from 2D and 3D RANS calculations with experimental data \cite{neuendorf_turbulent_1999}}
\label{fig:cp_3d}
\end{figure}

Following that methodology and by using the skin friction coefficient based on the streamwise wall shear stress on the cylinder, the separation angle is found to be $221.7^{\circ}$. This results is nearly equal to what was found for the 2D RANS were an angle of $221.9^{\circ}$ was found. The Figures \ref{fig:2d_against_3d_rans_1}, \ref{fig:2d_against_3d_rans_2} and \ref{fig:cp_3d} present a more thorough comparison with velocity profiles at different angles, the jet velocity decay, the jet half-thickness and the pressure coefficient around the cylinder. It appears that the 3D simulation is only capable of reproducing the results from the 2D simulation. This is disappointing especially since it has been demonstrated experimentally that 3D structures exist in the flow considered and therefore using a 3D domain should offer a significant improvement over 2D. To be more precise, spanwise and streamwise vortices have been observed in the flow has described by Han et al. \cite{han_streamwise_2004}, with the streamwise vortices thought to have the most impact on the flow development.

Those 3D results lead to believe that the set up used is unable to capture any of the 3-dimensional structures present in the flow. This could be due to the inability of the model to capture such complex phenomenon, by whether not capturing them at all or the structures being dissipated in the flow. To investigate that further the flow around the cylinder obtained will be compared to a version were longitudinal vortices were artificially forced by following the methodology presented by Gross et al. \cite{gross_coanda_2006}. In this paper, disturbances were introduced at the inlet boundary by disturbing the wall-normal and spanwise velocity, the same methodology is used here with the disturbance functions being

\begin{equation}
    v=\sum_{i=1}^2 A_i cos\left(\frac{2 \pi i z}{0.3} \right)sin\left(\frac{\pi y}{b} \right),
\end{equation}

\begin{equation}
    w=\sum_{i=1}^2 A_i sin\left(\frac{2 \pi i z}{0.3} \right)sin\left(\frac{\pi y}{b} \right).
\end{equation}

With $A_i$ being the forcing amplitude and chosen as $0.05$. Results using that forcing methodology cannot be compared to the experimental results directly as an important artificial phenomenon was introduced in the flow, it will, however, permit to see how the longitudinal vortices develop in the predicted flow.\\

In Figure \ref{fig:isocontourlines}, the isocontour lines of the velocity normal to the wall are presented on planes of constant angle $\theta$ for the results with and without forcing. It can first be clearly observed that the flow without forcing is 2 dimensional, this does not, however, permit to give any conclusion as if longitudinal vortices are impacting the flow. In fact, RANS is inherently steady when the streamwise vortices are as stated by Han et al. \cite{han_streamwise_2004} "non-stationary, meandering in both spanwise and radial directions. They may be observed (using flow visualization) at a given instant, but these observations could not  be  translated  into  statistically  meaningful  quantity." The forcing change that by fixing the location of the vortices, they can be well observed. At $\theta=90^{\circ}$ a longitudinal vortices, that appears to be two smaller vortices merging together can be observed and at $\theta=180^{\circ}$ the merging is complete. It can also be observed that as we are moving further downstream the longitudinal vortices increase in size.

\begin{figure}[H]
\centering
\begin{subfigure}{0.5\textwidth}
  \centering
  \includegraphics[width=0.99\textwidth]{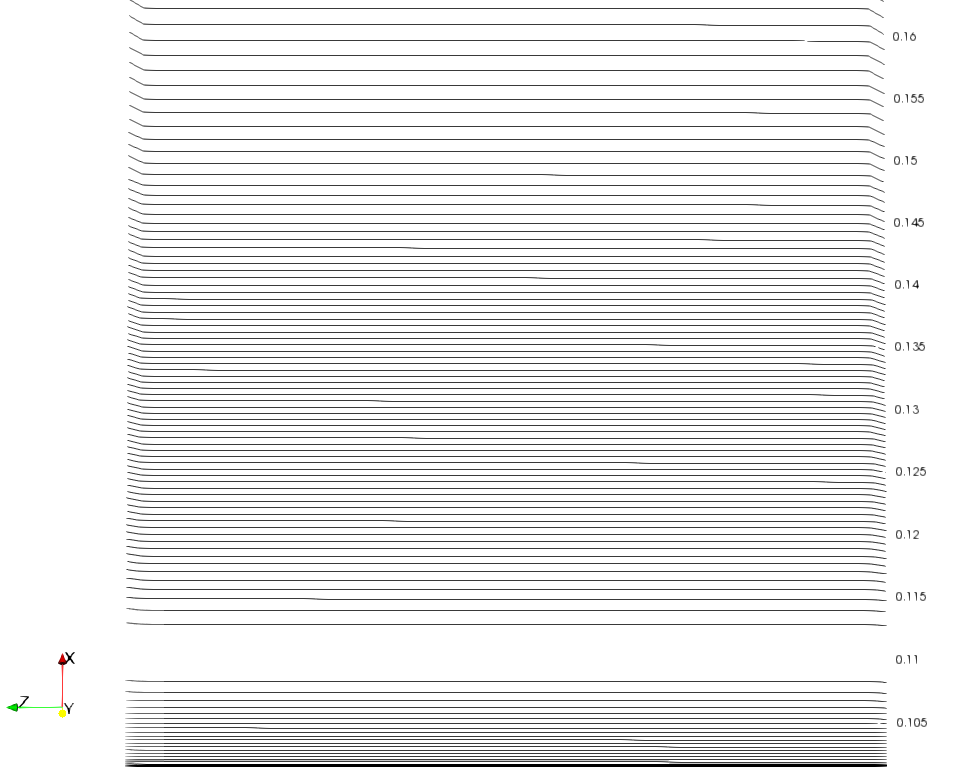}
  \caption{No forcing-$\theta=90^{\circ}$}
  \label{fig:sub1}
\end{subfigure}%
\begin{subfigure}{0.5\textwidth}
  \centering
  \includegraphics[width=0.99\textwidth]{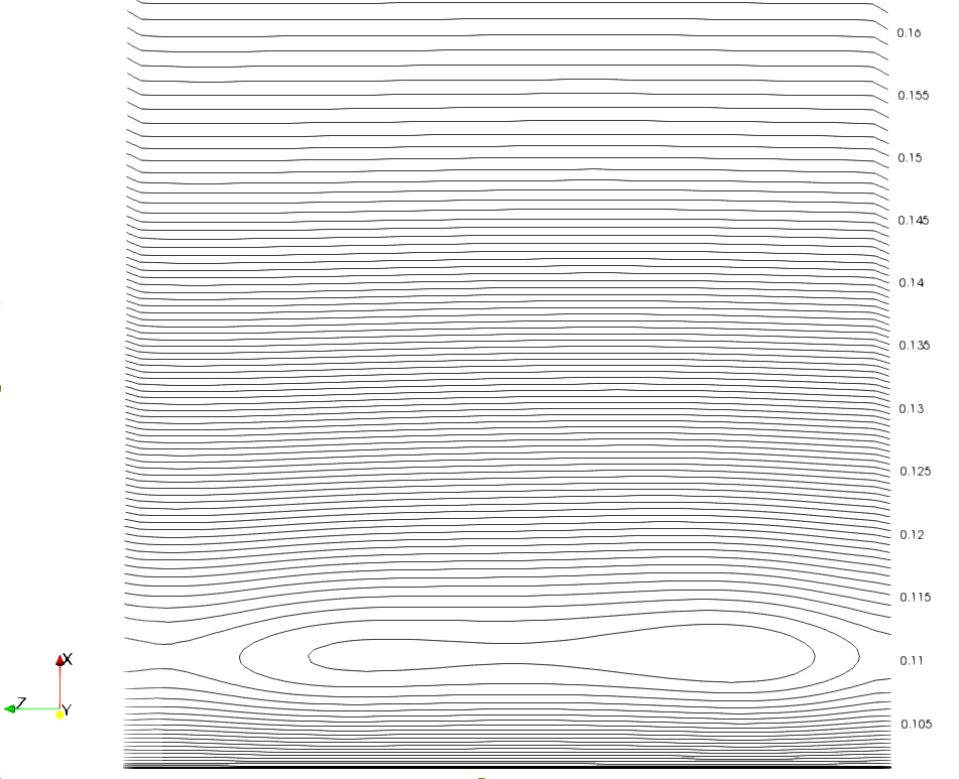}
  \caption{Forcing-$\theta=90^{\circ}$}
  \label{fig:sub2}
\end{subfigure}%

\caption{Isocontourlines of wall-normal velocity in planes of constant $\theta$ with contour level increment of $0.02\ m.s^{-1}$ for the Coand\u{a} flow around the cylinder obtained from 3D RANS calculations}
\label{fig:isocontourlines}
\end{figure}

\begin{figure}[H]
\ContinuedFloat
\begin{subfigure}{0.5\textwidth}
  \centering
  \includegraphics[width=0.99\textwidth]{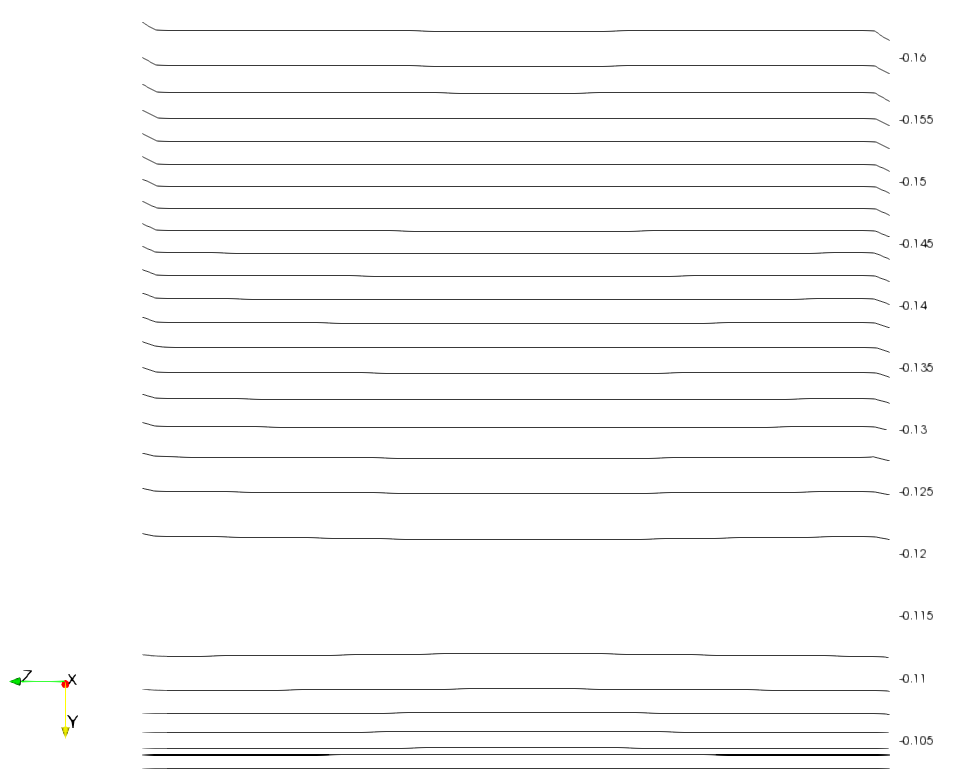}
  \caption{No forcing-$\theta=180^{\circ}$}
  \label{fig:sub1}
\end{subfigure}%
\begin{subfigure}{0.5\textwidth}
  \centering
  \includegraphics[width=0.99\textwidth]{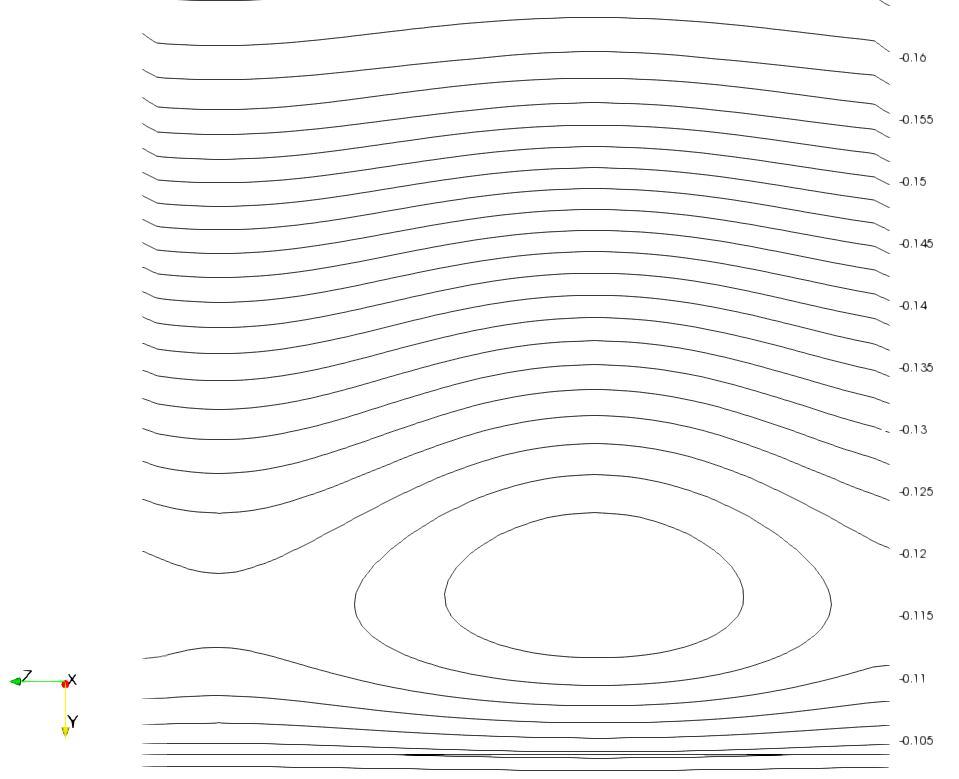}
  \caption{Forcing-$\theta=180^{\circ}$}
  \label{fig:sub2}
\end{subfigure}

\caption[]{Isocontourlines of wall normal velocity in planes of constant $\theta$ with contour level increment of $0.02m.s^{-1}$ for the Coand\u{a} flow around the cylinder obtained from 3D RANS calculations (Continued)}
\label{fig:isocontourlines2}
\end{figure}

\begin{figure}[H]
\centering
\begin{subfigure}{12cm}
  \centering
  \includegraphics[width=8.9cm]{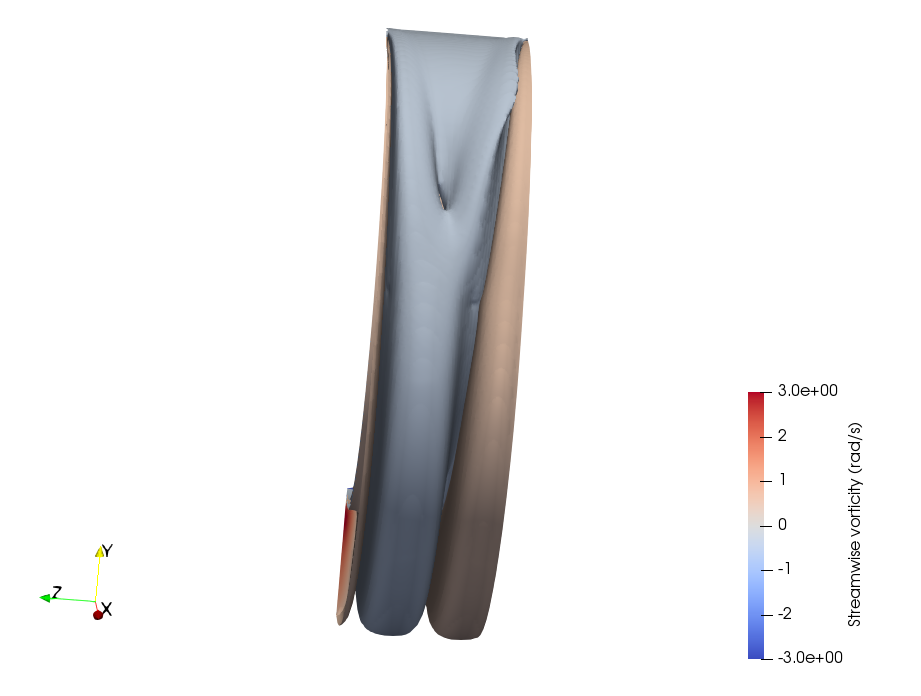}
\end{subfigure}%

\caption{Isosurfaces of streamwise vorticity for the first half of the cylinder obtained from 3D RANS calculations}
\label{fig:isosurface}
\end{figure}

To get a better view on the development of these vortices, the Figure \ref{fig:isosurface} presents the streamwise vorticity around the first half of the cylinder. The vortices merging can be well observed, as well as, their development. If these vortices are artificial and not the ones actually found naturally in the flow, they were created in an attempt to replicate them. When looking at the size and merging of the vortices, it might be possible that the 3D RANS simulation was unable to capture the impact of the streamwise vortices due to a too thin spanwise extent that does not let them develop.

\myparagraph{Conclusion on 3D RANS}
Due to the existence of spanwise and streamwise vortices in the flow, it was thought that using a 3D domain might permit to improve the results obtained with a 2D domain. The results obtained were however disappointing, being nearly similar to the ones obtained in 2D, despite the considerable increase in computational cost induced by adding a dimension. By forcing these structures it was however found that they can develop well using a RANS model. Therefore, a few hypotheses can be considered as to why using a 3D domain couldn't give any improvement in results. It could be because the extent of the spanwise domain is not wide enough to capture these structures accurately, or that more cells are needed in the spanwise direction. Another reason would be that the RANS model itself with its inherent averaging, might not be the appropriate model to capture the effect of such complex and more importantly non-stationary phenomenons. Therefore using a higher fidelity method such as LES might be a better alternative if 2D RANS results are not satisfactory. This is what is going to be discussed in the next section.

\subsubsection{Preliminary results for LES}
From the previous discussion, it appears that RANS might not be accurate enough to capture a Coand\u{a} flow around a cylinder accurately. A LES was therefore attempted, unfortunately, due to computational limitations, a fully developed and statistically stationary state could not be obtained. The methodology and preliminary results obtained will, however, be presented as they could be helpful for eventual future work on the subject. 

\myparagraph{Simulation strategy}
The same flow conditions used for the RANS simulations and presented in Section \ref{sec:2d_rans_strategy} are taken here. The equations are solved using the PISO algorithm as a transient flow is now considered. When it comes to the numerical schemes, they are mostly similar to what was used for RANS, but the turbulence quantities are calculated using the limited linear divergence scheme of OpenFOAM\textsuperscript{\textregistered}. The temporal discretization is done using the backward time scheme that is second order accurate and implicit. It was hoped that by using an implicit scheme, a higher courant number could be taken as the stability condition would be less strict, permitting to reduce the number of iterations. However if the scheme might permit to have stability it does not guarantee accuracy, and in practice using a courant number higher that one was found to give unphysical results. The time step was therefore dynamically chosen to get a maximum courant number in the domain smaller or equal to 0.8. When it comes to the subgrid-scale model, the dynamic $k_{sgs}$ equation model was used. 

\myparagraph{Mesh}
Due to the high computational cost of LES, a proper grid convergence study is generally not carried out. In Georgiadis et al. \cite{georgiadis_large-eddy_2010}, the following non-dimensional spacings for wall-resolved LES are recommended:

\begin{equation}
    50\le \Delta x^+ \le 150,\ \ \ 15\le\Delta z^+ \le 40,\ \ \ y^+ \le 1.
\end{equation}

These recommendations were followed and a structured hexahedral grid containing 41.6 million grids points was generated, the parameters of that grid are given in Table \ref{tab:grid_LES}. 

\begin{table}[H]
\centering
\begin{tabular}{|c|c|}
\hline
$y+$           &0.85 \\ \hline
Expansion rate &1.1  \\ \hline
$\Delta x^+$     &145   \\ \hline
$\Delta z^+$    &40    \\ \hline
\end{tabular}
\caption{Parameters for definition of the grids around the cylinder for the LES simulation}
\label{tab:grid_LES}
\end{table}

Similarly, as for the RANS simulations, the value of $y+$ was calculated from the flat-plate boundary layer theory for turbulent flow and $\Delta x+$ and $\Delta z+$ were defined based on the $y+$ value. The maximum $y+$ found on the cylinder at $t=2.5e-2\ s$ was 0.61, since this is not for a fully developed flow this might not be the final value, but it is a good indicator that the condition $y+<1$ is well respected.

\myparagraph{Boundary and initial conditions}
The boundary conditions are exactly the same as the ones used for the 3D RANS simulations. The inlet is therefore laminar, this was done following the work from \cite{wernz_numerical_2005}, were no turbulent forcing was used at the inlet and still captured interesting phenomenon, although the spanwise extent of $20mm$ was too small to capture the flow accurately.\\
In an attempt to save computational time while improving the stability of the simulation, the result from the 3D RANS simulation without forcing was used as an initial condition for the domain.

\myparagraph{Preliminary results}
The velocity contours at different times are given in Figure \ref{fig:LES_preliminary}. The fact that the jet is not fully developed and a statistically stationary state is not reached, does not allow to make any meaningful conclusion on the results obtained. They are presented as they prove the stability of the simulation, at least up to the last time step calculated and the development of a turbulent jet attaching to the cylinder surface shown in Figure \ref{fig:LES_preliminary_tf} is promising.

\begin{figure}[H]
\centering
\begin{subfigure}{0.5\textwidth}
  \centering
  \includegraphics[width=0.99\textwidth]{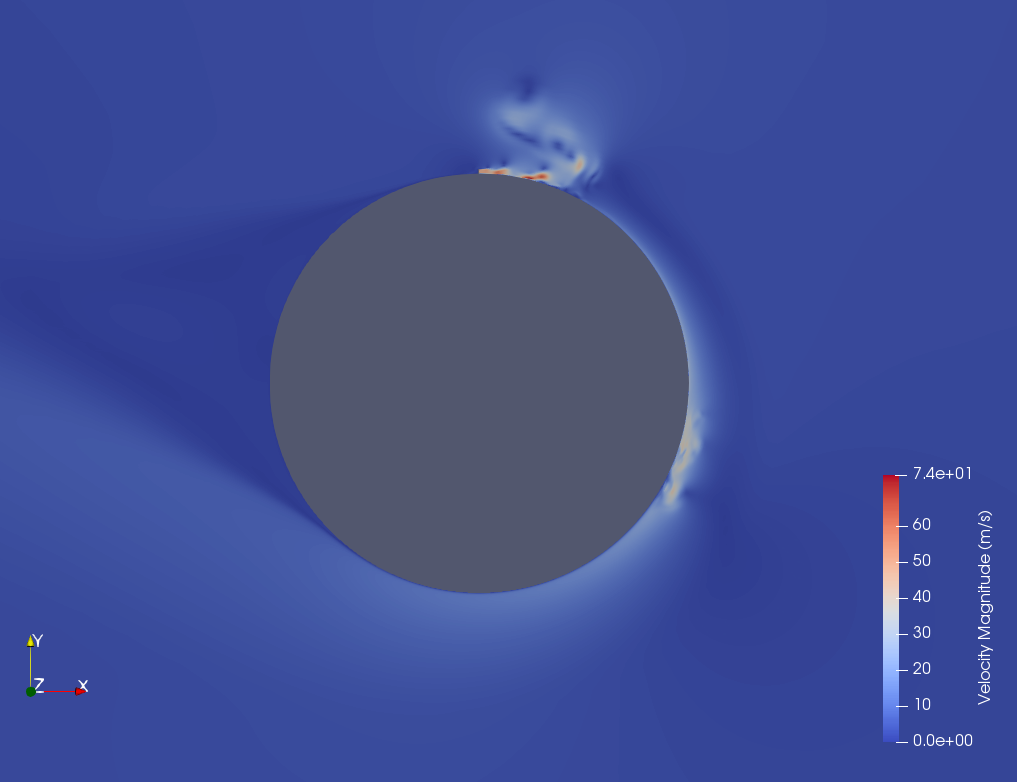}
  \caption{$t=0.5 \times 10^{-2}\ s$}
  \label{fig:sub1}
\end{subfigure}%
\begin{subfigure}{0.5\textwidth}
  \centering
  \includegraphics[width=0.99\textwidth]{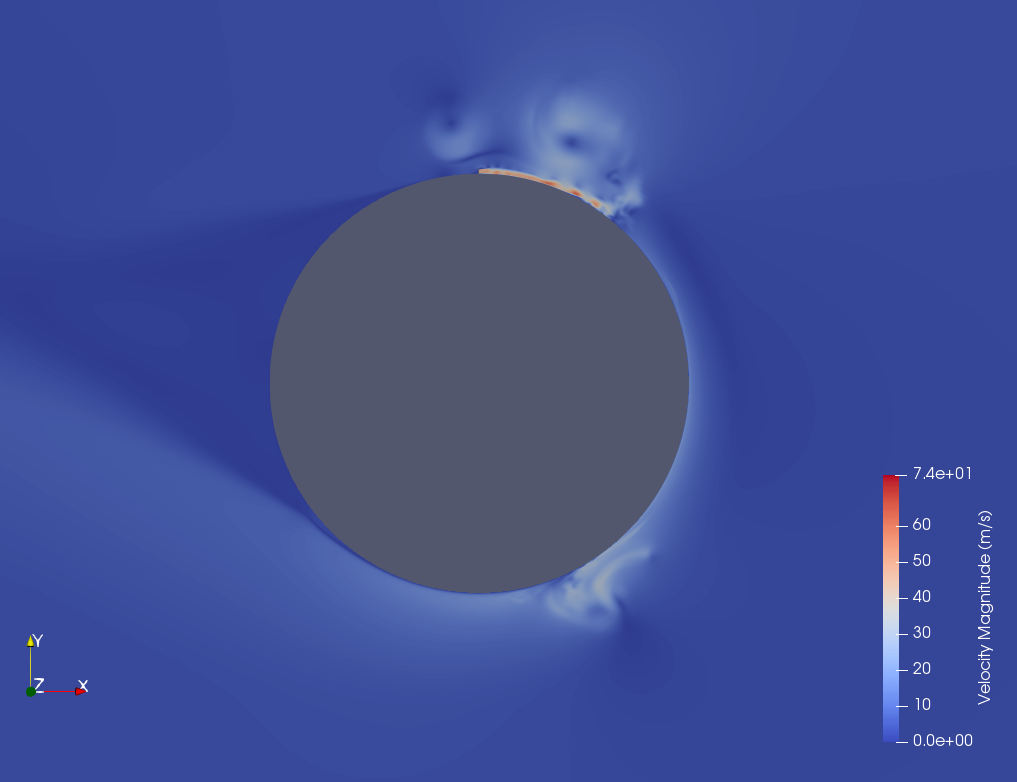}
  \caption{$t=1.0\times 10^{-2}\ s$}
  \label{fig:sub2}
\end{subfigure}%

\caption{Velocity contours around the cylinder from the LES simulation at different times}
\label{fig:LES_preliminary}

\end{figure}

\begin{figure}[H]
\ContinuedFloat
\centering
\begin{subfigure}{0.5\textwidth}
  \centering
  \includegraphics[width=0.99\textwidth]{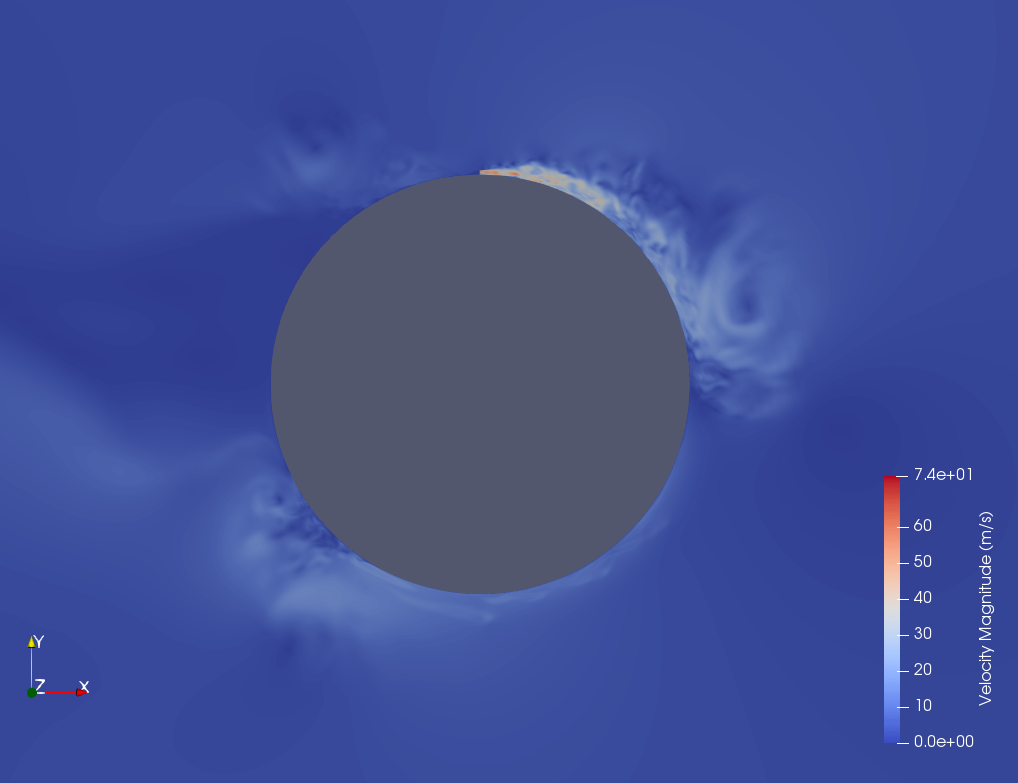}
  \caption{$t=2.5\times 10^{-2}\ s$}
  \label{fig:LES_preliminary_tf}
\end{subfigure}

\caption[]{Velocity contours around the cylinder from the LES simulation at different times (Continued)}
\label{fig:LES_preliminary2}
\end{figure}

\clearpage
\section{Conclusion}
The aim of this thesis was to assess the capability of CFD to capture a Coand\u{a} flow, particularly in the presence of an important streamline curvature.\\

At first, a more fundamental test case was considered with a jet blown on an offset flat plate reproducing the experiment from Gao and Ewing \cite{gao_experimental_2007}. Two-dimensional RANS simulations with both $k-\omega$ SST and $k-\epsilon$ turbulence model were carried out. The results obtained were generally in good agreement with the experiment, in particular, the reattachment location and the velocity profiles near the jet exit were really close to the experimental data. When considering the jet development, that was assessed by the decay of the maximum velocity and the wall jet spreading, the results were a little less accurate with a slight underprediction of these quantities. However, this underprediction did not exceed $20\%$ in all the domain considered. It is, therefore, safe to say that overall the simulation was capable to predict the experimental flow accurately. When comparing the two turbulence model used, $k-\omega$ SST was found to be slightly better and was therefore used for the future RANS simulations.\\

Since most applications of the Coand\u{a} effect are using curved surfaces and Coand\u{a} flows around them have been found challenging to capture numerically, the focus was then shifted on the experiment from Wygnanski et al. \cite{neuendorf_turbulent_2000,neuendorf_turbulent_1999, likhachev_streamwise_2001,cullen_role_2002, han_streamwise_2004, neuendorf_large_2004} of a jet blown tangentially to a cylinder. Again a two-dimensional RANS simulation was carried out with $k-\omega$ SST with and without a Curvature Correction used as turbulence models. The impact of streamline curvature on the accuracy of the simulation was found to be considerable, with a prediction of the experimental flow a lot poorer. Some of the flow characteristics were still well captured, such as the separation location or the velocity profiles for relatively low angles. However, the jet development,the velocity profiles near the separation location and the pressure coefficient on the wall were poorly predicted. When it comes to the benefit of the Curvature Correction, they could not be clearly proven as some flow quantities were more accurate with it (e.g: jet half-thickness) but others were rendered worse (e.g: separation location).\\

Because spanwise and streamwise vortices were found experimentally in the flow \cite{han_streamwise_2004,cullen_role_2002,neuendorf_large_2004}, a RANS simulation using a three-dimensional domain was attempted to improve the prediction of the flow. The results were however disappointing, with the 3D simulation giving results nearly exactly similar to the 2D ones, despite being considerably more computationally expensive. By forcing the longitudinal vortices, it was found that the simulation is capable of capturing them. It, therefore, appears that the poor results obtained might be due to a too thin spanwise extent or limitation of RANS modelling that cannot capture such complex and transient phenomenons. \\

Taking the assumption that RANS might be to limited too capture the flow around the cylinder perfectly, a LES simulation was finally attempted. Unfortunately due to time and computational restraints, meaningful results could not be obtained, but a methodology was proposed.\\

To conclude, it appears that if 2D RANS simulations are capable of capturing a Coand\u{a} flow on a flat plane accurately, if an important streamwise curvature is introduced this is not the case. Some of the flow features are however still captured accurately. Using a 3D domain was not found to permit any improvement despite a considerable higher computational cost. Based on these results it appears that the choice is whether  to use a 2D RANS simulation with a careful validation and by considering the shortcomings of the model or if more computational resources are available to use a turbulence model with a higher order of fidelity such as LES for example. Further work is however needed to first determine for sure that a 3D RANS simulation is unable to improve the flow prediction but also investigate the benefits that a higher order fidelity method could offer, this is discussed in more details in Section \ref{sec:future}. 

\section{Future work}
\label{sec:future}
A proposition on the future work that would be, in the author opinion, the most interesting to carry out are as follows:

\begin{itemize}[label=\textbullet]
    \item A LES simulation was attempted during this thesis, but due to time restrictions could not be completed. Carrying it could permit great improvement in the prediction of the flow, especially for its ability to capture the 3D transient structures in the flow.
    
    \item If LES is a promising option, the important computational cost of such a method is an issue, and thus even if it is capable of capturing the Coand\u{a} flow around a curved surface accurately. In an attempt to reduce computational cost, while still improving the results from RANS, a DES simulation could be attempted. In this method, RANS simulation is used close to the wall and LES far from it. This permits to drastically reduce computational cost, while still capturing instantaneous flow patterns and could be a viable option.
    
    \item Even if 3D RANS gave poor results in this thesis. It would be interesting to carry out such a simulation for a larger spanwise extent and assess if those poor results are due to limitation of RANS or a too thin spanwise extent.
    
    \item In order to assess more precisely the impact of the streamline curvature on the numerical results, a simulation for various curvatures could be carried out. 
    
    \item Finally, if the results from 2D RANS were far from perfect they still were able to capture some of the flow features. In this thesis, a wall-resolved approach was chosen in an attempt to get the best flow prediction possible. An attempt using wall functions could be interesting to see, as they are even cheaper.  \end{itemize}

\clearpage

%%%%%%%%%%%%%%%%%%%%%%%%%%%%%%%%%%%%%%%%%%%%%%%%%%%%%%%%%%%%%%%%%%%%

\clearpage
\section*{References}
\addcontentsline{toc}{section}{References}
\printbibliography[heading=none]

\clearpage
\appendix \label{Appendix}
\addcontentsline{toc}{section}{Appendix}

\section{OpenFOAM\textsuperscript{\textregistered} RANS set-up}
\label{appendixA}

The set-up given here is for the 2D RANS simulation around the cylinder with $k-\omega$ SST turbulence model. 

\subsection{"0" folder}

\subsubsection{initialConditions}

\begin{lstlisting}
/*--------------------------------*- C++ -*----------------------------------*\
| =========                 |                                                 |
| \\      /  F ield         | OpenFOAM: The Open Source CFD Toolbox           |
|  \\    /   O peration     | Version:  v1912                                 |
|   \\  /    A nd           | Website:  www.openfoam.com                      |
|    \\/     M anipulation  |                                                 |
\*---------------------------------------------------------------------------*/

flowVelocity         (0 0 0);
pressure             0;
turbulentKE          1.83e-6; 
turbulentOmega       212915; 

// ************************************************************************* //
\end{lstlisting}

\subsubsection{k}
\begin{lstlisting}
/*--------------------------------*- C++ -*----------------------------------*\
| =========                 |                                                 |
| \\      /  F ield         | OpenFOAM: The Open Source CFD Toolbox           |
|  \\    /   O peration     | Version:  v1912                                 |
|   \\  /    A nd           | Website:  www.openfoam.com                      |
|    \\/     M anipulation  |                                                 |
\*---------------------------------------------------------------------------*/
FoamFile
{
    version     2.0;
    format      ascii;
    class       volScalarField;
    object      k;
}
// * * * * * * * * * * * * * * * * * * * * * * * * * * * * * * * * * * * * * //

#include        "include/initialConditions"

dimensions      [0 2 -2 0 0 0 0];

internalField   uniform $turbulentKE;

boundaryField
{
    inlet
    {
    	type  fixedValue;
    	value $internalField;
    }

    outlet
    {
        type            zeroGradient;
    }

    cylinder
    {
        type            fixedValue;
        value           uniform 0;
    }

    BaseAndTop
    {
	type	empty;
    }
    
}

// ************************************************************************* //
\end{lstlisting}

\subsubsection{nut}
\begin{lstlisting}
/*--------------------------------*- C++ -*----------------------------------*\
| =========                 |                                                 |
| \\      /  F ield         | OpenFOAM: The Open Source CFD Toolbox           |
|  \\    /   O peration     | Version:  v1912                                 |
|   \\  /    A nd           | Website:  www.openfoam.com                      |
|    \\/     M anipulation  |                                                 |
\*---------------------------------------------------------------------------*/
FoamFile
{
    version     2.0;
    format      ascii;
    class       volScalarField;
    location    "0";
    object      nut;
}
// * * * * * * * * * * * * * * * * * * * * * * * * * * * * * * * * * * * * * //

dimensions      [0 2 -1 0 0 0 0];

 internalField   uniform 8.6e-12; //=C_mu^0.25 x k^0.5 x L = 8.66e-12


boundaryField
{
    inlet
    {
        type            calculated;
        value           $internalField;
    }

    outlet
    {
        type            calculated;
        value           $internalField;
    }

    cylinder
    {
        type            fixedValue;
 	value		$internalField;
    }

    BaseAndTop
    {
	type	empty;
    }
}

// ************************************************************************* //
\end{lstlisting}

\subsubsection{omega}

\begin{lstlisting}
/*--------------------------------*- C++ -*----------------------------------*\
| =========                 |                                                 |
| \\      /  F ield         | OpenFOAM: The Open Source CFD Toolbox           |
|  \\    /   O peration     | Version:  v1912                                 |
|   \\  /    A nd           | Website:  www.openfoam.com                      |
|    \\/     M anipulation  |                                                 |
\*---------------------------------------------------------------------------*/
FoamFile
{
    version     2.0;
    format      ascii;
    class       volScalarField;
    object      omega;
}
// * * * * * * * * * * * * * * * * * * * * * * * * * * * * * * * * * * * * * //

#include        "include/initialConditions"

dimensions      [0 0 -1 0 0 0 0];

internalField   uniform $turbulentOmega;

boundaryField
{
    inlet
    {
    	type  fixedValue;
    	value $internalField;
    }

    outlet
    {
        type            zeroGradient;
    }

    cylinder
    {
        type            fixedValue;
        value           uniform 9.2e8;
    }

    BaseAndTop
    {
	type	empty;
    }
}

// ************************************************************************* //
\end{lstlisting}

\subsubsection{p}

\begin{lstlisting}
/*--------------------------------*- C++ -*----------------------------------*\
| =========                 |                                                 |
| \\      /  F ield         | OpenFOAM: The Open Source CFD Toolbox           |
|  \\    /   O peration     | Version:  v1912                                 |
|   \\  /    A nd           | Website:  www.openfoam.com                      |
|    \\/     M anipulation  |                                                 |
\*---------------------------------------------------------------------------*/
FoamFile
{
    version     2.0;
    format      ascii;
    class       volScalarField;
    object      p;
}
// * * * * * * * * * * * * * * * * * * * * * * * * * * * * * * * * * * * * * //

#include        "include/initialConditions"

dimensions      [0 2 -2 0 0 0 0];

internalField   uniform $pressure;

boundaryField
{
    inlet
    {
        type            zeroGradient;
    }

   outlet
   {
        type            fixedValue;
        value           $internalField;
    }

    cylinder
    {
        type            zeroGradient;
    }

    BaseAndTop
    {
	type	empty;
    }
}

// ************************************************************************* //
\end{lstlisting}

\subsubsection{U}

\begin{lstlisting}
/*--------------------------------*- C++ -*----------------------------------*\
| =========                 |                                                 |
| \\      /  F ield         | OpenFOAM: The Open Source CFD Toolbox           |
|  \\    /   O peration     | Version:  v1912                                 |
|   \\  /    A nd           | Website:  www.openfoam.com                      |
|    \\/     M anipulation  |                                                 |
\*---------------------------------------------------------------------------*/
FoamFile
{
    version     2.0;
    format      ascii;
    class       volVectorField;
    location    "0";
    object      U;
}
// * * * * * * * * * * * * * * * * * * * * * * * * * * * * * * * * * * * * * //

#include        "include/initialConditions"

dimensions      [0 1 -1 0 0 0 0];

internalField   uniform $flowVelocity;

boundaryField
{
    inlet
    {
type fixedValue;
value nonuniform List<vector>
73
(
// The velocity at every cell centre on the inlet need to be given here
// Format: (u v w)
);
    }

    outlet
    {
        type            zeroGradient;
    }

    cylinder
    {
        type            noSlip;
    }

    BaseAndTop
    {
	type	empty;
    }
}

// ************************************************************************* //
\end{lstlisting}

\subsection{"constant" folder}

\subsubsection{transportProperties}

\begin{lstlisting}
/*--------------------------------*- C++ -*----------------------------------*\
| =========                 |                                                 |
| \\      /  F ield         | OpenFOAM: The Open Source CFD Toolbox           |
|  \\    /   O peration     | Version:  v1912                                 |
|   \\  /    A nd           | Website:  www.openfoam.com                      |
|    \\/     M anipulation  |                                                 |
\*---------------------------------------------------------------------------*/
FoamFile
{
    version     2.0;
    format      ascii;
    class       dictionary;
    object      transportProperties;
}
// * * * * * * * * * * * * * * * * * * * * * * * * * * * * * * * * * * * * * //

transportModel  Newtonian;

nu              1.5e-05;

// ************************************************************************* //
\end{lstlisting}

\subsubsection{turbulenceProperties}

\begin{lstlisting}
/*--------------------------------*- C++ -*----------------------------------*\
| =========                 |                                                 |
| \\      /  F ield         | OpenFOAM: The Open Source CFD Toolbox           |
|  \\    /   O peration     | Version:  v1912                                 |
|   \\  /    A nd           | Website:  www.openfoam.com                      |
|    \\/     M anipulation  |                                                 |
\*---------------------------------------------------------------------------*/
FoamFile
{
    version     2.0;
    format      ascii;
    class       dictionary;
    object      turbulenceProperties;
}
// * * * * * * * * * * * * * * * * * * * * * * * * * * * * * * * * * * * * * //

simulationType RAS;

RAS
{
    RASModel            kOmegaSST;

    turbulence          on;

    printCoeffs         on;
}

// ************************************************************************* //
\end{lstlisting}

\subsection{"system" folder}

\subsubsection{controlDict}

\begin{lstlisting}
/*--------------------------------*- C++ -*----------------------------------*\
| =========                 |                                                 |
| \\      /  F ield         | OpenFOAM: The Open Source CFD Toolbox           |
|  \\    /   O peration     | Version:  v1912                                 |
|   \\  /    A nd           | Website:  www.openfoam.com                      |
|    \\/     M anipulation  |                                                 |
\*---------------------------------------------------------------------------*/
FoamFile
{
    version     2.0;
    format      ascii;
    class       dictionary;
    object      controlDict;
}
// * * * * * * * * * * * * * * * * * * * * * * * * * * * * * * * * * * * * * //

application     simpleFoam;

startFrom       latestTime;

startTime       0;

stopAt          endTime;

endTime         330000;

deltaT          1;

writeControl    timeStep;

writeInterval   10000;

purgeWrite      0;

writeFormat     ascii;

writePrecision  6;

writeCompression off;

timeFormat      general;

timePrecision   6;

runTimeModifiable true;

// ************************************************************************* //
\end{lstlisting}

\subsubsection{decomposeParDict}

\begin{lstlisting}
/*--------------------------------*- C++ -*----------------------------------*\
| =========                 |                                                 |
| \\      /  F ield         | OpenFOAM: The Open Source CFD Toolbox           |
|  \\    /   O peration     | Version:  v1912                                 |
|   \\  /    A nd           | Website:  www.openfoam.com                      |
|    \\/     M anipulation  |                                                 |
\*---------------------------------------------------------------------------*/
FoamFile
{
    version     2.0;
    format      ascii;
    class       dictionary;
    object      decomposeParDict;
}
// * * * * * * * * * * * * * * * * * * * * * * * * * * * * * * * * * * * * * //

numberOfSubdomains 8;

 method          scotch;

// ************************************************************************* //
\end{lstlisting}

\subsubsection{fvSchemes}

\begin{lstlisting}
/*--------------------------------*- C++ -*----------------------------------*\
| =========                 |                                                 |
| \\      /  F ield         | OpenFOAM: The Open Source CFD Toolbox           |
|  \\    /   O peration     | Version:  v1912                                 |
|   \\  /    A nd           | Website:  www.openfoam.com                      |
|    \\/     M anipulation  |                                                 |
\*---------------------------------------------------------------------------*/
FoamFile
{
    version     2.0;
    format      ascii;
    class       dictionary;
    object      fvSchemes;
}
// * * * * * * * * * * * * * * * * * * * * * * * * * * * * * * * * * * * * * //

ddtSchemes
{
    default         steadyState;
}

gradSchemes
{
    default         Gauss linear;
}

divSchemes
{
    default         none;
    div(phi,U)      bounded Gauss linearUpwind grad(U);  
    div(phi,k)      bounded Gauss upwind;
    div(phi,omega)  bounded Gauss upwind;
    div((nuEff*dev2(T(grad(U))))) Gauss linear;
}

laplacianSchemes
{
   default         Gauss linear corrected;
}

interpolationSchemes
{
    default         linear;
}

snGradSchemes
{
    default         corrected;
}

wallDist
{
    method meshWave;
}

// ************************************************************************* //
\end{lstlisting}

\subsubsection{fvSolution}

\begin{lstlisting}
/*--------------------------------*- C++ -*----------------------------------*\
| =========                 |                                                 |
| \\      /  F ield         | OpenFOAM: The Open Source CFD Toolbox           |
|  \\    /   O peration     | Version:  v1912                                 |
|   \\  /    A nd           | Website:  www.openfoam.com                      |
|    \\/     M anipulation  |                                                 |
\*---------------------------------------------------------------------------*/
FoamFile
{
    version     2.0;
    format      ascii;
    class       dictionary;
    object      fvSolution;
}
// * * * * * * * * * * * * * * * * * * * * * * * * * * * * * * * * * * * * * //

solvers
{
    p
    {
          solver          GAMG;
          smoother        GaussSeidel;
          tolerance       1e-8;
          relTol          0.05;

          minIter         3;              // a minimum number of iterations
          maxIter         100;            // limitation of iterions number
          smoother        DIC;            // setting for GAMG
          nPreSweeps      1;              // 1 for pd, set to 0 for all other!
          nPostSweeps     2;              // 2 is fine
          nFinestSweeps   2;              // 2 is fine
          scaleCorrection true;           // true is fine
          directSolveCoarsestLevel false; // false is fine
          cacheAgglomeration on;          // on is fine; set to off, if dynamic
                                        // mesh refinement is used!
          nCellsInCoarsestLevel 500;      // 500 is fine,
                                        // otherwise sqrt(number of cells)
          agglomerator    faceAreaPair;   // faceAreaPair is fine
          mergeLevels     1;              // 1 is fine
   }

    Phi
    {
        $p;
    }

    U
    {
        solver          PBiCG;
	preconditioner	DILU;
        tolerance       1e-8;
        relTol          0.1;
        nSweeps         2;
    }

    k
    {
        solver          PBiCG;
	preconditioner	DILU;
        tolerance       1e-8;
        relTol          0.1;
        nSweeps         2;
    }

    omega
    {
        solver          PBiCG;
	preconditioner	DILU;
        tolerance       1e-10;
        relTol          0.1;
        nSweeps         2;
    }

}

SIMPLE
{
    nNonOrthogonalCorrectors 0;
    consistent yes;

    residualControl
    {
	p	1e-7;
	U	1e-7;
	k	1e-7;
    omega	1e-7;
     }
}

potentialFlow
{
    nNonOrthogonalCorrectors 10;
}

relaxationFactors
// Start the simulation with low values and increase them after to reduce number of 
// iterations needed for convergence while keeping the simulation stable.
{

    fields
    {
	default 0;
	p	0.7;
     }

    equations
    {
	default 0;
        U               0.8;
        k               0.8;
        omega           0.8;
    }
}

cache
{
    grad(U);
}

// ************************************************************************* //
\end{lstlisting}

\section{OpenFOAM\textsuperscript{\textregistered} LES set-up}
\label{appendixB}
This set-up is for the LES attempt to capture the Coand\u{a} around the cylinder.

\subsection{"0" folder}

\subsubsection{initialConditions/k/nut/p/U}
Please refer to the files given in Appendix \ref{appendixA}. The only difference is the use of cyclic boundary conditions instead of empty for top and bottom of the domain since a 3-dimensional domain is considered here.

\subsubsection{s}
\begin{lstlisting}
/*--------------------------------*- C++ -*----------------------------------*\
| =========                 |                                                 |
| \\      /  F ield         | OpenFOAM: The Open Source CFD Toolbox           |
|  \\    /   O peration     | Version:  v1912                                 |
|   \\  /    A nd           | Website:  www.openfoam.com                      |
|    \\/     M anipulation  |                                                 |
\*---------------------------------------------------------------------------*/
FoamFile
{
    version     2.0;
    format      ascii;
    class       volScalarField;
    object      s;
}
// * * * * * * * * * * * * * * * * * * * * * * * * * * * * * * * * * * * * * //

dimensions      [0 0 0 0 0 0 0];

internalField   uniform 0;

boundaryField
{
    inlet
    {
        type            fixedValue;
        value           uniform 1;
    }

    outlet
    {
        type            fixedValue;
        value           uniform 0;
    }

    cylinder
    {
        type            zeroGradient;
    }

    top2
    {
        type            cyclic;
    }

    bottom2
    {
        type            cyclic;
    }
    
}

// ************************************************************************* //
\end{lstlisting}

\subsection{"constant" folder}

\subsubsection{transportProperties}
Please refer to the file given Appendix \ref{appendixA}.

\subsubsection{turbulentProperties}

\begin{lstlisting}
/*--------------------------------*- C++ -*----------------------------------*\
| =========                 |                                                 |
| \\      /  F ield         | OpenFOAM: The Open Source CFD Toolbox           |
|  \\    /   O peration     | Version:  v1912                                 |
|   \\  /    A nd           | Website:  www.openfoam.com                      |
|    \\/     M anipulation  |                                                 |
\*---------------------------------------------------------------------------*/
FoamFile
{
    version     2.0;
    format      ascii;
    class       dictionary;
    object      turbulenceProperties;
}
// * * * * * * * * * * * * * * * * * * * * * * * * * * * * * * * * * * * * * //

simulationType LES;

LES
{
    LESModel            dynamicKEqn;

    turbulence          on;

    printCoeffs         on;

    delta		cubeRootVol;

    dynamicKEqnCoeffs
    {
        filter simple;
    }

    cubeRootVolCoeffs
    {
        deltaCoeff      1;
    }

}
	
// ************************************************************************* //
\end{lstlisting}

\subsection{"system" folder}

\subsubsection{controlDict}

\begin{lstlisting}
/*--------------------------------*- C++ -*----------------------------------*\
| =========                 |                                                 |
| \\      /  F ield         | OpenFOAM: The Open Source CFD Toolbox           |
|  \\    /   O peration     | Version:  v1912                                 |
|   \\  /    A nd           | Website:  www.openfoam.com                      |
|    \\/     M anipulation  |                                                 |
\*---------------------------------------------------------------------------*/
FoamFile
{
    version     2.0;
    format      ascii;
    class       dictionary;
    object      controlDict;
}
// * * * * * * * * * * * * * * * * * * * * * * * * * * * * * * * * * * * * * //

application     pimpleFoam;
//piso algorithm is actually used by using 1 for nOuterCorrectors in fvSolution 
//file. Using the pimpleFoam application permits to use a dynamic time step based on 
//the courant number.

startFrom       latestTime;

startTime       0;

stopAt          endTime;

endTime         4.0e-2;

deltaT          3e-6;

writeControl    timeStep;

writeInterval   50;

purgeWrite      1;

writeFormat     ascii;

writePrecision  6;

writeCompression off;

timeFormat      general;

timePrecision   6;

runTimeModifiable true;

adjustTimeStep	yes;

maxCo		0.8; 

maxDeltaT       3e-6;

functions
{
    probes
    {
        type            probes;
        libs            (sampling);
	name		probes;
        writeControl    timeStep;
        writeInterval   1;

        fields
        (
            p
	        U
        );

        probeLocations
        (
            (0.1167 0 0.01)
	        (0.1167 0 0.03)
            (0.1167 0 0.05)
            (0 -0.1622 0.01)
            (0 -0.1622 0.03)
	        (0 -0.1622 0.05)
	        (-0.2239 -0.18795 0.01)
            (-0.2239 -0.18795 0.03)
	        (-0.2239 -0.18795 0.05)
        );

    }
}

// ************************************************************************* //
\end{lstlisting}

\subsubsection{decomposePar}

\begin{lstlisting}
/*--------------------------------*- C++ -*----------------------------------*\
| =========                 |                                                 |
| \\      /  F ield         | OpenFOAM: The Open Source CFD Toolbox           |
|  \\    /   O peration     | Version:  v1912                                 |
|   \\  /    A nd           | Website:  www.openfoam.com                      |
|    \\/     M anipulation  |                                                 |
\*---------------------------------------------------------------------------*/
FoamFile
{
    version     2.0;
    format      ascii;
    class       dictionary;
    object      decomposeParDict;
}
// * * * * * * * * * * * * * * * * * * * * * * * * * * * * * * * * * * * * * //

numberOfSubdomains 64;

 method          scotch;

// ************************************************************************* //
\end{lstlisting}

\subsubsection{fvSchemes}

\begin{lstlisting}
/*--------------------------------*- C++ -*----------------------------------*\
| =========                 |                                                 |
| \\      /  F ield         | OpenFOAM: The Open Source CFD Toolbox           |
|  \\    /   O peration     | Version:  v1912                                 |
|   \\  /    A nd           | Website:  www.openfoam.com                      |
|    \\/     M anipulation  |                                                 |
\*---------------------------------------------------------------------------*/
FoamFile
{
    version     2.0;
    format      ascii;
    class       dictionary;
    object      fvSchemes;
}
// * * * * * * * * * * * * * * * * * * * * * * * * * * * * * * * * * * * * * //

ddtSchemes
{
    default         backward;
}

gradSchemes
{
    default         Gauss linear;
}

divSchemes
{
    default         none;
    div(phi,U)      bounded Gauss linearUpwind grad(U);  
    div(phi,k)      Gauss limitedLinear 1;
    div(phi,s)      bounded Gauss limitedLinear 1;
    div((nuEff*dev2(T(grad(U))))) Gauss linear;
}

laplacianSchemes
{
   default         Gauss linear corrected;
}

interpolationSchemes
{
    default         linear;
}

snGradSchemes
{
    default         corrected;
}

wallDist
{
    method meshWave;
}

// ************************************************************************* //
\end{lstlisting}

\subsubsection{fvSolution}

\begin{lstlisting}
/*--------------------------------*- C++ -*----------------------------------*\
| =========                 |                                                 |
| \\      /  F ield         | OpenFOAM: The Open Source CFD Toolbox           |
|  \\    /   O peration     | Version:  v1912                                 |
|   \\  /    A nd           | Website:  www.openfoam.com                      |
|    \\/     M anipulation  |                                                 |
\*---------------------------------------------------------------------------*/
FoamFile
{
    version     2.0;
    format      ascii;
    class       dictionary;
    object      fvSolution;
}
// * * * * * * * * * * * * * * * * * * * * * * * * * * * * * * * * * * * * * //

solvers
{

      p
      {
          solver          PCG;
          preconditioner  DIC;
          tolerance       1e-06;
          relTol          0.05;
      }
      
// Surprisingly PCG solver was found to be faster than GAMG.

    pFinal
    {
        $p;
        smoother        DICGaussSeidel;
        tolerance       1e-05;
        relTol          0;
	    maxIter         1000;
    }	

    U
    {
        solver          PBiCG;
	    preconditioner	DILU;
        tolerance       1e-10;
        relTol          0;
        nSweeps         2;
    }

    UFinal
    {
        $U;
    }

    k
    {
        solver          PBiCG;
	    preconditioner	DILU;
        tolerance       1e-10;
        relTol          0;
        nSweeps         2;
    }

    kFinal
    {
        $k;
    }

    s
    {
        solver          PBiCG;
	    preconditioner	DILU;
        tolerance       1e-8;
        relTol          0;
        nSweeps         2;
    }

    sFinal
    {
        $s;
    }

}

PIMPLE
{
    nCorrectors     2;
    nNonOrthogonalCorrectors 0;
    nOuterCorrectors 1;
}

potentialFlow
{
    nNonOrthogonalCorrectors 10;
}

cache
{
    grad(U);
}

// ************************************************************************* //
\end{lstlisting}

\end{document}